\documentclass[%
reprint,
superscriptaddress,
 amsmath,amssymb,
 aps,
prd,
]{revtex4-2}

\usepackage{graphicx}
\usepackage{dcolumn}
\usepackage{bm}

\begin{document}


\title{Numerical simulation of sky localization for LISA-TAIJI joint observation}

\author{Gang Wang}
\email[Gang Wang: ]{gwang@shao.ac.cn, gwanggw@gmail.com}
\affiliation{Shanghai Astronomical Observatory, Chinese Academy of Sciences, Shanghai 200030, China}
\affiliation{School of Astronomy and Space Science, University of Chinese Academy of Sciences, Beijing 100049, China}
\author{Wei-Tou Ni}
\email[Wei-Tou Ni: ]{weitou@gmail.com}
\affiliation{National Astronomical Observatories, Chinese Academy of Sciences, Beijing, China}
\affiliation{State Key Laboratory of Magnetic Resonance and Atomic and Molecular Physics, Wuhan Institute of Physics and Mathematics, Chinese Academy of Sciences, Wuhan 430071, China}
\affiliation{Department of Physics, National Tsing Hua University, Hsinchu, Taiwan, 30013, ROC}

\author{Wen-Biao Han}
\email[Wen-Biao Han: ]{wbhan@shao.ac.cn}
\affiliation{Shanghai Astronomical Observatory, Chinese Academy of Sciences, Shanghai 200030, China}
\affiliation{School of Astronomy and Space Science, University of Chinese Academy of Sciences, Beijing 100049, China}
\author{Shu-Cheng Yang}
\affiliation{Shanghai Astronomical Observatory, Chinese Academy of Sciences, Shanghai 200030, China}
\affiliation{School of Astronomy and Space Science, University of Chinese Academy of Sciences, Beijing 100049, China}
\author{Xing-Yu Zhong}
\affiliation{Shanghai Astronomical Observatory, Chinese Academy of Sciences, Shanghai 200030, China}
\affiliation{School of Astronomy and Space Science, University of Chinese Academy of Sciences, Beijing 100049, China}

\date{\today}

\begin{abstract}
LISA is considered to be launched alongside the Athena to probe the energetic astrophysical processes. LISA can determine the direction of sources for Athena's follow-up observation. As another space gravitational wave mission, TAIJI is expected to be launched in the 2030s. The LISA-TAIJI network would provide abundant merits for sources understanding. In this work, we simulate the joint LISA-TAIJI observations for gravitational waves from coalescing supermassive black hole binaries and monochromatic sources. By using the numerical mission orbits, we evaluate the performances of sky localization for various time-delay interferometry channels. For 30 days observation until coalescence, the LISA-TAIJI network in optimal operation can localize all simulated binary sources, $(10^7,\ 3.3 \times 10^6)\ M_\odot$, $(10^6,\ 3.3 \times 10^5)\ M_\odot$ and $(10^5,\ 3.3 \times 10^4)\ M_\odot$ at redshift $z=2$, in 0.4 deg$^2$ (field of view of Wide Field Imager on Athena). The angular resolution can be improved by more than 10 times comparing to LISA or TAIJI single detector at a given percentage of population. The improvements for monochromatic sources at 3 mHz and 10 mHz are relatively moderate in one-year observation. The precision of sky localization could be improved by a factor of 2 to 4 comparing to single LISA at a given percentage of sources. For a simulated 90 days observation for monochromatic waves, the LISA-TAIJI network still represents a considerable localization advantage which could be more than 10 times better.

\end{abstract}


\maketitle


\section{Introduction \label{sec:level1}}

Gravitational wave (GW) astronomy was unveiled by the first detection of advanced LIGO and Virgo -- GW150914 \cite{Abbott:2016blz}. Tens of detections have been confirmed during the advanced LIGO and Virgo O1 and O2 runs including the additional new candidates \cite[and references therein]{TheLIGOScientific:2016pea,LIGOScientific:2018mvr,Venumadhav:2019lyq,Nitz:2019hdf}. The GW detection, GW170817, and electromagnetic counterparts observations from binary neutron stars merger opened a new multi-messenger era \cite[and references therein]{TheLIGOScientific:2017qsa,GBM:2017lvd}. Scores of candidates have been preliminarily identified by advanced LIGO and Virgo with upgraded sensitivity during the O3 run from April 2019 to March 2020 \cite{gracedb}, and a new binary neutron stars detection, GW190425, has been confirmed \cite{Abbott:2020uma}. KAGRA started its observation from February 25th, 2020 \cite{KAGRA}, and ground-based network  would enter the four interferometers era.

The space missions targeting for low-frequency GW detections are expected to be launched in the 2030s. With LISA Pathfinder successfully demonstrated the drag-free technology and GRACE follow-on testing the laser metrology, the essential technologies for LISA mission are reaching maturity and paving the path for its launch in the 2030s \cite{2017arXiv170200786A,Armano:2016bkm,Armano:2018kix,Abich:2019cci}. In China, two space missions are proposed to detect the GW in low frequency band -- TAIJI \cite{Hu:2017mde} and TianQin \cite{Luo:2015ght}. The TAIJI mission is considered to use a LISA-like formation in a heliocentric orbit, and TianQin is considered to be in a geocentric orbit. TAIJI and TianQin launched their respective pathfinders, TAIJI-1 in August \cite{Luo:2020} and TianQin-1 in December \cite{Tianqin-1} of 2019, testing the first stage of development of their accelerometers, metrology, and others technologies in space.

As the L3 mission of ESA's Cosmic Vision 2015–25 plan, LISA is considered to be promoted to an earlier launch time alongside the L2 mission -- Athena \cite{ESAnews}. Athena is a space mission with new generation spatially-resolved X-ray spectroscopy and deep wide-field X-ray spectral imaging system, and targeting to observe the energetic processes in the universe \cite{2013arXiv1306.2307N}. The joint observations from LISA and Athena can boost the understanding of the basic physics and astrophysics of the universe \cite{Colpi:2019}. One apparent scenario could be that LISA detects the GWs from sources and identify their sky locations for Athena's follow-up observations \cite{2020NatAs.tmp....4M}.

The angular resolutions of LISA for binary black holes and monochromatic sources have been studied since the original LISA proposed \cite{Cutler:1997ta,1997CQGra..14.1507P,1998AIPC..456...95C}, with more detailed studies thereafter \cite[and references therein]{Vecchio:2004ec,2007PhRvD..76j4016A,2008arXiv0806.1591B,2016PhRvL.116w1102S}. The investigations usually adopted the averaged sensitivity and treated LISA as two independent interferometers. This approach should be enough to estimate the average performance. 
\citet{Vallisneri:2012np} pointed out that SNR from an average sensitivity is not accurate for individual sources, and parameter estimation could be imprecise due to response to a GW signal vary with frequency and orientation. On the other hand, time-delay interferometry (TDI) is required for LISA-like missions to suppress the laser frequency noise. The response of a TDI combination to a GW signal is formed by combining the measurements from time shifted laser links. \citet{2010PhRvD..81f4014M,2011PhRvD..84f4003M} investigated the TDI responses, especially in the optimal-A, E and T channels, for GW detections and parameter estimations.

To simulate the TDI effects in the LISA measurements, multiple simulators were developed.
The Synthetic LISA is developed to simulate the LISA science process at the level of scientific and technical requirements \cite{Vallisneri:2004bn}. 
LISACode is developed to bridge the gap between the basic principles of LISA and sophisticated simulator \cite{2008PhRvD..77b3002P}, and its successor LISANode is newly developed to adapt to the updated LISA measurements \cite{2019PhRvD..99h4023B}.

Besides the angular resolution investigations for solo LISA mission, \citet{Crowder:2005nr,2016PhRvD..94h1101T,Tinto:2017kre} also explored the sky localization improvement with LISA and other presumed detector(s).
TAIJI is expected to be launched in the 2030s and could have observation period overlapped with LISA and Athena. It is considered to be a heliocentric orbit mission in front of Earth by around 20$^\circ$. With large distance separation, LISA and TAIJI could form a network and bring abundant merits for the GW observations.
By using the average sensitivity, \citet{Ruan:2019tje} estimated that the joint observation of LISA and TAIJI could significantly improve the sky localization precision for the supermassive black hole (SMBH) binaries. 

In this work, by using the numerical orbit achieved and a simulator developed, we numerically calculate the sky- and polarization- averaged sensitivities for various TDI channels of LISA and TAIJI based on the updated missions' requirements. Then we investigate the sky localization merits from LISA and TAIJI joint observation for the coalescing SMBH binaries and monochromatic sources considering the response of specific TDI configurations.
We organize our paper as follows. In Section \ref{sec:orbit}, we introduce the numerical orbits we achieved for LISA and TAIJI missions. In Section \ref{sec:TDI}, we describe the noise assumptions, TDI response to GW signals in TDI channels and their corresponding sensitivities. In Section \ref{sec:SMBH}, we report the simulation on the sky localization performance of LISA-TAIJI joint observation for coalescing SMBH binaries. In Section \ref{sec:mono}, we present the simulation of sky localization for the monochromatic sources by using the LISA and TAIJI network. And we recapitulate our conclusions and discussions in Section \ref{sec:conclusions}.

\section{Numerical Mission Orbit} \label{sec:orbit}

\begin{figure}[htb]
\includegraphics[width=0.45\textwidth]{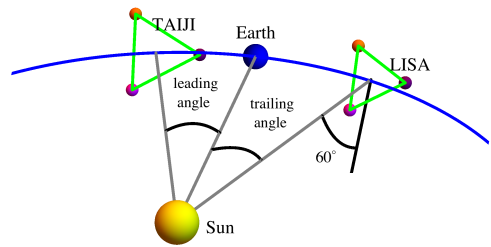}
\caption{\label{fig:orbit-scheme} The schematic of LISA and TAIJI orbit.}
\end{figure}

LISA mission originally proposed an equilateral triangle constellation with $5 \times 10^6$ km laser links to detect the low-frequency GWs \cite{LISA2000}. The formation has a $60^\circ$ inclination angle with respect to the ecliptic plane and trails the Earth by about $20^\circ$. After NASA's withdrawal from ESA-NASA LISA collaboration in 2011, the LISA mission evolved to a down-scaled mission, so called eLISA/NGO, which was supported by the European countries (France, Germany, Italy, the Netherlands, Spain, Switzerland and the UK) and ESA. In 2017, LISA team proposed a new LISA configuration with $2.5 \times 10^6$ km arm length and $20^\circ$ trailing angle \cite{2017arXiv170200786A}. The schematic is shown in Fig. \ref{fig:orbit-scheme}.

TAIJI program is a LISA-like mission proposed by Chinese Academy of Sciences \cite{Hu:2017mde,Luo:2020}. Three spacecraft (S/C) form a triangle constellation with $3 \times 10^6$ km arm length. The constellation could be leading or trailing the Earth by $20^\circ$ on the heliocentric orbit. Considering the merits of LISA-TAIJI joint observation with large separation, the current mission trend is to be localized in front of Earth as shown in Fig. \ref{fig:orbit-scheme} \citet{Ruan:2019tje}.

The studies of LISA(-like) orbit have been going on for more than two decades. \citet{Folkner:1997hn} estimated the LISA orbit stability by using an analytical method. \citet{Dhurandhar:2004rv} and \citet{Nayak:2006zm} formulated and optimized the LISA(-like) orbit formation analytically by using Clohessy-Wiltshire frame. \citet{Wu:2019thj} expanded the orbital equations in \citet{Dhurandhar:2004rv} to the higher order of the eccentricity and applied to the TAIJI mission orbit. In two companion papers, \citet{Yi:2008zza} employed coorbital restricted problem to design the LISA orbit and \citet{Li:2008al} introduced an algorithm to optimize the orbit numerically.

In our previous works \cite{Wang:2011,Wang:2014aea,Wang:2012ce,Wang:2012te,Dhurandhar:2011ik,Wang:2014cla,Wang:2017aqq,Wang:2019ipi}, we developed a workflow to design and optimize the drag-free orbit of a GW space mission by using an ephemeris framework, and calculate the time difference of the TDI paths. In our recent work \citep{Wang:2017aqq}, we worked out a LISA mission orbit based on the new requirements in its proposal \citep{2017arXiv170200786A}. Furthermore, we applied the procedures to TAIJI mission and achieved an optimized orbit at the same trailing angle and epoch \cite{Wang:2017aqq}. In this work, along the workflow we developed, we obtain an optimized orbit for TAIJI mission in front of the Earth by $20^\circ$ to study the merits of the LISA-TAIJI joint observation.

We refer the specific formulation and optimization method for LISA-like orbits to our previous works \cite{Wang:2012ce,Dhurandhar:2011ik,Wang:2017aqq}. In this section, we briefly summarize our procedures of workflow as follow: 1) determine the starting time of a mission; 2) obtain the initial condition by using the formula (e.g. Equation (2.1)-(2.6) in \cite{Wang:2012ce}); 3) put the initial conditions of the celestial bodies and S/C to the ephemeris framework and calculate the orbit by using numerical integration; 4) adjust the orbital periods and eccentricities of the S/C iteratively to meet the mission requirements; and 5) expand the achieved orbit to the backward time direction and truncate the time period which satisfied the requirements that can prolong the effective mission time.

To calculate the orbit accurately, the interactions considered in our ephemeris framework CGC3.0 include, 1) the Newtonian and first-order post-Newtonian interaction between Sun, major planets, Pluto, Moon, Ceres, Pallas and Vesta; 2) the figure interactions between Sun/Earth/Moon and others as point mass bodies; 3) the perturbations from selected 340 asteroids and 4) the tide effects on Moon from the Earth. For a starting time, the CGC3.0 firstly read constants, positions and velocities of the celestial bodies from DE430, and then integrate the orbits of the celestial bodies/spacecraft by using the equations of motion considered. The heliocentric distance of Earth calculated by CGC3.0 is less than 0.3 m in 10 years comparing to the ephemeris DE430 \cite{DE430}.

\begin{figure*}[htb]
\includegraphics[width=0.48\textwidth]{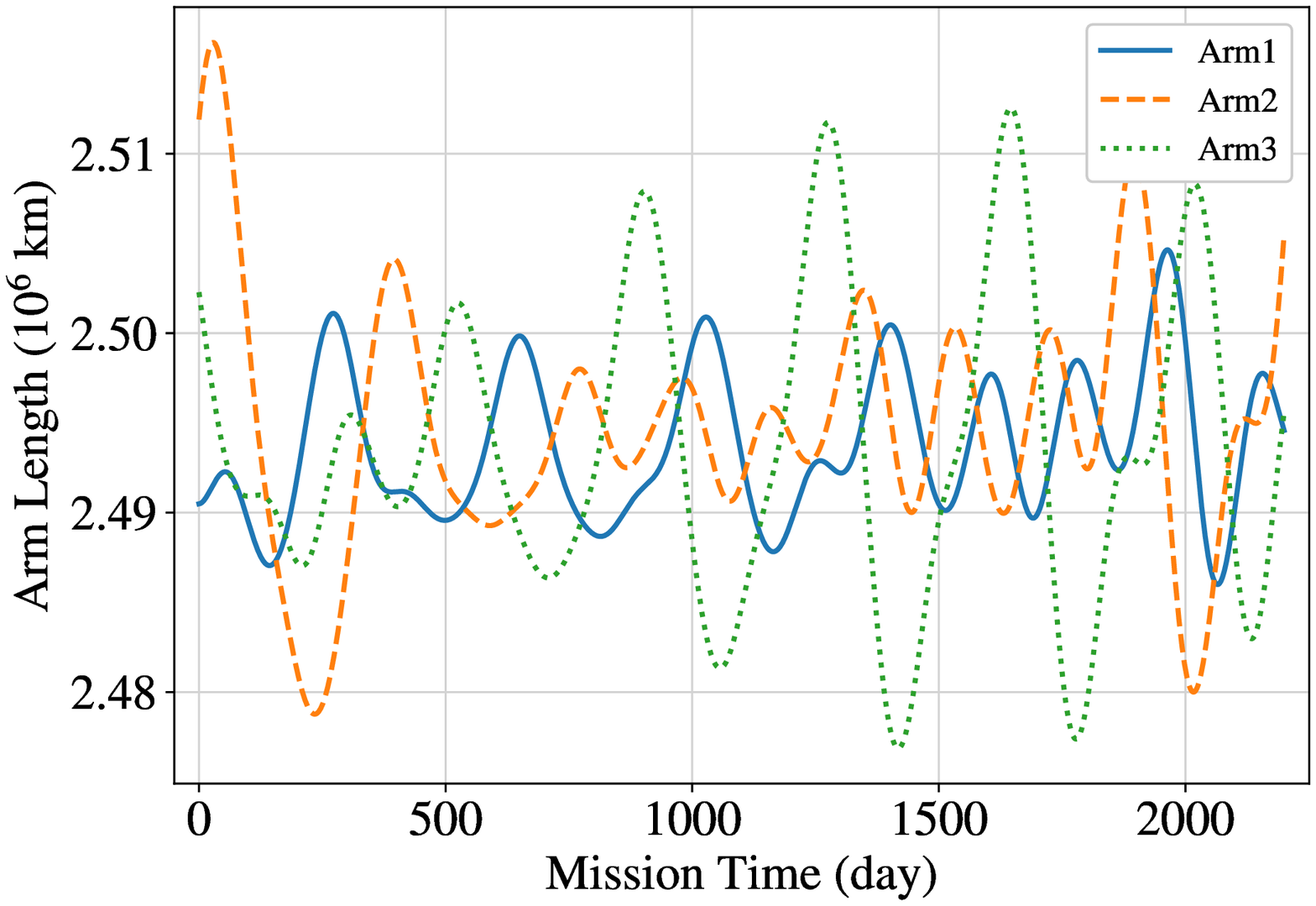} 
\includegraphics[width=0.48\textwidth]{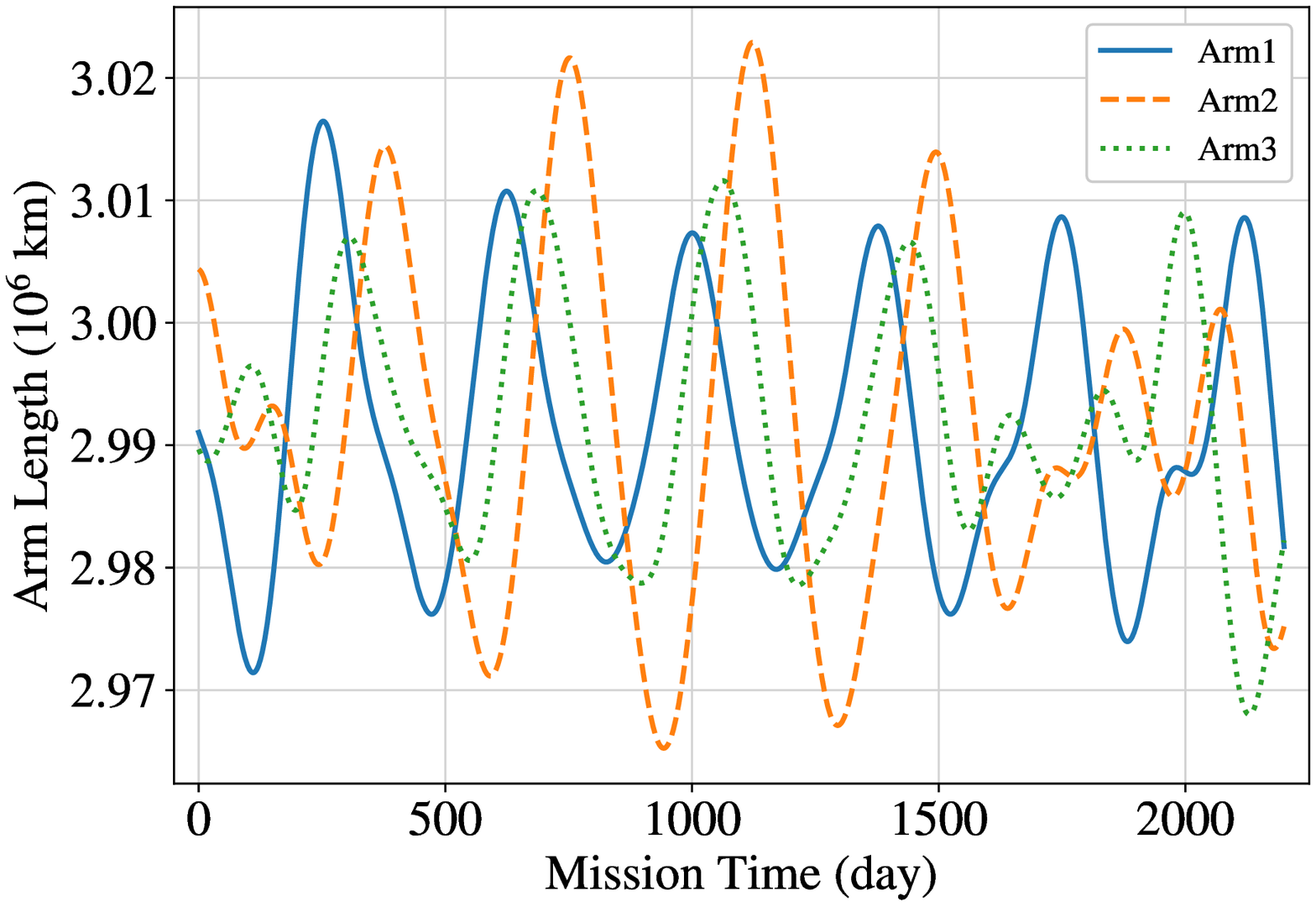} \\
\includegraphics[width=0.48\textwidth]{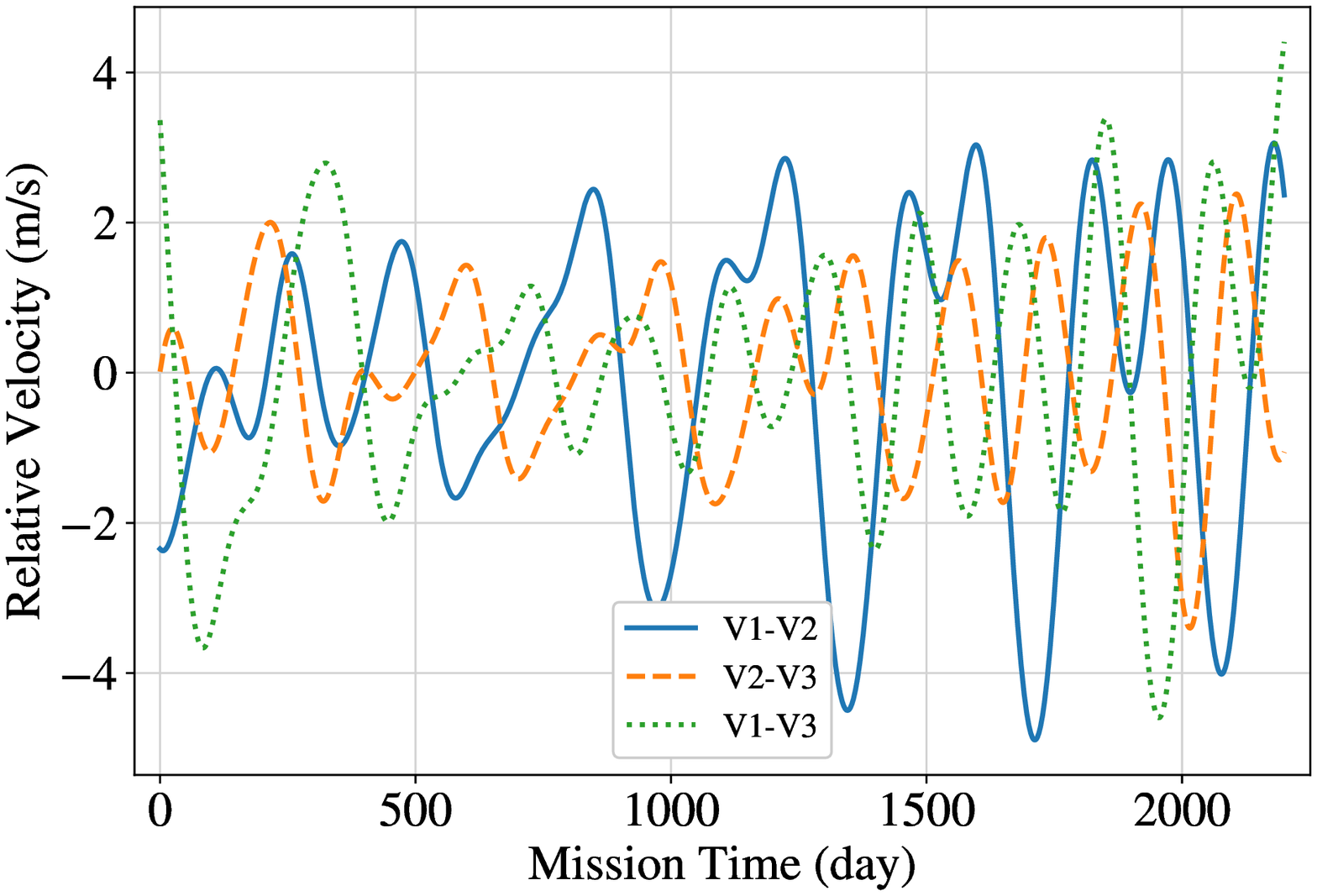} 
\includegraphics[width=0.48\textwidth]{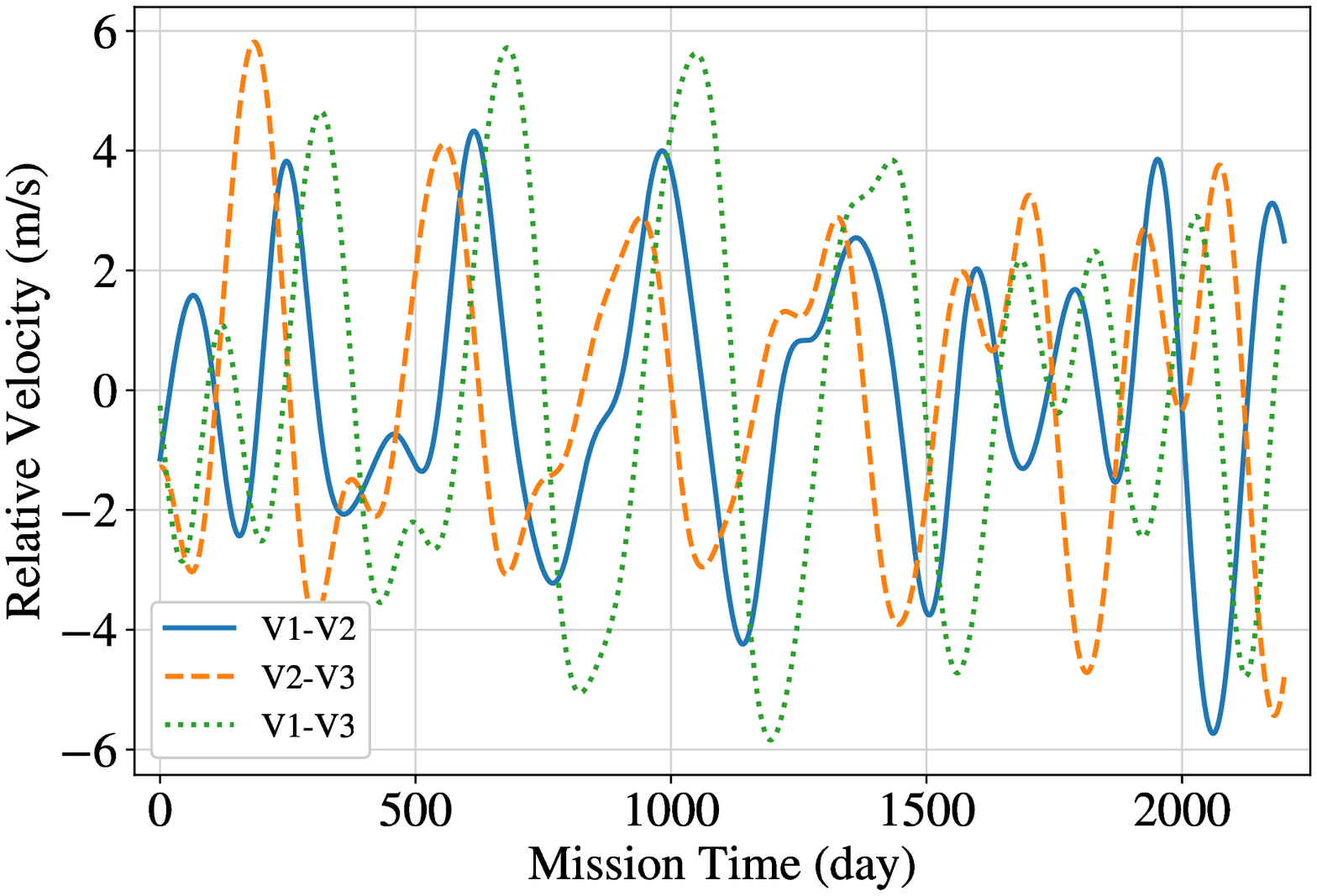}
\caption{\label{fig:orbit} The numerical LISA and TAIJI orbit achieved and used in the hereafter simulation. The arm length variations with mission time are shown in the upper row (upper left for LISA and right for TAIJI). The relative velocities between S/Cs are shown in the lower row (lower left for LISA and lower right for TAIJI).}
\end{figure*}

Based on the orbital requirements for new LISA \cite{2017arXiv170200786A}, we set the optimization criteria: 1) the relative velocities between S/C should be smaller than 5 m/s; and 2) the effective mission duration should be longer than 4 years. For TAIJI mission, due to the larger arm length, we set the criteria of the relative velocities to be less than 6 m/s. The optimized orbits achieved for LISA and TAIJI missions are shown in Fig. \ref{fig:orbit}. These numerical orbits start on March 22nd, 2028 (JD2461853.0) and can maintain in required status for 2200 days ($\sim$6 years). The relevant equations for the ephemeris calculation and for choosing the initial conditions of S/C are assembled in Appendix A.

It is needed to emphasize that the orbit shown in Fig. \ref{fig:orbit} are the geodesics of three S/C in the solar system barycentric (SSB) coordinate system without considering an orbital maneuver. \citet{Bender:2013mma} proposed the periodic maneuver control for eLISA mission orbit and evaluated the requirement for the thruster. \citet{Halloin:2017tzi} presented an optimization method for the numerical orbit including the periodic orbital maneuvers. A group of orbital maneuvers could be implemented to our achieved orbit to extend the mission duration. In our recent work, we also explored the possibility that to maintain a constant arm triangular constellation by using thruster propulsion \cite{Wang:2019ipi}.

\section{Time-Delay Interferometry} \label{sec:TDI}

For a LISA-like space mission, TDI is essential to achieve the sensitivity goal. The principle of TDI is to properly time shift and combine the data streams to suppress the laser frequency noise and preserve the GW signals. Two generations of TDI combinations were proposed and studied for LISA mission depending on their demands to cancel the laser frequency noise \cite[and references therein]{1999ApJ...527..814A,2000PhRvD..62d2002E,2001CQGra..18.4059A,Prince:2002hp,Dhurandhar:2002zcl,Cornish:2003tz,2003PhRvD..67l2003T,Vallisneri:2005ji,Tinto:2014lxa,Vallisneri:2007xa,Vallisneri:2012np}. The first-generation TDI combination can cancel the frequency noise in a stationary unequal-arm interferometry, and the second-generation TDI combinations are targeting to further cancel the frequency noise in a moving interferometer. In this work, we focus on the investigation of the first-generation TDI combinations.

\begin{figure*}[htb]
\includegraphics[width=0.22\textwidth]{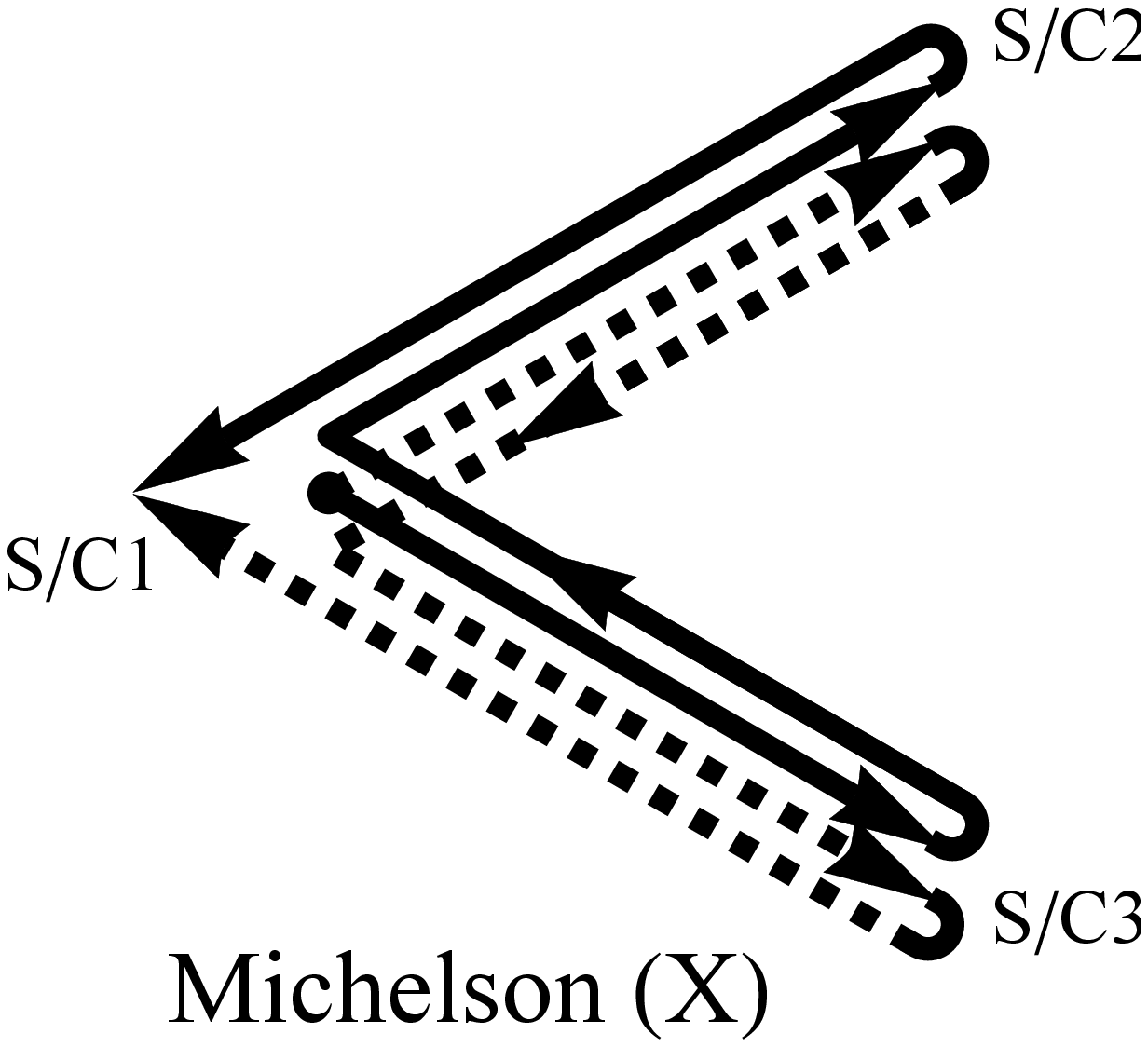}
\includegraphics[width=0.2\textwidth]{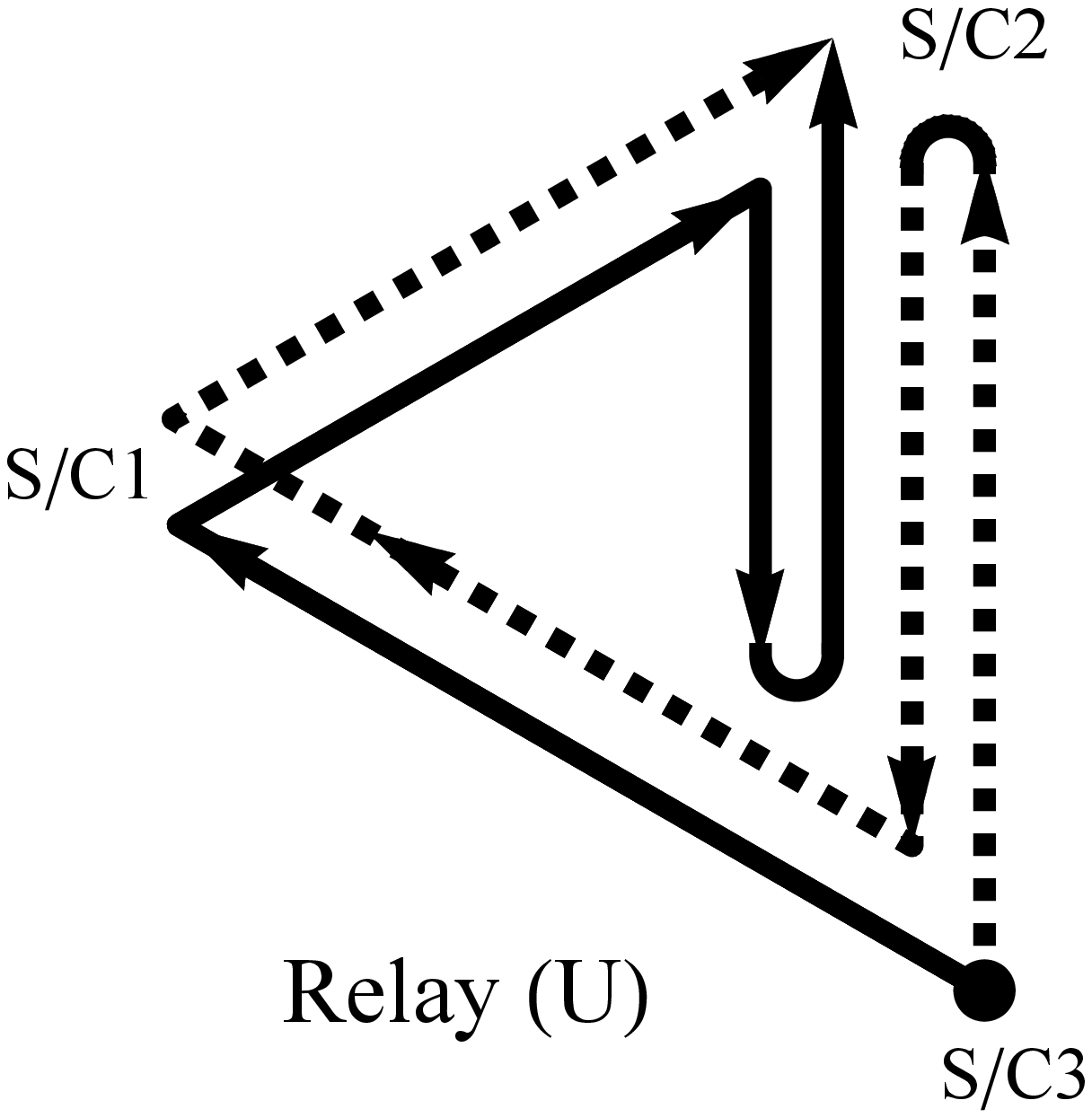}
\includegraphics[width=0.22\textwidth]{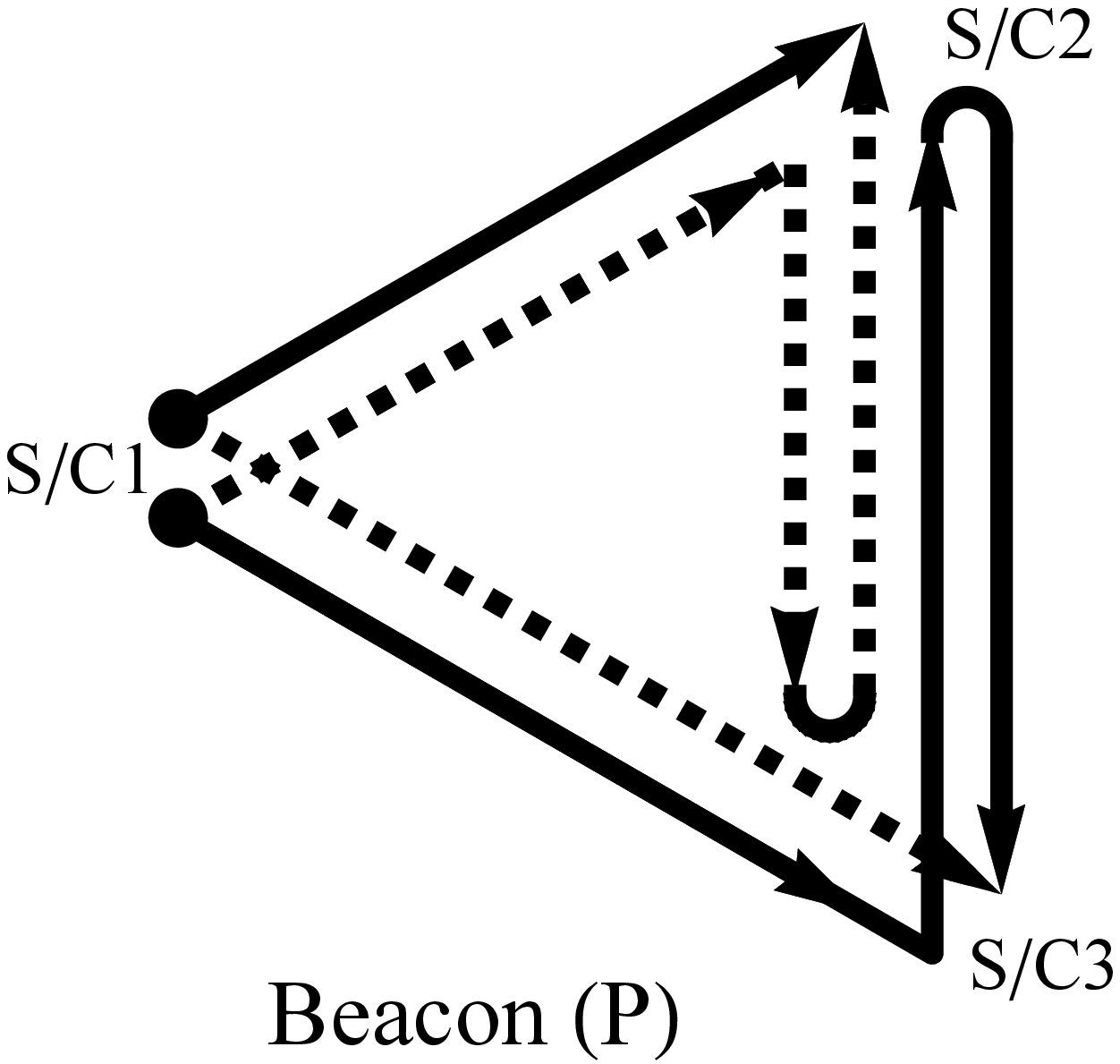}
\includegraphics[width=0.2\textwidth]{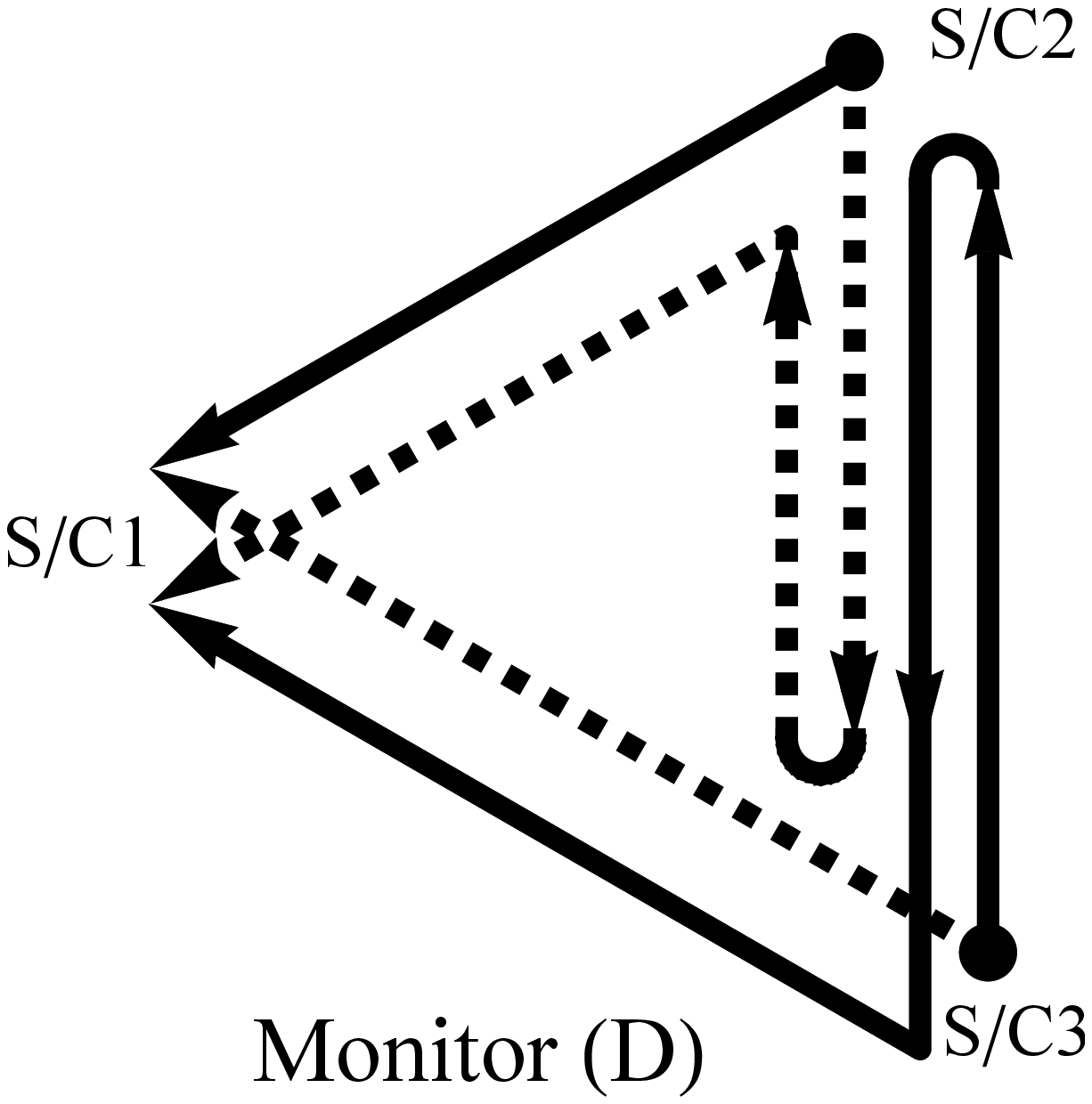}
\caption{\label{fig:TDI1st_path} The diagram of selected channels (Michelson-X, Relay-U, Beacon-P and Monitor-D) from first-generation TDI configurations \cite{Vallisneri:2005ji}. }
\end{figure*}

The first-generation TDI combinations are classified as five configurations which are Sagnac $(\alpha, \beta, \gamma)$, unequal-arm Michelson (X, Y, Z), Relay (U, V, W), Beacon (P, Q, R) and Monitor (E, F, G), plus symmetric Sagnac $\zeta$. To distinguish from the optimal TDI channel E hereinafter, we use the (D, F, G) to indicate the Monitor channels instead. The three channels in each configuration are obtained by cyclical permutation of the spacecraft indexes. Except for the Sagnac, the first channel for each configuration are shown in Fig. \ref{fig:TDI1st_path}.

\subsection{Noise PSDs of TDI channels} \label{secsub:TDI_noise}

\begin{figure}[htb]
\includegraphics[width=0.48\textwidth]{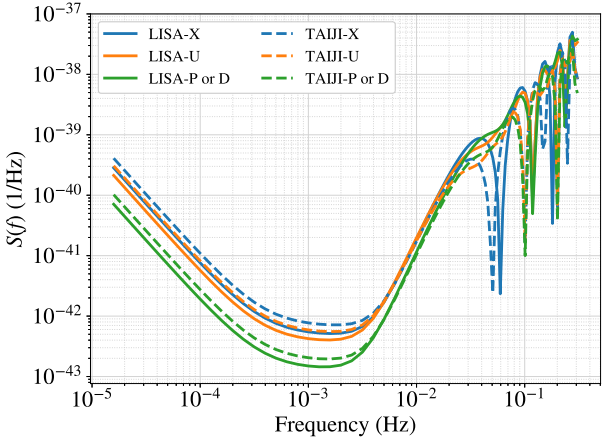} \\
\includegraphics[width=0.48\textwidth]{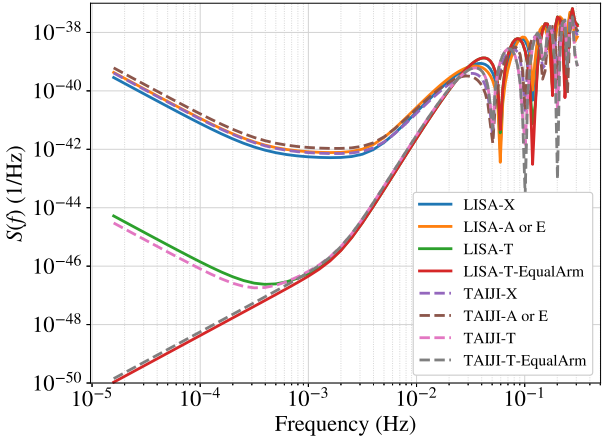}
\caption{\label{fig:TDI1st_noise}The PSD of selected first-generation TDI channels (Michelson-X, Relay-U, Beacon-P, and Monitor-D) for LISA and TAIJI (upper panel), the PSD of the optimal TDI channels (A, E and T) and X channel (lower panel). The mission-T curves show the PSD of T channel at the starting time of the numerical orbits, and the mission-T-EqualArm curves show the PSD of T channel by using the Eq. \eqref{eq:Sn_AET} which is for fully equal arm configuration. (The curves of P and D channels are fully overlapped in the upper panel, as well as the A and E channel in the lower panel.) }
\end{figure}

Under the assumption that laser frequency noise has been canceled by TDI combinations, only acceleration noise and optical path noise are considered in our simulation. The current requirements of acceleration noise $S_{\rm acc}$ for LISA and TAIJI are the same which are \cite{2017arXiv170200786A,Luo:2020},
\begin{equation}
 S^{1/2}_{\rm acc} \leq 3 \times 10^{-15} \frac{\rm m/s^2}{\sqrt{\rm Hz}} \sqrt{1 + \left(\frac{0.4 {\rm mHz}}{f} \right)^2 }  \sqrt{1 + \left(\frac{f}{8 {\rm mHz}} \right)^4 }.
\end{equation}
The requirement of optical path noise $S_{\rm op}$ for LISA and TAIJI mission are slightly different:
\begin{equation}
\begin{aligned}
 S^{1/2}_{\rm op, LISA} &\leq 10 \times 10^{-12} \frac{\rm m}{\sqrt{\rm Hz}} \sqrt{1 + \left(\frac{2 {\rm mHz}}{f} \right)^4 },  \\
S^{1/2}_{\rm op, TAIJI} &\leq 8 \times 10^{-12} \frac{\rm m}{\sqrt{\rm Hz}} \sqrt{1 + \left(\frac{2 {\rm mHz}}{f} \right)^4 }.
 \end{aligned}
\end{equation}
By assuming there is no correlation between the different test masses and optical benches, the power spectral density (PSD) functions of selected first-generation TDI configurations/channels are \cite{1999ApJ...527..814A,2000PhRvD..62d2002E,2001CQGra..18.4059A,Vallisneri:2007xa,Vallisneri:2012np}
\begin{equation} \label{eq:Sn_XUD}
\begin{aligned}
S_{\rm X} (f) = & 16 S_{\rm op} (f) \sin^2 x \\ &+ 16 S_{\rm acc} (f) (3 + \cos 2x ) \sin^2 x , \\
S_{\rm U} (f)  = & 8  S_{\rm op}  (4 + 4  \cos x +  \cos 2x )  \sin^2 (x/2) \\ &+ 16  S_{\rm acc}  (5 + 5   \cos x + 2   \cos 2x )  \sin^2 (x/2) , \\
S_{\rm D} (f) = & S_{\rm P} (f) =  8  S_{\rm op}  (3 + 2  \cos x)  \sin^2 (x/2) \\ &+ 16  S_{\rm acc}  (3 +  \cos x )  \sin^2 (x/2) ,
\end{aligned}
\end{equation}
where $x = 2 \pi f L$. The PSD curves of these four TDI channels for LISA and TAIJI are shown in Fig. \ref{fig:TDI1st_noise} upper panel.

A group of optimal TDI channels could be obtained by linear combinations of the three channels in one TDI configuration, (e.g. Sagnac TDI configuration used in \citet{Prince:2002hp}). In our investigation, we apply the Michelson TDI configuration, (X, Y, Z), to compose the optimal TDI configurations as used in \citet{Vallisneri:2007xa},
\begin{equation} \label{eq:optimalTDI}
 {\rm A} =  \frac{ {\rm Z} - {\rm X} }{\sqrt{2}} , \quad {\rm E} = \frac{ {\rm X} - 2 {\rm Y} + {\rm Z} }{\sqrt{6}} , \quad {\rm T} = \frac{ {\rm X} + {\rm Y} + {\rm Z} }{\sqrt{3}}.
\end{equation}
The corresponding PSD functions of noise for these three channels are
\begin{equation} \label{eq:Sn_AET}
\begin{aligned}
 S_{\rm A} =  S_{\rm E} = & 8  S_{\rm op}  (2+\cos x ) \sin^2 x \\ & + 16  S_{\rm acc}  (3+ 2 \cos x + \cos 2x )  \sin^2 x, \\
 S_{\rm T} = & 16  S_{\rm op} (1 - \cos x ) \sin^2 x \\ & + 128  S_{\rm acc} \sin^2 x \sin^4(x/2).
\end{aligned}
\end{equation}
The PSD curves for LISA and TAIJI are shown in Fig. \ref{fig:TDI1st_noise} lower panel. It's needed to emphasize that the PSD function in Eq. \eqref{eq:Sn_XUD} and \eqref{eq:Sn_AET} are achieved by assuming the fully equal arm configuration. For an unequal arm configuration, the PSD of T channel could be divergent from expected at lower frequency band  as shown by the curves of mission-T and mission-T-EqualArm.

\subsection{Response of TDI to GW} \label{secsub:TDI_response}

\citet{1975GReGr...6..439E} formulated the response to a GW signal in a single link Doppler measurement. A TDI channel is connected by multiple laser links, the final response to a GW signal is combined by each single link. Furthermore, the response of each TDI configuration to a same GW source could vary with relative position, orientation and frequency. Referring to the formulas in \citet{Vallisneri:2012np}, we briefly state the response functions applied in our simulation. Different from using their analytical orbit, we adopt the numerical orbit achieved in Section \ref{sec:orbit}.

For a GW source locate at direction $(\lambda, \beta)$, where $\lambda $ and $\beta$ is the ecliptic longitude and latitude in the SSB coordinates, the propagation vector is described as
\begin{equation}
 \hat{k}  = -( \cos \lambda \cos \beta, \sin \lambda \cos \beta ,  \sin \beta ).
\end{equation} 
The polarization tensors of plus and cross of GW signal are
\begin{equation}
\begin{aligned}
{\rm e}_{+} & \equiv \mathcal{O}_1 \cdot 
\begin{pmatrix}
1 & 0 & 0\\
0 & -1 & 0 \\
0 & 0 & 0
\end{pmatrix}
\cdot \mathcal{O}^T_1 ,
\ \ 
{\rm e}_{\times} & \equiv \mathcal{O}_1 \cdot 
\begin{pmatrix}
0 & 1 & 0\\
1 & 0 & 0 \\
0 & 0 & 0
\end{pmatrix}
\cdot \mathcal{O}^T_1,
\end{aligned}
\end{equation}
with
\begin{widetext}
\begin{equation}
\mathcal{O}_1 =
\begin{pmatrix}
\sin \lambda \cos \psi - \cos \lambda \sin \beta \sin \psi & -\sin \lambda \sin \psi - \cos \lambda \sin \beta \cos \psi & -\cos \lambda \cos \beta  \\
     -\cos \lambda \cos \psi - \sin \lambda \sin \beta \sin \psi & \cos \lambda \sin \psi - \sin \lambda \sin \beta \cos \psi & -\sin \lambda \cos \beta  \\
         \cos \beta \sin \psi & \cos \beta \cos \psi & -\sin \beta 
\end{pmatrix},
\end{equation}
where $\psi$ is the polarization angle. 

The response in a single link measurement from S/C$i$ to $j$ is described by \cite{Vallisneri:2007xa}
\begin{equation}
\begin{aligned}
y^{\rm{GW}}_{ij} (f) =&  \frac{ (1 + \cos^2 \iota ) \hat{n}_{ij} \cdot {\rm e}_+ \cdot \hat{n}_{ij} + i (- 2 \cos \iota ) \hat{n}_{ij} \cdot {\rm e}_\times \cdot \hat{n}_{ij} }{4 (1 - \hat{n}_{ij} \cdot \hat{k} ) } 
 \times \left[  \exp( 2 \pi i f (L_{ij} + \hat{k} \cdot p_i ) ) -  \exp( 2 \pi i f  \hat{k} \cdot p_j )  \right] ,
\end{aligned}
\end{equation}
\end{widetext}
where $\hat{n}_{ij}$ is the unit vector from SC$i$ to $j$, $L_{ij}$ is the arm length along the SC$i$ and $j$, $p_i$ is the position of the S/C$i$ in the SSB coordinates. and $\iota$ is the inclination of the GW source from the line of sight.

\begin{figure}[htb]
\includegraphics[width=0.48\textwidth]{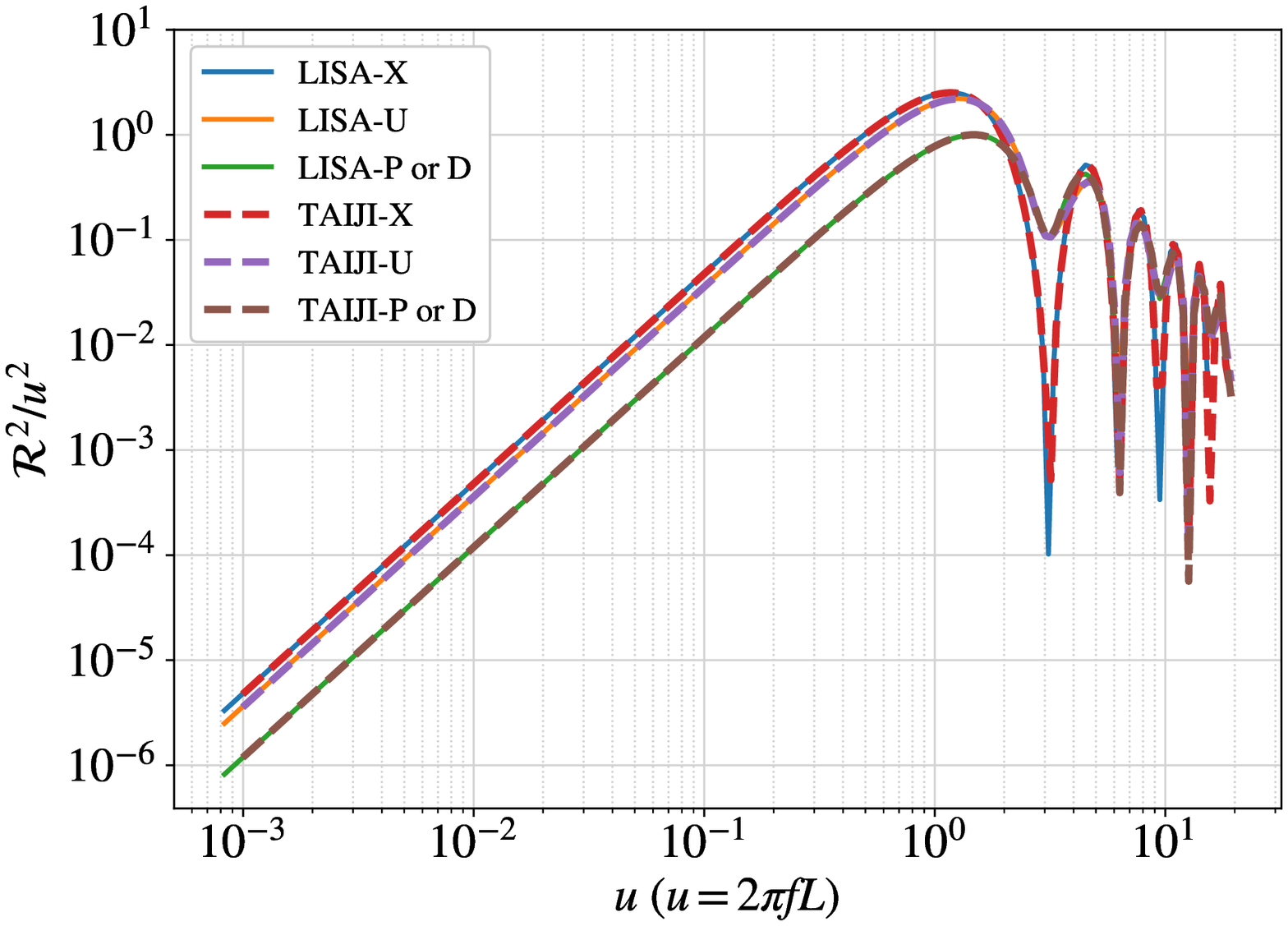}
\includegraphics[width=0.48\textwidth]{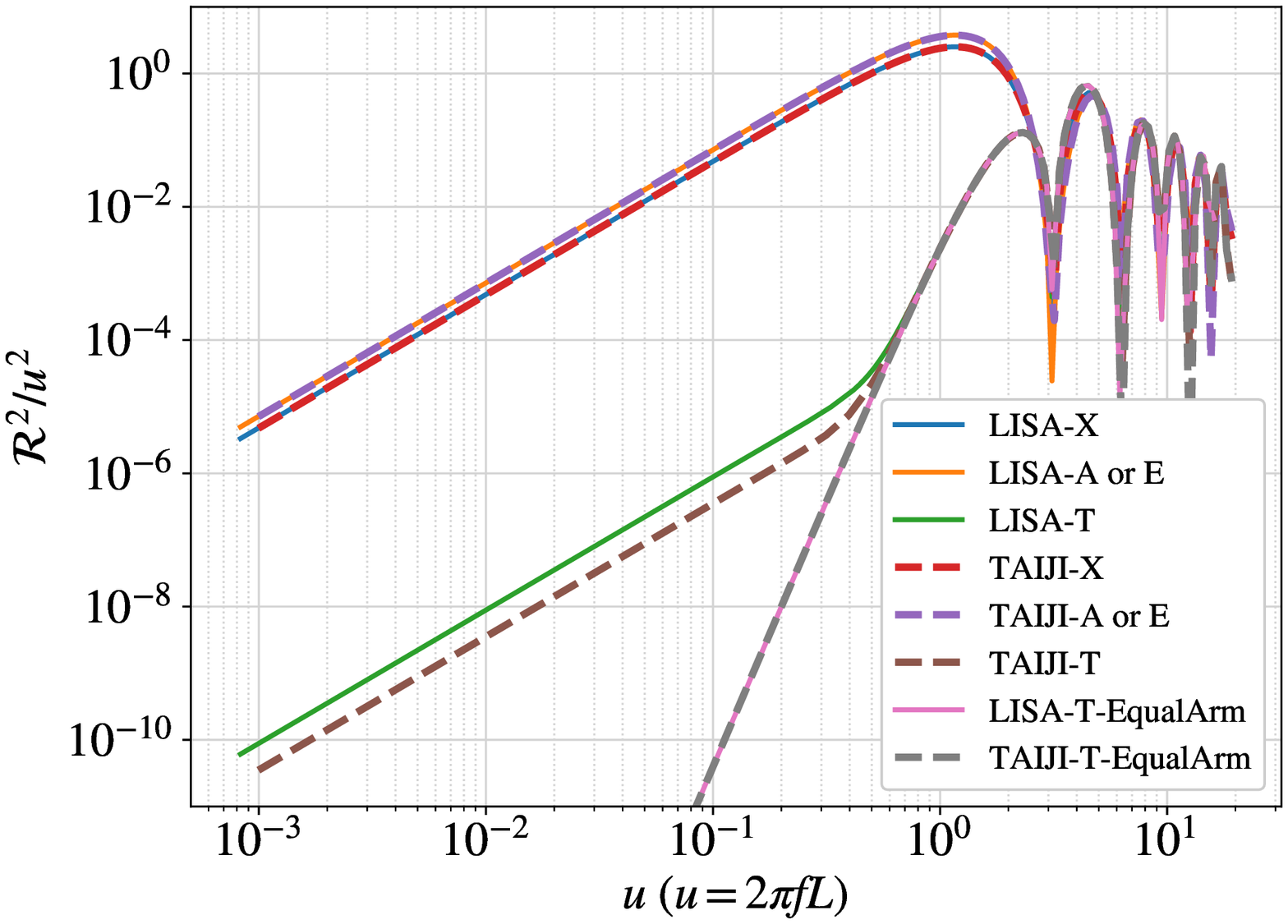}
\caption{\label{fig:response_1st} The average responses of selected TDI channels at different frequencies. The regular TDI channels (Michelson-X, Relay-U, Beacon-P and Monitor-D) are plotted in the upper panel, and optimal-A, E and T channels with Michelson-X are shown in the lower panel. (The curves of P and D channels are fully overlapped in the upper panel, as well as the A and E channel in the lower panel). For comparison, the response of T channel for fully equal arm configuration are shown by curves LISA-T-EqualArm and TAIJI-T-EqualArm.}
\end{figure}

For the four first-generation TDI channels (Michelson-X, Relay-U, Beacon-P and Monitor-D), the response expressions in frequency-domain could be described by linear combination of each link measurement with/without time delay factor(s),
\begin{widetext}
\begin{equation}
\begin{aligned}
 F^{\rm{GW}}_{ \rm X} (f) =& (-\Delta_{21} + \Delta_{21}  \Delta_{13}  \Delta_{31})  y^{\rm GW}_{12}
         + (-1 + \Delta_{13}  \Delta_{31} )  y^{\rm GW}_{21} 
         + (\Delta_{31} - \Delta_{31}  \Delta_{12}  \Delta_{21})  y^{\rm GW}_{13}
         + ( 1 - \Delta_{12}  \Delta_{21} )  y^{\rm GW}_{31}, \\
F^{\rm{GW}}_{ \rm U} (f) =& (1 - \Delta_{32}  \Delta_{21}  \Delta_{13})  y^{\rm GW}_{23}
         + (\Delta_{23} - \Delta_{21}  \Delta_{13})  y^{\rm GW}_{32} 
         + (\Delta_{32}  \Delta_{23} - 1)  y^{\rm GW}_{13}
         + (\Delta_{13} \Delta_{23}  \Delta_{32} - \Delta_{13})  y^{\rm GW}_{21}, \\
F^{\rm{GW}}_{ \rm P} (f) =& (\Delta_{13} - \Delta_{12}  \Delta_{23})  y^{\rm GW}_{32}
         + (\Delta_{13}  \Delta_{32} - \Delta_{12})  y^{\rm GW}_{23} 
         + (\Delta_{13}  \Delta_{32}  \Delta_{23} - \Delta_{13})  y^{\rm GW}_{12}
         + (\Delta_{12} - \Delta_{12}  \Delta_{23}  \Delta_{32})  y^{\rm GW}_{13}, \\
F^{\rm{GW}}_{ \rm D} (f) =& (1 - \Delta_{23}  \Delta_{32})  y^{\rm GW}_{21}
         + (\Delta_{21} - \Delta_{31}  \Delta_{23})  y^{\rm GW}_{32} 
         + (\Delta_{21}  \Delta_{32} - \Delta_{31})  y^{\rm GW}_{23}
         + (\Delta_{23}  \Delta_{32} - 1)  y^{\rm GW}_{31} ,
\end{aligned}
\end{equation}
\end{widetext}
where $\Delta_{ij} = \exp(2 \pi i f L_{ij})$.
To evaluate the detectability of a TDI channel, its response could be averaged over sky and polarization at a given frequency by
\begin{equation} \label{eq:resp_averaged}
\begin{aligned}
 \mathcal{R}^2_{\rm TDI} (f) =& \frac{1}{4 \pi^2}  \int^{2 \pi}_{0} \int^{\frac{\pi}{2}}_{-\frac{\pi}{2}} \int^{\pi}_{0} |F^{\rm{GW}}_{ \rm TDI} (f, \iota=0)|^2 \cos \beta {\rm d} \psi {\rm d} \beta {\rm d} \lambda.
\end{aligned}
\end{equation}
The averaged responses of four TDI channels (Michelson-X, Relay-U, Beacon-P, and Monitor-D) at different frequencies are shown in Fig. \ref{fig:response_1st} upper panel. As we can read from the plot, the unequal-arm Michelson-X combination has the best response in these four cases, and the curves of Beacon-P and Monitor-D channels are identical.
The lower panel of Fig. \ref{fig:response_1st} shows the optimal channels (A, E, T) comparing to the Michelson-X. The responses of optimal-A and E are identical and slightly higher than the Michelson-X channel. Although the averaged responses for optimal-A and E are the same, the instant sensitive directions of them could be different.
In the lower frequency band, the response of optimal-T channel is several orders lower than other TDI channels. On the other hand, considering the arm lengths vary with time, the average response of a TDI channel also can fluctuate. The curves shown in Fig. \ref{fig:response_1st} are calculated at the starting time of the mission orbits. For comparison, the response of T channel for fully equal arm configuration is shown by the curves LISA-T-EqualArm and TAIJI-T-EqualArm in Fig. \ref{fig:response_1st} lower panel.

\subsection{Sensitivities of TDI channels}

\begin{figure}[htb]
\includegraphics[width=0.48\textwidth]{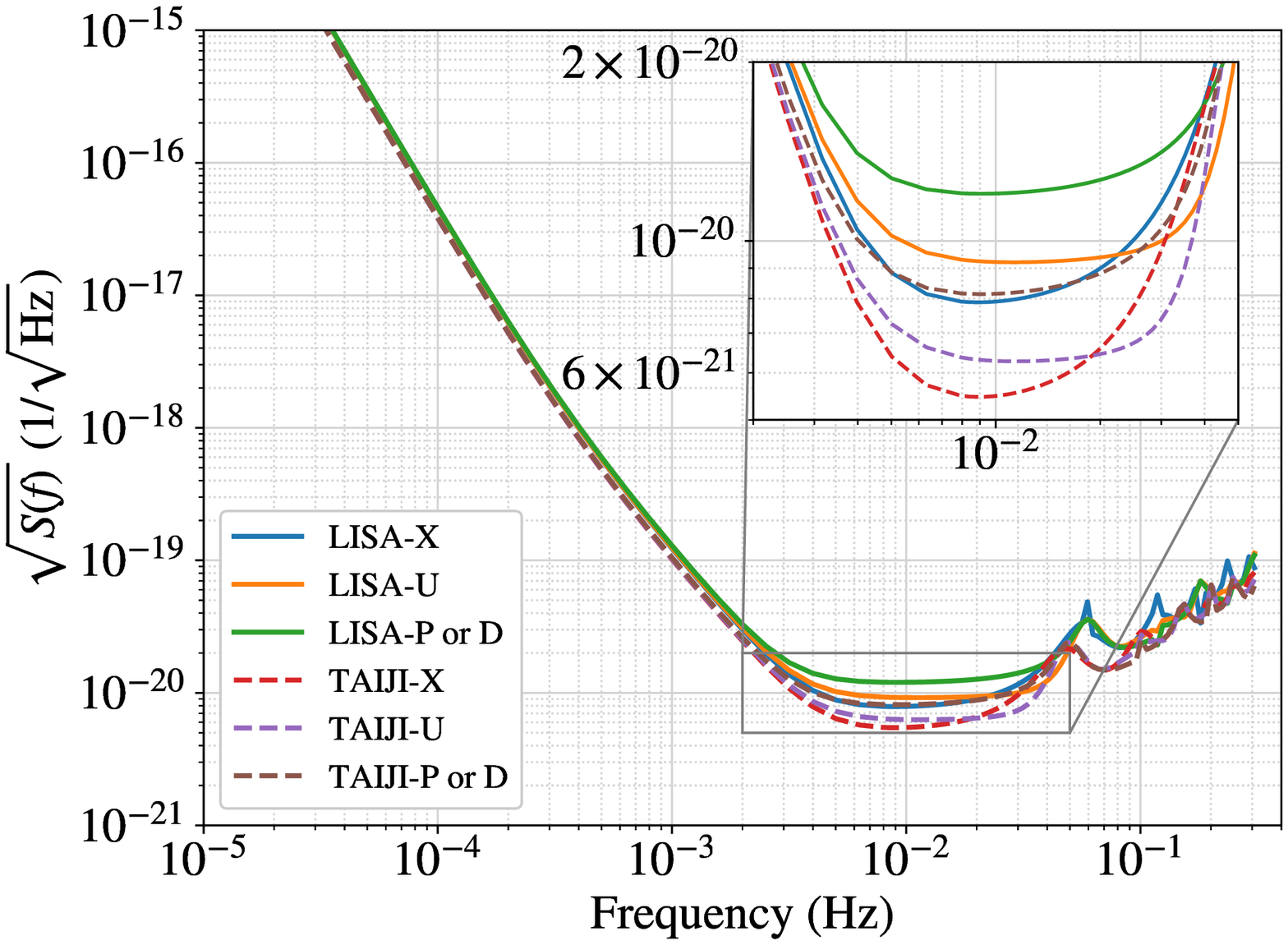}
\includegraphics[width=0.48\textwidth]{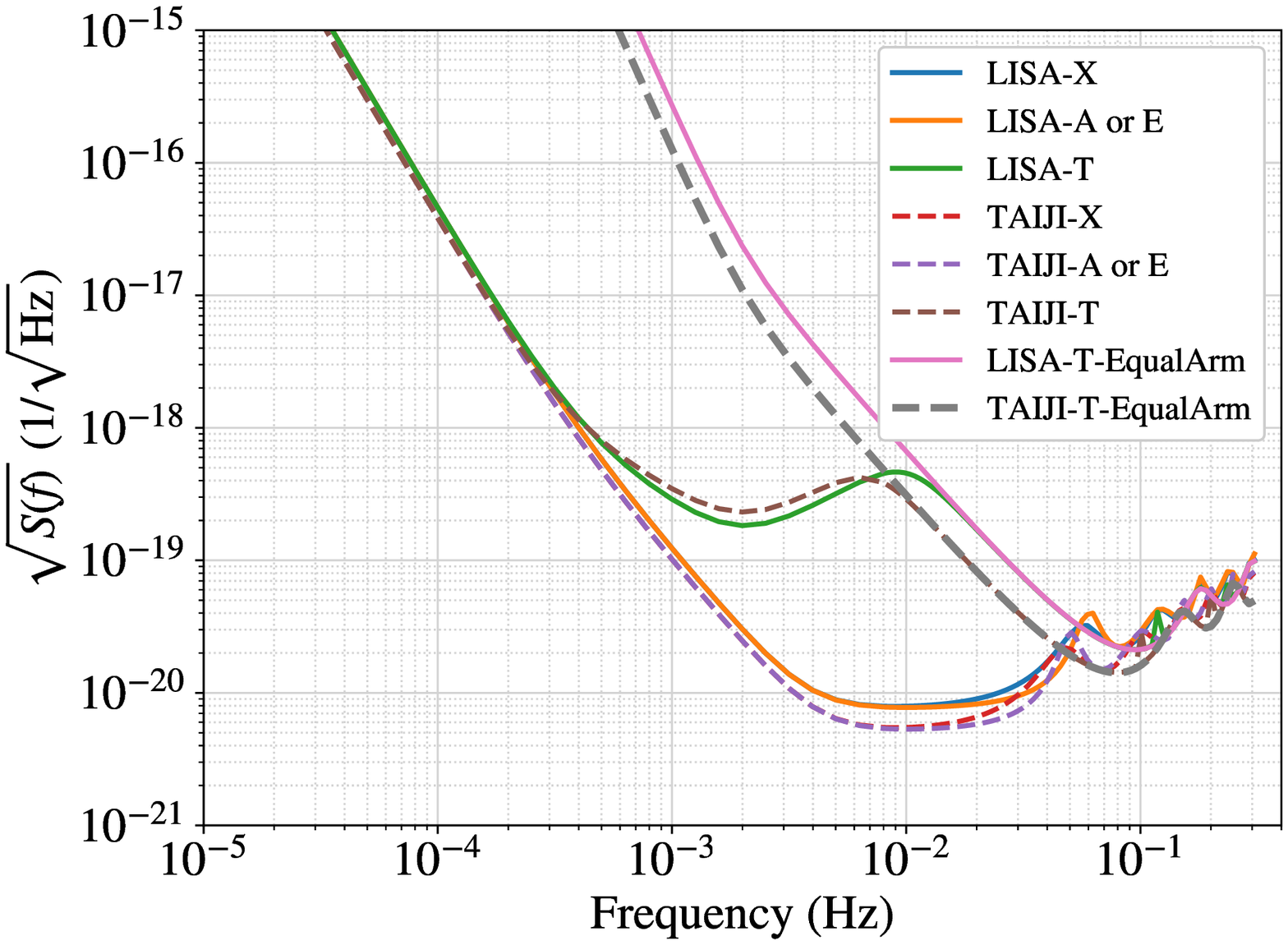}
\caption{\label{fig:Sensitivity_1st} The average sensitivities of selected TDI channels at starting time. The regular TDI channels (Michelson-X, Relay-U, Beacon-P and Monitor-D) are plotted in the upper panel, and optimal-A, E and T channels  with Michelson-X are shown in the lower panel. The sensitivity of T channel for fully equal arm configuration are shown by curves LISA-T-EqualArm and TAIJI-T-EqualArm, and this sensitivity is consistent with the T channel results described in \cite{Prince:2002hp,Vallisneri:2007xa}. (The curves of P and D channels are fully overlapped in the upper panel, as well as the A and E channel in the lower panel. The confusion noise is not considered in sensitivity estimations).}
\end{figure}

Based on the noise assumptions and average response, the PSD of average sensitivity in one TDI channel could be obtained by weighting the PSD of noise by the averaged response, $S_{\rm avg} = {S_{\rm n}}/{\mathcal{R}^2}$. 
The sensitivities of LISA and TAIJI on various TDI channels at the starting time are shown in Fig. \ref{fig:Sensitivity_1st}. Because of lower optical path noise requirement and longer arm length, TAIJI achieves slightly better sensitivities than LISA. For sensitivities from different TDI channels, the main divergences appear at the most sensitivity band, $\sim$[2, 50] mHz as shown in the amplified figures. The optimal-T channel has obvious divergence from others in the frequency band $\sim$[0.5, 50] mHz, and tends to have the same sensitivity as other channels when the frequency is lower than 0.5 mHz. This is different from the T channel results in \cite{Prince:2002hp,Vallisneri:2007xa} which assuming the equal arm.

Our further investigations show that the response of T channel to GW is proportional to the square of arm length differences at low frequency ($2 \pi f L \ll 1$) \cite{Wang:2020ongoing},
\begin{equation}
 \mathcal{R}^2_{\rm T} \propto \sum^{3}_{i=1} (L_{ij} - L_{ik})^2. 
\end{equation}
When the arm lengths are fully equal, the response to a GW signal could be canceled and make T channel as a quasi-null stream. However, in a realistic geodesic orbit, the arms to construct the TDI paths are not perfectly equal. In this scenario, the T channel can still respond to a GW signal even it could be $\sim$5 orders worse than X channels at the lower frequency as shown in Fig. \ref{fig:response_1st}. Since the response of X channel is proportional to $L^{2}$ at low frequency, this orders difference could be explained by the unequal arms from numerical orbit,
\begin{equation}
 \frac{\mathcal{R}^2_{\rm T}}{\mathcal{R}^2_{\rm X}} \simeq \frac{\delta L^2}{L^2} \simeq 10^{-5}.
\end{equation}
On the other side, the PSD of T channel is lower than X channels by $\sim$5 orders at $10^{-5}$ Hz as shown in Fig. \ref{fig:TDI1st_noise}.   
Consequently, the sensitivity of T channel could be equivalent to X channel as shown in Fig. \ref{fig:Sensitivity_1st}. 
There is a caveat that only acceleration and optical path noises are considered here in the PSD of T channel, the imperfect TDI due to path difference would upraise laser frequency noise, and further make sensitivity deteriorate. We commit ourselves to this study in the upcoming work \citep{Wang:2020ongoing}.

As the figures show, the average sensitivities of Beacon-P and Monitor-D channels are identical, as well as the pair of the optimal optimal-A and E channels. Even so, their instant antenna patterns and sensitivities to a same source could be different. Therefore, in the simulations included in the following two sections, the sensitivity of a TDI channel to a GW signal is calculated with variables including time, frequency, orientation and polarization, etc.

\section{Binary Supermassive Black Hole Mergers} \label{sec:SMBH}

The SMBH binary coalescence is one of the most important sources for LISA and TAIJI missions. In this section, we investigate the angular resolutions of LISA, TAIJI and joint network for the coalescing SMBH binary system.

\subsection{The source of SMBH binary}

LISA is expected to detect the SMBH binaries in the mass range of $10^5-10^7 M_\odot$ up to redshift $z \simeq 20$ \cite{2017arXiv170200786A}, as well as the TAIJI mission with the equivalent sensitivity. A mass ratio of $q=1/3$ is supposed to be the typical for these sources \cite{Colpi:2014poa}. Those coalescences could associate with the X-ray counterpart which could be detected by the Athena in redshift $z \lesssim 2$. \citet{Colpi:2019} investigated the sky localization of LISA for typical SMBH binary coalescences at the redshift $z=0.5,\ 1$ and $2$ which could trigger the efficient Athena follow-up observations.

\begin{figure}[htb]
\includegraphics[width=0.48\textwidth]{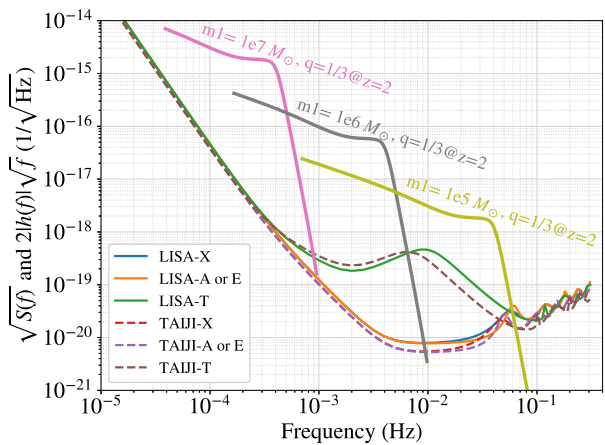}
\caption{\label{fig:SMBH_and_sensitivity} The amplitude of frequency-domain GW waveform from facing on/off three SMBH binaries at redshift $z=2$ and averaged sensitivity curves of (X, A, E, T) TDI channels at starting time. The GW signals start from the frequency at 30 days before the coalescences. }
\end{figure}

TAIJI is expected to be launched in the 2030s and could have overlapped observation time with LISA and Athena. The joint observation of LISA and TAIJI could significantly improve the sky localization precision for the SMBH binaries \cite{Ruan:2019tje}. To quantify this improvement, referring to the assumptions in \cite{Colpi:2019}, the sources considered in our simulation are following, 1) $m_1 = 10^7 \ M_\odot, m_2 = 3.3 \times 10^6 \ M_\odot$, 2) $m_1 = 10^6 \ M_\odot, m_2 = 3.3 \times 10^5 \ M_\odot$ and 3) $m_1 = 10^5 \ M_\odot, m_2 = 3.3 \times 10^4 \ M_\odot$ in source frame at redshift $z = 2$. The GW observations are simulated for 30 days to the coalescences. The redshift effect on the GW waveforms are included by adopting the cosmological parameters from the Planck 2015 results \cite{Ade:2015xua}. The redshifted amplitudes of three optimal orientated sources in frequency-domain are shown in Fig. \ref{fig:SMBH_and_sensitivity}, as well as the averaged sensitivity curves of TDI channels.

The approximant IMRPhenomPv2 is employed to represent the GW waveform which includes the inspiral-merger-ringdown phases \cite{Khan:2015jqa}. The LALSuite and PyCBC are utilized to implement the numerical waveform \cite{lalsuite,pycbc}. Eight parameters are considered to describe a GW signal with respect to the SSB coordinates which are ecliptic longitude $\lambda$, ecliptic latitude $\beta$, GW polarization angle $\psi$, inclination $\iota$, luminosity distance $D$, coalescence phase $\phi_c$, total mass of binary $M$ and mass ratio $q$. In our Monte Carlo simulation, the $(\lambda, \beta)$ is randomly sampled in the sphere, $\psi$ is uniformly sampled in $[0, 2 \pi]$ and $\cos \iota$ is sampled uniformly in $[-1, 1]$, and the starting time of GW signal is uniform in one year. 2000 binaries are generated for each kind of SMBH binaries.

\subsection{Fisher information matrix method} \label{subsec:FIM_FD}

The Fisher Information Matrix (FIM) is widely employed to determine the uncertainty of parameter measurements for GW observation \cite{1994PhRvD..49.2658C,Cutler:1997ta,Vallisneri:2007ev,Kuns:2019upi}. For multiple detectors, the joint FIM is calculated by summing up the FIM of each individual detectors,
\begin{equation}
\Gamma_{ij}  = \sum_{\rm det} \left( \frac{\partial h }{ \partial {\theta_i} } \bigg\rvert \frac{\partial h }{ \partial {\theta_j} }  \right)_{\rm det},
\end{equation}
with 
\begin{equation}
\left( g | h\right)_{\rm det} = 4 \mathrm{Re} \int^{\infty}_0 \frac{g^{\ast} (f) h(f)}{S^{\rm det} (f) } \mathrm{d} f ,
\end{equation}
where $h$ is the GW waveform in frequency domain, $\theta_i$ is the $i$-th parameter measured, and $S^{\rm det} (f)$ is the noise PSD of one LISA/TAIJI TDI channel. From the FIM, the variance-covariance matrix of the parameters is obtained by
\begin{equation}
\begin{aligned}
\sigma_{ij} &= \left\langle \Delta \theta_i \Delta \theta_j  \right\rangle = \left( \Gamma^{-1} \right)_{ij} + \mathcal{O}({\rho}^{-1}) \overset{{\rho} \gg 1}{\simeq } \left( \Gamma^{-1} \right)_{ij} . \\
\end{aligned}
\end{equation}
For a detected source with a significant SNR ($\rho > 7$), the angular uncertainty of the sky localization is evaluated by
\begin{equation} \label{eq:delta_Omega}
 \Delta \Omega \simeq 2 \pi | \cos \beta | \sqrt{ \sigma_{\lambda \lambda} \sigma_{\beta \beta} -  \sigma^2_{\lambda \beta}  }.
\end{equation}

The TDI channels are treated as detectors to calculate the FIM individually and cooperatively.
The first-generation TDI configurations can make observation even when dysfunction happened in any two of the six laser links. When the full links are available, the three channels for each TDI configuration can operate simultaneously which can composite to three optimal TDI channels as shown in Eq. \eqref{eq:optimalTDI}. Four observation scenarios are considered in the simulation for a simulated GW signal: 1) only one regular TDI channel (X, U, P and D) of two missions is operating, (without losing the representativity, only one of three channels in each TDI configuration is included.), 2) optimal TDI channels (A, E and T) are available for a mission when full links are functional, 3) the joint observation of one mission's regular TDI channel with another's optimal TDI channels, and 4) the joint of two mission's optimal TDI channels.

\subsection{The results of sky localization}

The angular resolutions of LISA, TAIJI and joint network are evaluated by the described Monte Carlo simulations. For each simulated source, the uncertainties of the localization are calculated by Eq. \eqref{eq:delta_Omega} for the TDI channels, and their values are shown by cumulative histograms.
To represent the results concisely, we select the results of X, A, T and joint AET channels for a single mission, and results of one mission's AET with another's regular or AET channels for joint network.
Considering the field of view (FoV) of Wide Field Imager (WFI) on Athena is designed to be 0.4 deg$^2$ (40 arcmin $\times$ 40 arcmin) \cite{2013arXiv1306.2307N}, it would be rather easy for Athena's follow-up observation when the uncertainties are constrained in 1 deg$^2$. 

\begin{figure*}[htb]
\includegraphics[width=0.48\textwidth]{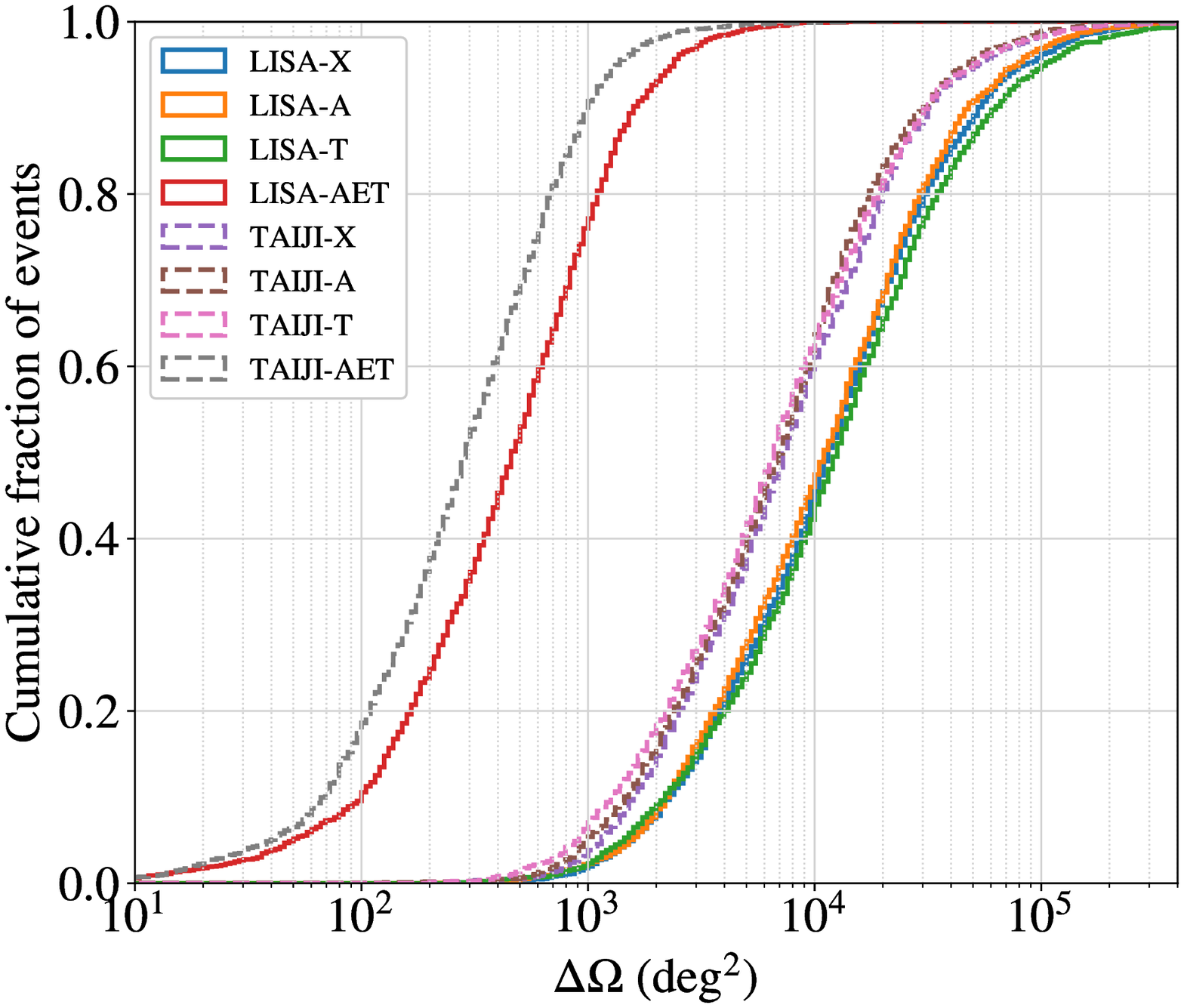}
\includegraphics[width=0.48\textwidth]{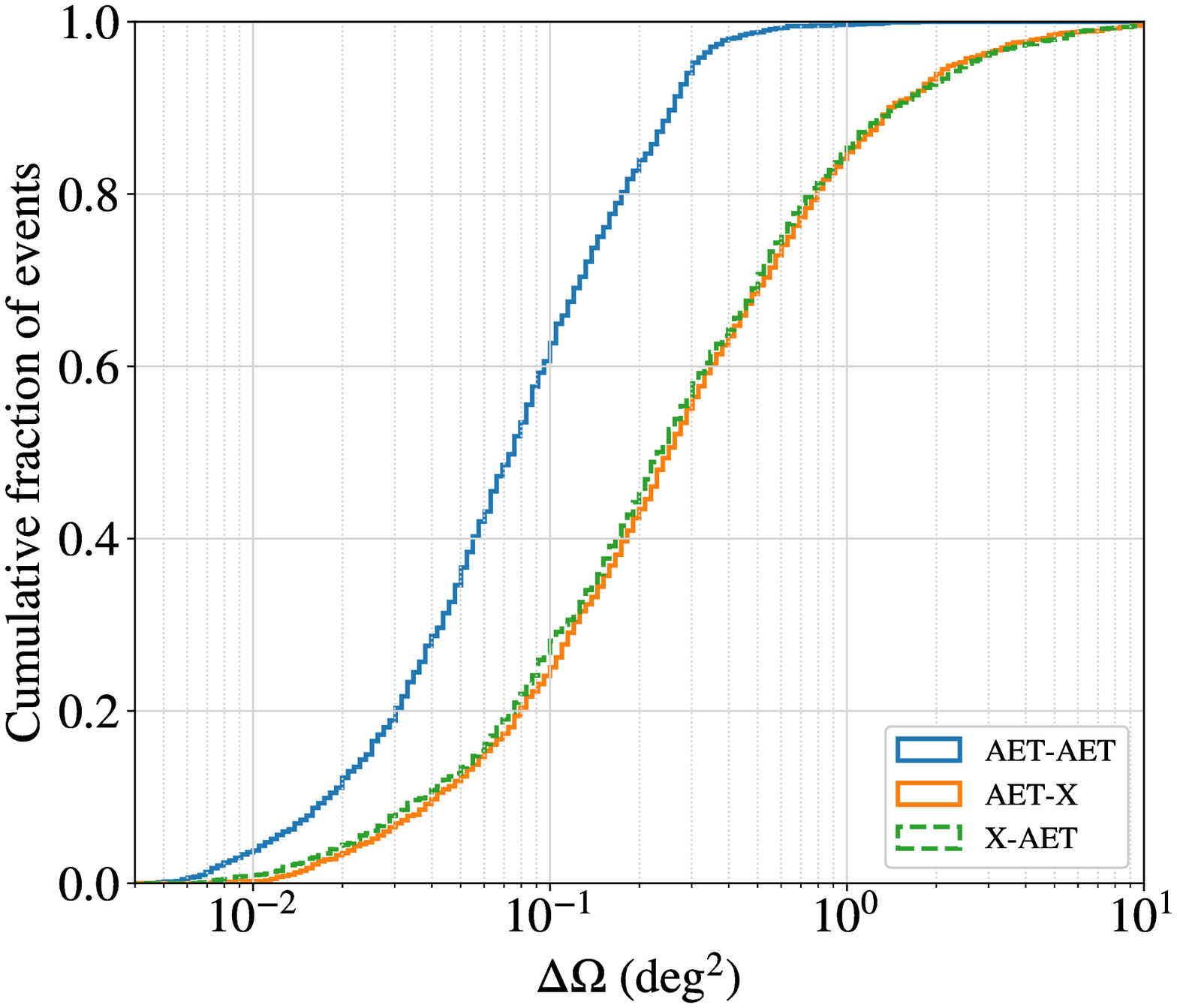}
\caption{\label{fig:SMBH_m1_1e7} The histograms of sky localization uncertainties of LISA, TAIJI, and LISA-TAIJI network for $(10^7, \ 3.3\times 10^6 )\ M_\odot$ binaries at redshift $z=2$. The left panel shows the results of selected single (X, A, and T) and joint optimal (AET) TDI channels in individual missions, and the right panel shows the results of joint TDI channels in LISA and TAIJI missions. The channels from a single mission are denoted by mission-TDI. For the curves with label TDI-TDI, the first TDI label represent the channel of LISA, the second TDI label indicates the channel of TAIJI, e.g.  AET-X means the joint observation of LISA-AET and TAIJI-X channel.}
\end{figure*}

The simulation results of $(10^7,\ 3.3\times 10^6)\ M_\odot$ binaries are shown in Fig \ref{fig:SMBH_m1_1e7}. The angular resolutions of TAIJI mission are slightly better than LISA in general. For the performance of individual missions, as shown in Fig \ref{fig:SMBH_m1_1e7} left panel, both LISA and TAIJI have lower angular resolution due to the relatively poor sensitivity in this corresponding band. Around $10\%$ and $20\%$ of the source could be constrained in 100 deg$^2$ by LISA and TAIJI's joint AET channels, respectively. The angular resolution of every single channel is closely identical due to their equivalent average sensitivities.
For the joint LISA-TAIJI observation, as shown in the right panel of Fig. \ref{fig:SMBH_m1_1e7}, we consider that a least one mission is in fully optimal TDI operation. When another mission is in TDI-X mode, more than $85\%$ of the sources could be localized in 1 deg$^2$. And all sources could be well localized in 0.4 deg$^2$ when both missions are in optimal operation.

\begin{figure*}[htb]
\includegraphics[width=0.48\textwidth]{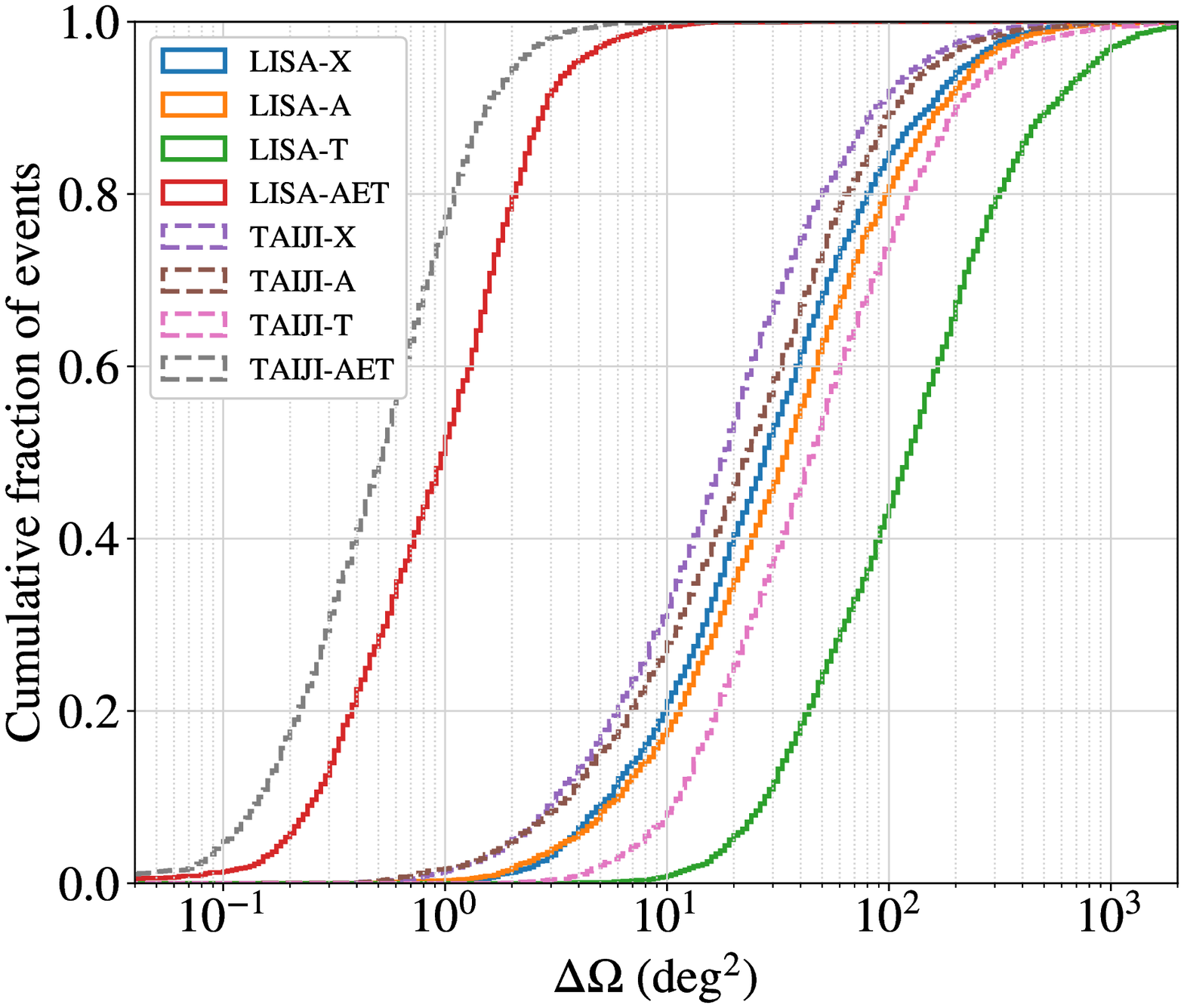}
\includegraphics[width=0.48\textwidth]{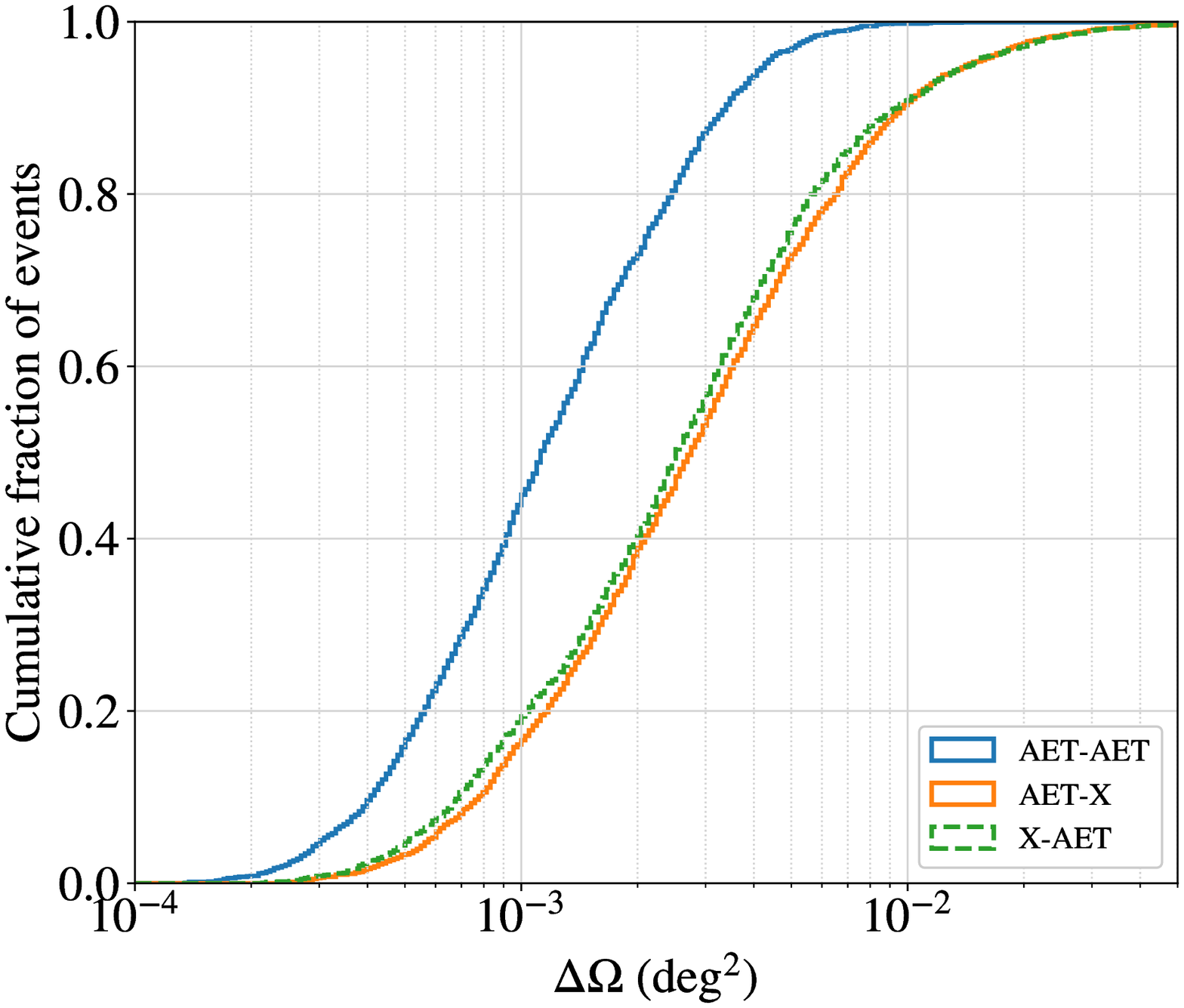}
\caption{\label{fig:SMBH_m1_1e6} The histograms of sky localization uncertainties of LISA, TAIJI, and LISA-TAIJI network for $(10^6,\ 3.3 \times 10^5)\ M_\odot$ binaries at redshift $z=2$. The left panels show the results of selected individual (X, A, and T) and AET channels, and the right panels show the results of joint TDI channels from two missions. TDI-TDI labels indicate the joint LISA (first TDI) and TAIJI (second TDI behind the dash) channels, e.g AET-X means the joint LISA-AET and TAIJI-X observation.}
\end{figure*}

The simulation results of $(10^6,\ 3.3\times 10^5)\ M_\odot$ binaries are shown in Fig \ref{fig:SMBH_m1_1e6}. The left panel shows the histograms of selected TDI channels for individual missions. As we can read from the curves, the optimal AET channels can effectively improve the sky localization by around $1-2$ orders comparing to a single channel; $75\%$ and $50\%$ of the source could be localized in 1 deg$^2$ by TAIJI's and LISA's AET channels, respectively; the sky localization of optimal-T channel become worse than X and A channels due to its sensitivity bulge in the frequency band [0.5, 50] Hz.
The histograms of joint LISA-TAIJI observations are shown in the right panel of Fig. \ref{fig:SMBH_m1_1e6}. Following the same strategy, we keep at least one mission in optimal operation for joint observation. Although another mission's TDI-X mode degrades the localization ability comparing to the double fully optimal mode, all the joint the observation can constrain the source in 0.05 deg$^2$.

\begin{figure*}[htb]
\includegraphics[width=0.48\textwidth]{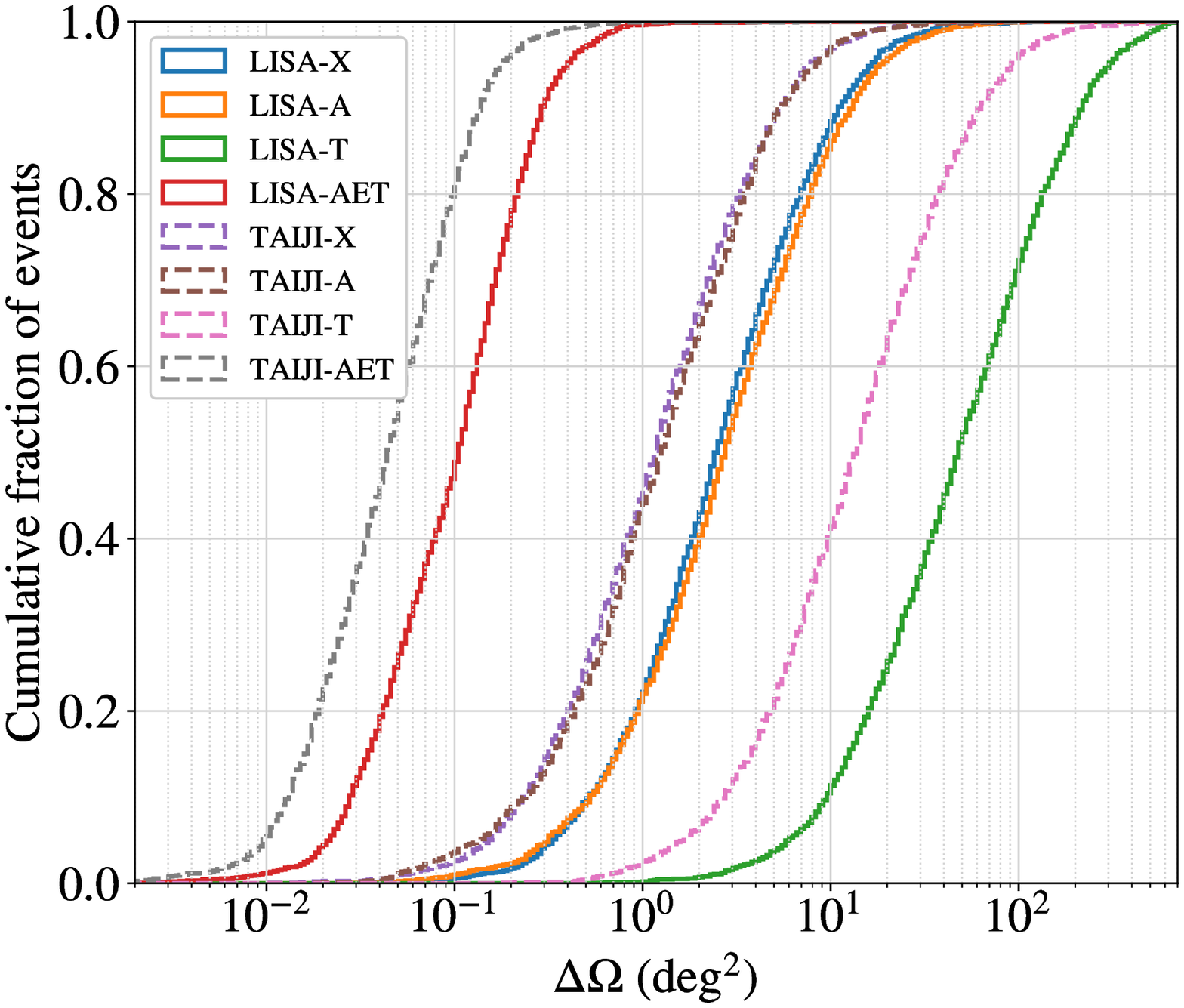}
\includegraphics[width=0.48\textwidth]{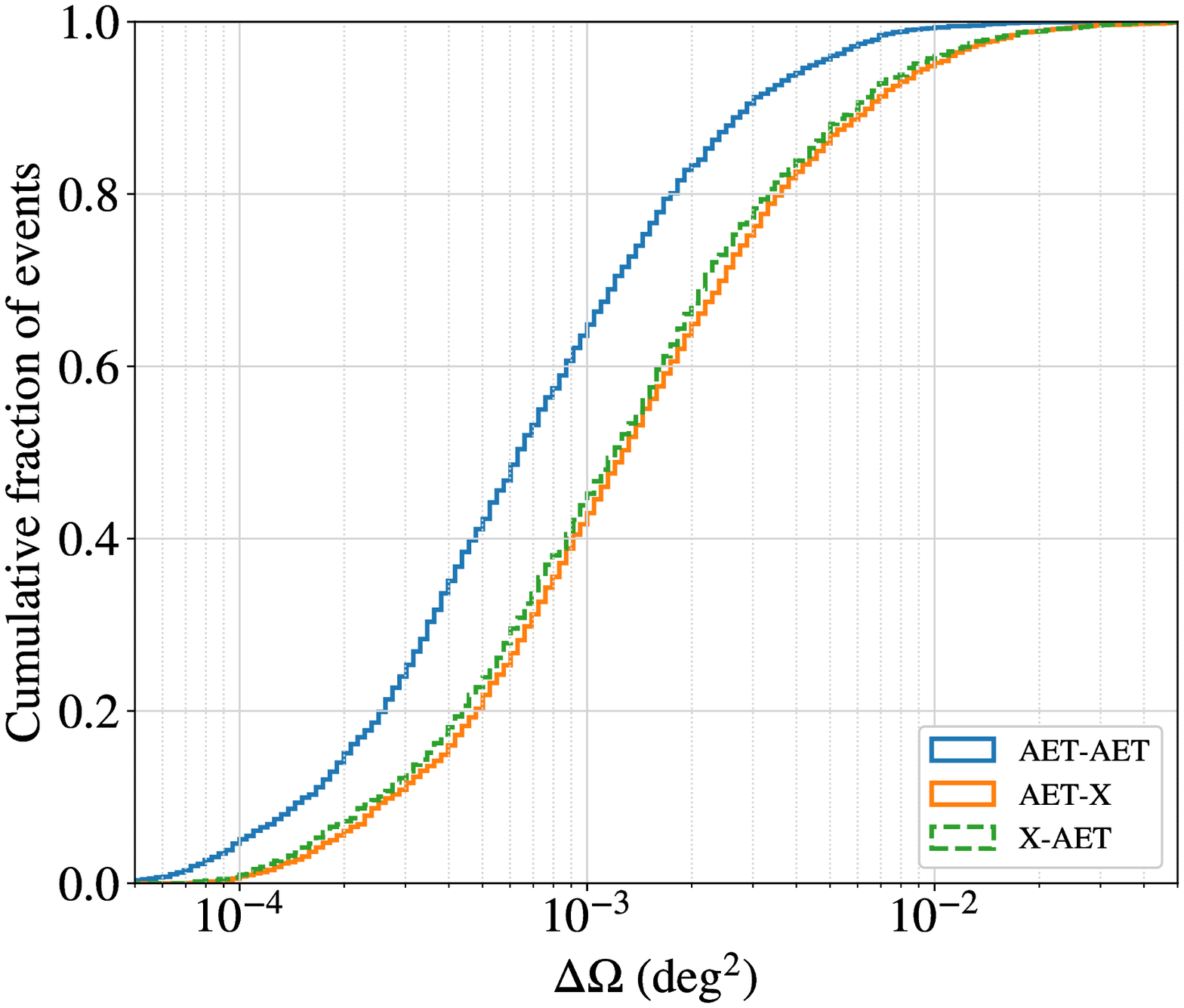}
\caption{\label{fig:SMBH_m1_1e5} The histograms of sky localization uncertainties of the LISA, TAIJI, and LISA-TAIJI network for $(10^5, \ 3.3\times 10^4)\ M_\odot$ binaries at redshift $z=2$. The left panel shows the results of the individual (X, A, and T) and AET TDI channels, and the right panel shows the results of joint TDI channels from LISA and TAIJI missions. TDI-TDI labels indicate the joint LISA (first TDI) and TAIJI (second TDI behind the dash) channels, e.g AET-X means the joint LISA-AET and TAIJI-X observation.}
\end{figure*}

The simulation results of $(10^5,\ 3.3\times 10^4)\ M_\odot$ binaries are shown in Fig. \ref{fig:SMBH_m1_1e5}. The left panel shows the histograms of TDI channels for individual missions. The optimal AET channels can effectively improve the sky localization by order(s) comparing to a single channel; almost all the sources could be localized in 1 deg$^2$ by TAIJI's and LISA's AET channels; the performance of optimal-T channel becomes much worse than X and A channels.
The histograms of joint LISA-TAIJI observations are shown in the right panel of Fig. \ref{fig:SMBH_m1_1e6}. Similar to the results of $(10^6,\ 3.3\times 10^5)\ M_\odot$ binaries, one mission's full optimal operation with another's TDI-X channel slightly decrease the localization performance compared to the double full optimal mode, all sources could be localized within 0.03 deg$^2$ by the LISA-TAIJI network.

The performances of the individual TDI channels, more or less, could be reflected in Fig. \ref{fig:SMBH_and_sensitivity}, especially for optimal-T channel. For the SMBH binaries $(10^7,\ 3.3\times 10^6)\ M_\odot$ at redshift $z=2$, the frequency range in the 30 days coalescing is in $[0.04, 1]$ mHz where the T channel has the equal average sensitivity to detect GW signals to other TDI channels. The frequency range shifts to $\sim[0.15, 10]$ mHz for $(10^6,\ 3.3\times 10^5)\ M_\odot$ binaries. In this band, the detectability of the T channel starts to decline comparing to other channels. As for $(10^5,\ 3.3\times 10^4)\ M_\odot$ binaries, their frequency band changes to $[0.7, 80]$ mHz where the sensitivity of the T channel is much worse than others. Consequently, the ability of the T channel to detect and localize the sources becomes poor.

\section{Monochromatic Gravitational Waves Source} \label{sec:mono}

Monochromatic GW could be generated by compact binaries in the early inspiral and its frequency evolution is negligible in years of observation.
Galactic compact binaries containing a white dwarf or neutron star could emit quasi-monochromatic GWs in the low-frequency band which are detectable for the LISA and TAIJI. The GW signals from this population would overlap and form an unresolved foreground around a few mHz. Based on the different population assumptions, the foreground was evaluated in \cite{Nissanke:2012eh} and references therein. However, there are also known galactic binaries that could be resolved by space detectors, for instance, AM CVn (with an orbital period of $\sim$1000 seconds), HM Cnc ( $\sim$320 seconds) and V407 Vul ($\sim$570 seconds) \cite{Colpi:2019}. In this section, we investigate the angular resolution of the LISA-TAIJI network for the resolvable monochromatic sources.

\subsection{Fisher information method for monochromatic source}

The FIM formula is also employed to evaluate the uncertainties of sky localization for monochromatic sources. Considering there is almost no intrinsic frequency evolution (the observed frequency could be modulated with the detector's motion), the equations applied in Section \ref{subsec:FIM_FD} are not suitable for the monochromatic case. By using the Parseval's theorem applied in \citep{Cutler:1997ta,Vecchio:2004ec}, the FIM formula could be modified to
\begin{equation}
\begin{aligned}
\Gamma_{ij} &= \sum_{\rm det} \left( \frac{\partial h }{ \partial {\theta_i} } \bigg\rvert \frac{\partial h }{ \partial {\theta_j} }  \right)_{\rm det} \\
	&= \sum_{\rm det} \left[  \frac{4}{S^{\rm det} (f_0)}  \int^{\infty}_0 \partial_i h^\ast(f) \partial_j h(f) {\rm d} f  \right] \\
	&= \sum_{\rm det} \left[  \frac{2  }{S^{\rm det} (f_0)}  \int^{\infty}_0 \partial_i h(t) \partial_j h(t) {\rm d} t  \right].
\end{aligned}
\end{equation}
The waveform of a monochromatic signal is described by,
\begin{equation}
 h_{\rm} (t) = F^{\rm GW}_{\rm TDI} (f_0, t) \times \mathcal{A} \exp \left( 2 \pi f_0 t + \phi_0 \right),
\end{equation}
and the amplitude $\mathcal{A}$ of waveform in time-domain is \cite{Maggiore:2007}
\begin{equation}
 \mathcal{A} = \frac{4}{D} \left( \frac{G \mathcal{M}_c}{c^2} \right)^{5/3} \left( \frac{\pi f_0}{c} \right)^{2/3},
\end{equation}
where $F^{\rm GW}_{\rm TDI} (f_0, t)$ is the TDI response function at frequency $f_0$ and time $t$, $\phi_0$ is the initial phase, $D$ is the luminosity distance of the source, $\mathcal{M}_c = {(m_1 m_2)^{3/5} }/{(m_1+m_2)^{1/5}}$ is chirp mass and $c$ is the speed of light.
In this case, seven parameters are used to describe a monochromatic source which are longitude $\lambda$ and latitude $\beta$ of the source in SSB ecliptic coordinates, polarization angle $\psi$, inclination $\iota$, amplitude $\mathcal{A}$, initial phase $\phi_0$ and frequency $f_0$. 

\subsection{Simulation and results for monochromatic source}

In the Monte Carlo simulation, the first step is to generate the sources in the galactic coordinate system and uniformly in a cylindrical volume with radial distance in [10, 50] kpc, azimuthal angle in $[0, 2 \pi]$ and height in $[-0.15, 0.15]$ kpc. Then the source locations are transformed to the SSB ecliptic coordinates by using the Python package -- Astropy (\url{https://www.astropy.org}). The amplitude $\mathcal{A}$ is generated assuming $m_1=m_2=1 \ M_\odot$. The polarization angle $\psi$ is randomly in $[0, 2 \pi]$ and inclination function $ \cos \iota$ is sampled uniformly in $[-1, 1]$. One-year observation is run with 1000 sources at frequency 3 mHz or 10 mHz. For comparison, we run an extra simulation at 10 mHz for the first 90 days observation.
In our current simulation, the sources are generated for the investigation of sky localization rather than specific for modeled galactic binary population. 

\begin{figure}[htb]
\includegraphics[width=0.48\textwidth]{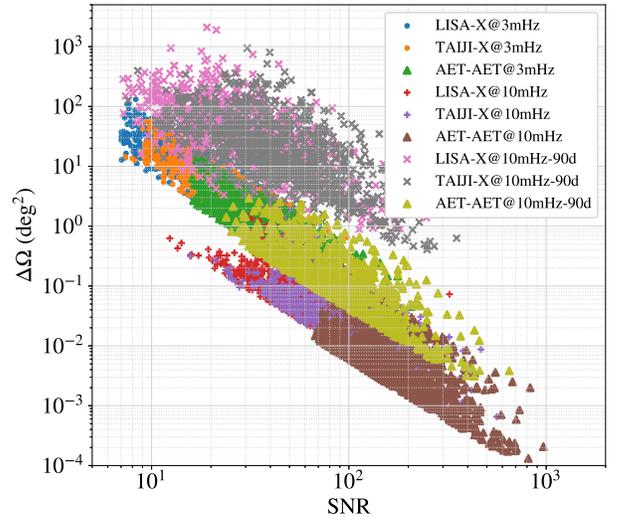}
\caption{\label{fig:Mono_SNR_SkyLocation} The sky localization uncertainties of the LISA, TAIJI, and LISA-TAIJI network with SNR for monochromatic sources. Three scenarios are included which are: 1) observing monochromatic wave at 3 mHz for one year (markers with @3mH); 2) observing monochromatic wave at 10 mHz for one year (markers with @10mH) and 3) observing monochromatic wave at 10 mHz for the first 90 days (markers with @10mH-90d). The AET-AET markers indicate the joint LISA-AET and TAIJI-AET observation results.}
\end{figure}

To keep the estimations from FIM sensible, signals are removed when SNR from LISA-X or TAIJI-X channel is lower than 7 \cite{Vallisneri:2007ev}. The uncertainties of sky localization with SNR are shown in Fig. \ref{fig:Mono_SNR_SkyLocation}. The SNRs at 10 mHz is generally higher than SNR at 3 mHz due to the better sensitivity at 10 mHz than 3 mHz. And the sky localization is more precisely with the higher SNR. The bands formed by the detected signal in log-log plots trend to have the same slopes. As we can see the results from the first 90 days at 10 mHz which are shown by the markers with 90d, the single detector could localize the source with poor precision even with the same SNR comparing to 3 mHz. It could be due to the contribution of the Doppler modulation from orbit motion is largely missing. Even so, their LISA-TAIJI joint observation can improve the angular resolution by around one order. This may verify that the LISA-TAIJI network contributes more significant advantages to the short duration signals.

\begin{figure*}[htb]
\includegraphics[width=0.48\textwidth]{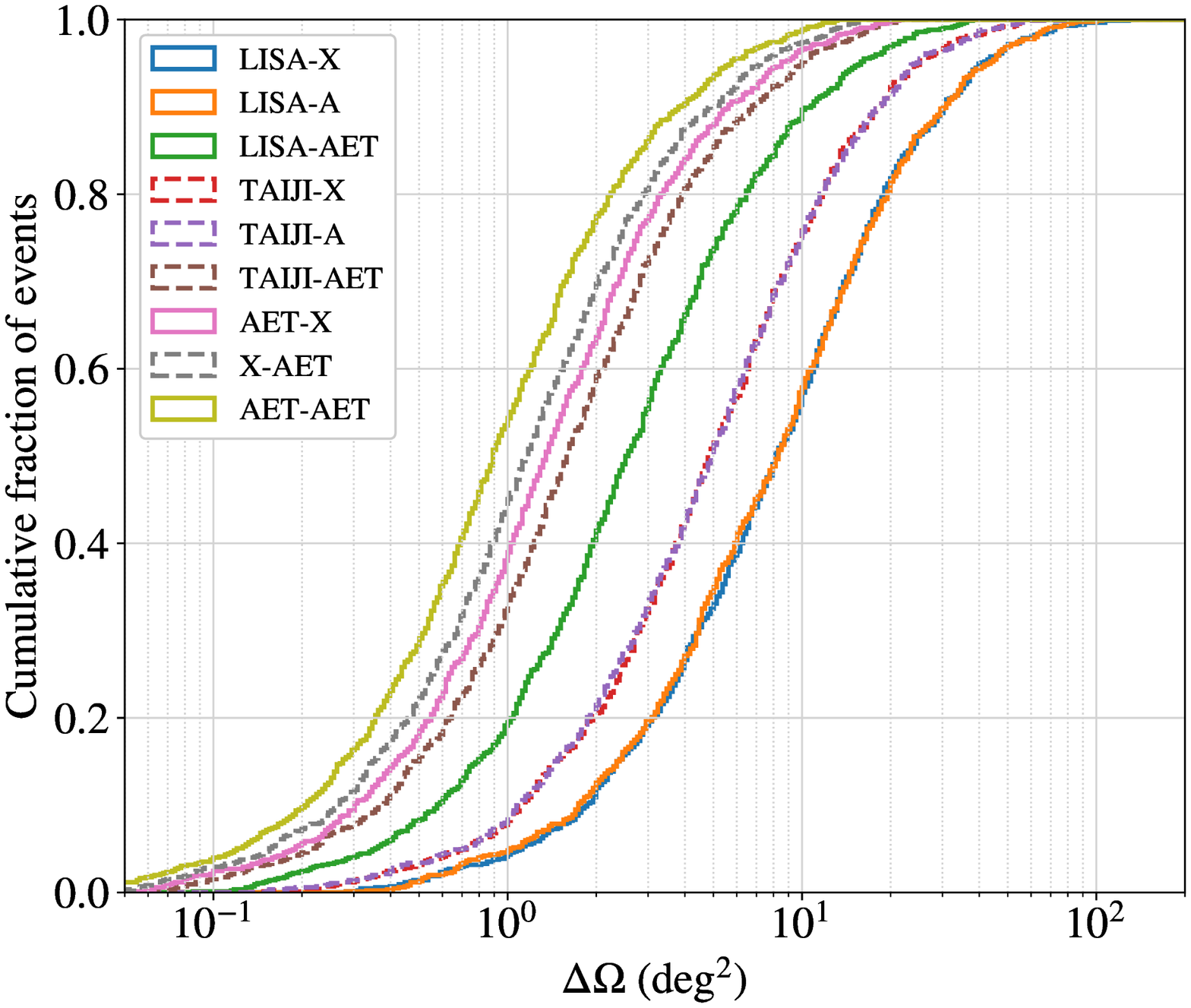}
\includegraphics[width=0.48\textwidth]{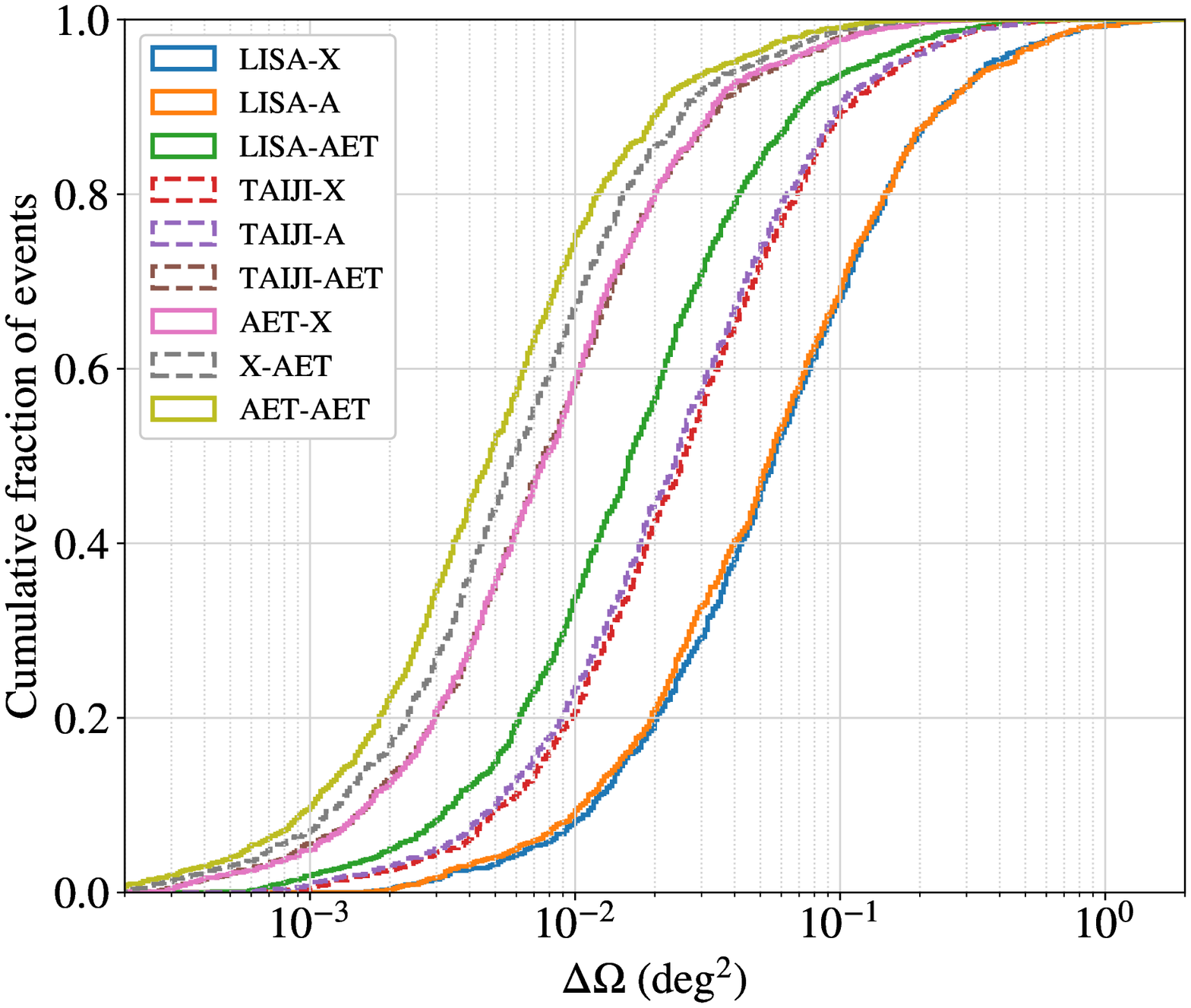}
\includegraphics[width=0.48\textwidth]{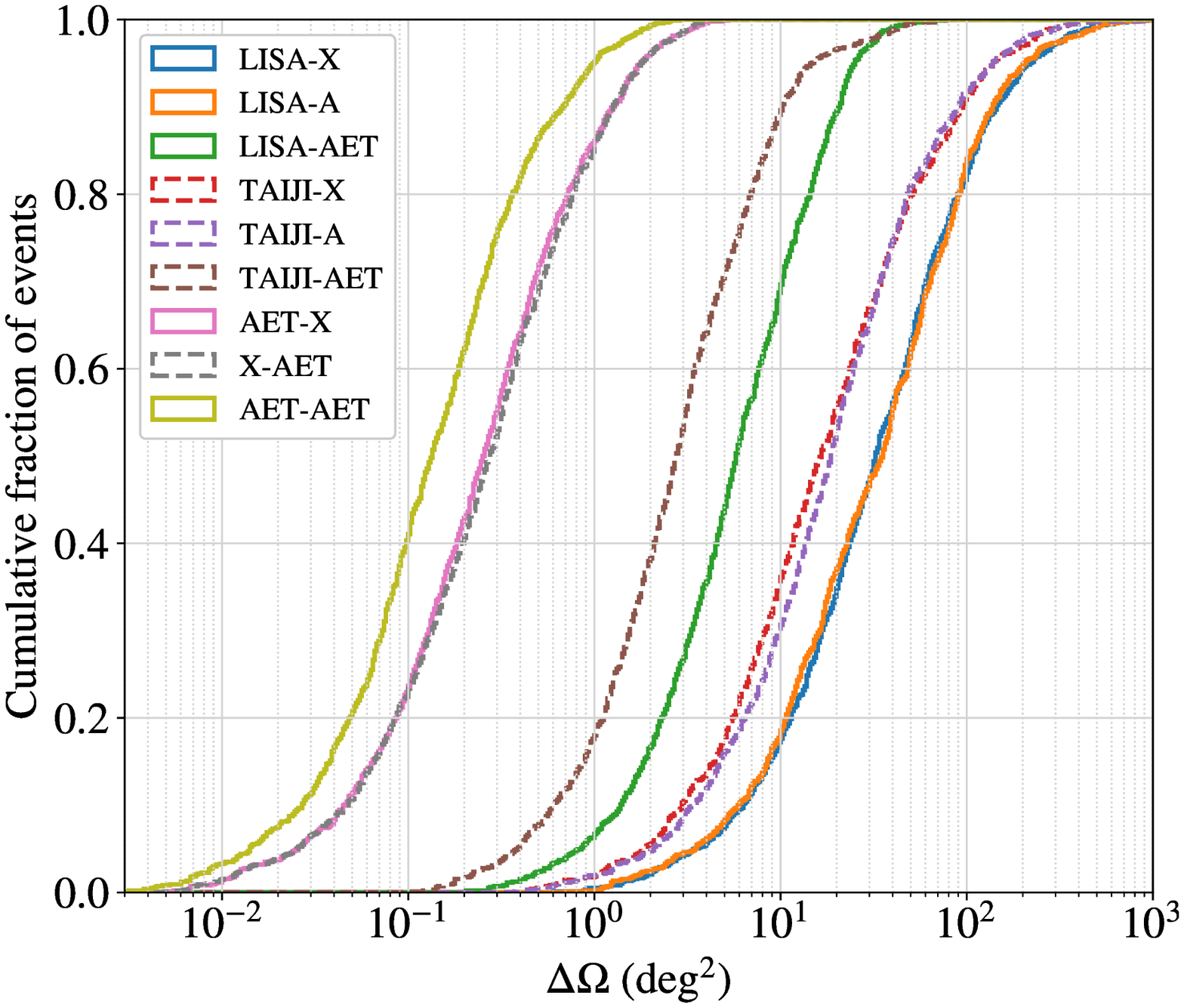}
\caption{\label{fig:Mono_SkyLocalization} The histograms of sky localization uncertainties of the LISA, TAIJI, and LISA-TAIJI network for the monochromatic sources at 3 mHz (upper left panel) and 10 mHz (upper right panel) for one year observation, as well as at 10 mHz for the first 90 days observation (lower panel). The mission-TDI labels denote the results of channel from single mission. TDI-TDI labels indicate the joint LISA (first TDI) and TAIJI (second TDI behind dash) channels, e.g AET-X means the LISA-AET joint with TAIJI-X observation. }
\end{figure*}

The histogram plots of localization uncertainties are shown in Fig. \ref{fig:Mono_SkyLocalization}. 
TAIJI mission has slightly better resolution than LISA due to the better sensitivity. The detectability of TDI X and A channels are equivalent at both 3 mHz and 10 mHz. The joint observation can bring a relatively moderate improvement compared to a single LISA or TAIJI mission for one year observation.
For a given percentage, e.g. 80\%, LISA-AET channel (green solid line) could localize the sources at 3 mHz in $\sim 6.5$ deg$^2$, and joint LISA-AET and TAIJI-X channels, AET-X (pink solid line), can localize them within $3.5$ deg$^2$. The double optimal operations, AET-AET (yellow solid line), can constrain them in 2.3 deg$^2$. The joint observation also can improve the LISA-AET's angular resolution by 1 to 3 times in one-year observation for 80\% of the sources at 10 mHz.
For the percentages constrained in 1 deg$^2$, LISA-AET channel can localize the 20\% of simulated sources at 3 mHz, and AET-AET network can promote it to more than 50\%. For the sources at 10 mHz, almost all the sources could be localized within 1 deg$^2$ in one-year observation, and most of them could be localized in 1 deg$^2$ in the first 90 days by LISA-TAIJI network.

The performances of TDI channels tend to identical for the channels with the equal averaged sensitivities at the corresponding frequency, as the X and A channels shown in Fig. \ref{fig:Sensitivity_1st}. The combined AET channel is effectively contributed by the optimal-A and E channels because of the poor sensitivity of the T channel at the 3 mHz and 10 mHz.
In addition, the amplitude of the monochromatic waveform is generated from assumed the galactic solar-mass compact binaries, and the FIM calculations take the amplitude as one parameter. Therefore, the results may represent simulations for more massive compact binaries from a larger distance as well.

\section{Conclusions and Discussions} \label{sec:conclusions}

In this work, we introduced the numerical orbit we achieved for LISA and TAIJI missions. To obtain the mission orbits, we employed an ephemeris framework to calculate S/C's geodesic including major gravitational interactions in the solar system. By implementing the orbital design and optimization workflow we created, the achieved LISA and TAIJI mission orbits could maintain in required status for 6 years without maneuver. And we assumed the starting time for LISA observation is on March 22nd, 2028 as a possible early schedule. If the observation starts from March 2030, there are still 4 years of valid mission orbit. If orbital maneuvers could be implemented, the orbit duration may be extended up to 10 years which could meet the LISA optimistic perspective \cite{2017arXiv170200786A}.

To understand the effects of TDI on GW observations, we examined four selected regulars (Michelson-X, Relay-U, Beacon-P, and Monitor-D) and three Michelson-type optimal (A, E, and T) TDI channels. We estimated the sky- and polarization- averaged sensitivities by using the current requirements on acceleration noise and optical path noise for LISA and TAIJI \cite{2017arXiv170200786A,Luo:2020}. And these sensitivities are time dependent and vary with the arm lengths. On the other hand, the calculations are based on the assumption that laser frequency noise is fully canceled by the first-generation TDI. However, the second-generation may be required for a realistic orbit to suppress the laser frequency noise under the acceleration and optical path noise \cite{Dhurandhar:2002zcl,Cornish:2003tz}, and their responses to GW could be depressed in this case especially for the lower frequency band \cite{Vallisneri:2004bn}. We would like to work on it in our future studies.

In order to investigate the angular resolutions of the LISA-TAIJI network, we run the Monte Carlo simulations for coalescing SMBH binaries and monochromatic sources. For each TDI channels, we calculated their response functions with variables including time, frequency, orientation, polarization, etc. For three SMBH populations at redshift $z=2$, $(10^7,\ 3.3 \times 10^6)\ M_\odot$, $(10^6,\ 3.3 \times 10^5)\ M_\odot$ and $(10^5,\ 3.3 \times 10^4)\ M_\odot$, the results showed the LISA-TAIJI joint observation significantly improve the angular resolution comparing to solo LISA or TAIJI mission. In an optimal scenario, all the simulated SMBH binaries could be determined in 1 deg$^2$ by the LISA-TAIJI network which would bring great merits for Athena's observations. The improvements are expected to take effect especially for short duration signals which are required to be identified quickly and precisely. Furthermore, we can deduce that SMBH binaries with comparable chip mass within redshift $z<2$ would be localized even more precisely. Another interesting result was shown by the detectability of the optimal-T channel to the SMBH binaries. With an irregular sensitivity curve comparing to the previous expectations \cite{Prince:2002hp,Vallisneri:2007xa}, its performance is rather different for the different masses binaries. 

The GW waveforms for coalescing SMBH binaries are represented by the IMRPhenomPv2 approximant. And only dominant quadrupole ($(l, m) = (2, \pm2)$) modes are included in the simulation. The waveform with higher harmonics could substantially increase the angular resolution for smaller mass ratio \cite{2007PhRvD..76j4016A,2010PhRvD..81f4014M}. The newly developed higher mode PhenomHM with spin effect may worth applying to the smaller mass ratio in the future for the LISA-TAIJI network \cite{London:2017bcn}.

For the monochromatic sources in one-year observation, the improvement of angular resolution by the LISA-TAIJI network is relatively moderate compared to single LISA mission. The network can improve the localization by a factor of 2 to 4 for a selected percentage of sources because longer observation of a single mission is supposed to compensate the disadvantage and reach a certain accuracy comparing to the two detectors. For a shorter observation, for instance, 90 days we simulated, the network still represents a considerable advantage.

In our current simulation, the confusion-foreground noise from galactic binaries is not considered. We suppose the confusion noise can decrease SNR by a limited amount and bring an insignificant impact on the SMBH binary simulations. It may change the results to a certain extent for the monochromatic sources. However, at a given SNR, we suppose it would not significantly change the conclusions about the relative improvement of sky localization by the LISA-TAIJI network. Their joint observation may also help to resolve the confusion-foreground, and we commit it as another study in the future.

\begin{acknowledgments}
This work was supported by NSFC No. 11773059, Key Research Program of Frontier Sciences, Chinese Academy of Science, No. QYZDB-SSW-SYS016 and the Strategic Priority Research Program of the Chinese Academy of Sciences under grant Nos. XDA1502070102, XDA15020700 and XDB21010100. and by the National Key Research and Development Program of China under Grant Nos. 2016YFA0302002 and 2017YFC0601602.
This work made use of the High Performance Computing Resource in the Core Facility for Advanced Research Computing at Shanghai Astronomical Observatory. The authors would like to thank the anonymous referees for their valuable comments and suggestions.
\end{acknowledgments}

\appendix

\section{}

\subsection{Interactions in Ephemeris Framework}

\subsubsection{Newtonian and first-order post-Newtonian interactions}

The point mass Newtonian and first-order post-Newtonian interactions between major celestial bodies (the Sun, major planets, Pluto, Moon, Ceres, Pallas and Vesta) are included in the ephemeris framework. The acceleration of one body/spacecraft due to this interactions is \cite{Brumberg1991}
\begin{equation} \label{equ:PPN} 
\ddot{\bm{r}}_i=-\sum_{j\neq{i}}\frac{GM_{j}}{r_{ij}^3}
{\bm{r}}_{ij} + \sum_{j\neq{i}} m_{j} (A_{ij} {\bm{r}}_{ij} + B_{ij}
\dot{\bm{r}}_{ij})\,,
\end{equation}
\begin{widetext}
\begin{equation}
\begin{split}
 A_{ij} =& \frac{{\dot{\bm{r}}_i}^2}{r^3_{ij}} -(\gamma+1)\frac{\dot{\bm{r}}^2_{ij}}{r_{ij}^3}+\frac{3}{2 r_{ij}^5}({\bm{r}}_{ij}\cdot\dot{\bm{r}}_j)^2  + G[(2\gamma+2\beta+1)M_i+(2\gamma+2\beta)M_j]\frac{1}{r_{ij}^4} \\ 
 & + \displaystyle\sum_{k\neq{i,j}} GM_k [(2\gamma+2\beta)\frac{1}{r_{ij}^3 r_{ik}} + (2\beta-1)\frac{1}{r_{ij}^3 r_{jk}} + \frac{2(\gamma+1)}{r_{ij} r_{jk}^3} -
 (2\gamma+\frac{3}{2})\frac{1}{r_{ik}
 r_{jk}^3}-\frac{1}{2r_{jk}^3}\frac{{\bm{r}}_{ij}\cdot{\bm{r}}_{ik}}{r_{ij}^3}]\,,
\end{split}
\end{equation}
\end{widetext}
\begin{equation}
B_{ij} = \frac{1}{r_{ij}^3}[(2\gamma+2)({\bm{r}}_{ij}\cdot\dot{\bm{r}}_{ij})+{\bm{r}}_{ij}\cdot\dot{\bm{r}}_{j}]\,,
\end{equation}
where $\gamma = \beta = 1$, $G$ is gravitational constant, $m_j = GM_j/c^2$, $\bm{r}_i$ is the position of body $i$ in the SSB coordinates, $\bm{r}_{ij}$ is the relative position between body $i$ and $j$, the $\dot{\bm{r}}$ and $\ddot{\bm{r}}$ represent the velocity and acceleration, respectively.

\subsubsection{Interaction with extended bodies}

The Sun, Earth and Moon are treated as extended bodies in the ephemeris framework. The acceleration due to an extended body is given by  \cite{DE430}

\begin{widetext}
\begin{equation}
\begin{split}
\begin{bmatrix} \ddot{\xi} \\ \ddot{\eta} \\ \ddot{\zeta} \end{bmatrix}
=&  - \frac{GM}{r^2}  \left\lbrace
\sum^{n_1}_{n=2} J_n \left( \frac{R}{r} \right)^n
\begin{bmatrix}
 (n+1) P_n ( \sin \varphi ) \\
 0 \\
 - \cos \varphi P^\prime_n ( \sin \varphi )
\end{bmatrix} \right.
  \\ 
 & \left. 
 + \sum^{n_2}_{n=2} \left( \frac{R}{r} \right)^n \sum^{n}_{m=1} 
 \begin{bmatrix}
  -(n+1) P^m_n ( \sin \varphi ) [ + C_{nm} \cos m \lambda + S_{nm} \sin m \lambda ] \\
  m / \cos \varphi P^m_n ( \sin \varphi ) [ - C_{nm} \sin m \lambda + S_{nm} \cos m \lambda ]  \\
  \cos \varphi P^\prime_n ( \sin \varphi )  [ + C_{nm} \cos m \lambda + S_{nm} \sin m \lambda ] 
 \end{bmatrix}
 \right\rbrace
 \end{split}
\end{equation}
\end{widetext}
where $r$ is the distance between the two bodies; $P_n (\sin \varphi )$ is the Legendre polynomial of degree $n$, and $P^m_n (\sin \varphi )$ is the associated Legendre function of degree $n$ and order $m$; $n_1$ and $n_2$  are the maximum degrees of the zonal and tesseral expansions; $J_n$ is the zonal harmonic coefficient and $C_{nm}, S_{nm}$ are the tesseral harmonic coefficients; $R$ is the equatorial radius of the extend body; $(\lambda, \varphi)$ are the direction of the point mass in the body-fixed coordinate system. 

By taking the coefficients from DE430, the accelerations due to the $J_2$ of the Sun, $J_2 - J_4$ of the Earth, and the Moon's zonal and tesseral harmonics up to degree of 6 are included in the framework.

\subsubsection{Perturbation from asteroids}

There are 340 asteroids selected to calculate the Newtonian perturbations on major celestial bodies and S/C  in the framework. The list is taken from Table 13 of \citet{DE430} (exclude the Ceres, Pallas and Vesta).

\subsubsection{Interaction on Moon from Earth tides}

The motion of the Moon will be affected by the Earth tides raised by the Sun and Moon. This tidal effects would slightly impact on the relative motion between the Earth and Moon, and indirectly impact on the mission orbit in this work trivially.  
We incorporate the tides effect to keep the integrity of the ephemeris framework, and for the more significant effect on the launch/transfer orbit.

\subsection{Initial Condition for Mission Orbit}

For a LISA-like mission with nominal arm length $\lambda \times 10^6$ km and starting at $t_0$ for observation, we generate the initial conditions of three S/C following \citet{Dhurandhar:2004rv},
 \begin{equation}
 \begin{split}
   X_k & = R ( \cos \psi_k + e) \cos \epsilon \\
   Y_k & = R \sqrt{ 1 - e^2} \sin \psi_k \qquad (k = 1, 2, 3) \\
   Z_k & = R( \cos \psi_k + \epsilon ) \sin \epsilon
 \end{split}
\end{equation}
where $e \simeq 0.001925 \times \lambda$; $\epsilon \simeq 0.00333 \times \lambda$; $R = 1$ AU, and $\psi_k$ is defined implicitly by
 \begin{equation}
   \psi_k + e \sin \psi_k = \Omega (t - t_0) - (k-1) \frac{2\pi}{3},
 \end{equation}
where $\Omega = 2 \pi/{\rm yr}$. Then $x_k , y_k, z_k $ could be calculated by
 \begin{equation}
 \begin{split}
   x_k & = X_k \cos \left[ \frac{2\pi}{3} (k-1) + \varphi_0  \right] - Y_k \sin \left[\frac{2\pi}{3} (k-1) + \varphi_0 \right], \\
   y_k & = X_k \cos \left[ \frac{2\pi}{3} (k-1) + \varphi_0 \right] + Y_k \sin \left[\frac{2\pi}{3} (k-1) + \varphi_0 \right], \\
   z_k & = Z_k,
 \end{split}
\end{equation}
where $\varphi_0 = \psi_E - \theta$, $\psi_E$ is the position angle of the Earth with respect to X-axis at $t_0$, and $\theta$ is the trailing angle of the constellation.

The initial position and velocity of S/C $k$ in the heliocentric coordinates are
\begin{equation} \label{eq:orbit-init}
 \begin{split}
  \vec{\mathcal{R}}_{\mathrm{S/Ck}} &= [x_k , y_k , z_k],  \\
  \vec{\mathcal{V}}_{\mathrm{S/Ck}} &= [\dot{x}_k, \dot{y}_k, \dot{z}_k ].
 \end{split}
\end{equation}
Then the initial conditions are transformed to the SSB coordinates and put into the ephemeris frame to calculate the mission orbit. We optimize the mission orbit by adjusting/trimming the heliocentric distance and velocities iteratively.

\nocite{*}
\bibliography{apssamp}

\providecommand{\noopsort}[1]{}\providecommand{\singleletter}[1]{#1}%
\begin{thebibliography}{81}%
\makeatletter
\providecommand \@ifxundefined [1]{%
 \@ifx{#1\undefined}
}%
\providecommand \@ifnum [1]{%
 \ifnum #1\expandafter \@firstoftwo
 \else \expandafter \@secondoftwo
 \fi
}%
\providecommand \@ifx [1]{%
 \ifx #1\expandafter \@firstoftwo
 \else \expandafter \@secondoftwo
 \fi
}%
\providecommand \natexlab [1]{#1}%
\providecommand \enquote  [1]{``#1''}%
\providecommand \bibnamefont  [1]{#1}%
\providecommand \bibfnamefont [1]{#1}%
\providecommand \citenamefont [1]{#1}%
\providecommand \href@noop [0]{\@secondoftwo}%
\providecommand \href [0]{\begingroup \@sanitize@url \@href}%
\providecommand \@href[1]{\@@startlink{#1}\@@href}%
\providecommand \@@href[1]{\endgroup#1\@@endlink}%
\providecommand \@sanitize@url [0]{\catcode `\\12\catcode `\$12\catcode
  `\&12\catcode `\#12\catcode `\^12\catcode `\_12\catcode `\%12\relax}%
\providecommand \@@startlink[1]{}%
\providecommand \@@endlink[0]{}%
\providecommand \url  [0]{\begingroup\@sanitize@url \@url }%
\providecommand \@url [1]{\endgroup\@href {#1}{\urlprefix }}%
\providecommand \urlprefix  [0]{URL }%
\providecommand \Eprint [0]{\href }%
\providecommand \doibase [0]{https://doi.org/}%
\providecommand \selectlanguage [0]{\@gobble}%
\providecommand \bibinfo  [0]{\@secondoftwo}%
\providecommand \bibfield  [0]{\@secondoftwo}%
\providecommand \translation [1]{[#1]}%
\providecommand \BibitemOpen [0]{}%
\providecommand \bibitemStop [0]{}%
\providecommand \bibitemNoStop [0]{.\EOS\space}%
\providecommand \EOS [0]{\spacefactor3000\relax}%
\providecommand \BibitemShut  [1]{\csname bibitem#1\endcsname}%
\let\auto@bib@innerbib\@empty
\bibitem [{\citenamefont {Abbott}\ \emph
  {et~al.}(2016{\natexlab{a}})\citenamefont {Abbott} \emph
  {et~al.}}]{Abbott:2016blz}%
  \BibitemOpen
  \bibfield  {author} {\bibinfo {author} {\bibfnamefont {B.~P.}\ \bibnamefont
  {Abbott}} \emph {et~al.} (\bibinfo {collaboration} {{LIGO Scientific
  Collaboration and Virgo Collaboration}}),\ }\bibfield  {title} {\bibinfo
  {title} {{Observation of Gravitational Waves from a Binary Black Hole
  Merger}},\ }\href {https://doi.org/10.1103/PhysRevLett.116.061102} {\bibfield
   {journal} {\bibinfo  {journal} {Phys. Rev. Lett.}\ }\textbf {\bibinfo
  {volume} {116}},\ \bibinfo {pages} {061102} (\bibinfo {year}
  {2016}{\natexlab{a}})},\ \bibinfo {note} {and references therein},\ \Eprint
  {https://arxiv.org/abs/1602.03837} {arXiv:1602.03837 [gr-qc]} \BibitemShut
  {NoStop}%
\bibitem [{\citenamefont {Abbott}\ \emph
  {et~al.}(2016{\natexlab{b}})\citenamefont {Abbott} \emph
  {et~al.}}]{TheLIGOScientific:2016pea}%
  \BibitemOpen
  \bibfield  {author} {\bibinfo {author} {\bibfnamefont {B.~P.}\ \bibnamefont
  {Abbott}} \emph {et~al.} (\bibinfo {collaboration} {{LIGO Scientific
  Collaboration and Virgo Collaboration}}),\ }\bibfield  {title} {\bibinfo
  {title} {{Binary Black Hole Mergers in the first Advanced LIGO Observing
  Run}},\ }\href {https://doi.org/10.1103/PhysRevX.6.041015,
  10.1103/PhysRevX.8.039903} {\bibfield  {journal} {\bibinfo  {journal} {Phys.
  Rev.}\ }\textbf {\bibinfo {volume} {X6}},\ \bibinfo {pages} {041015}
  (\bibinfo {year} {2016}{\natexlab{b}})},\ \bibinfo {note} {[erratum: Phys.
  Rev.X8,no.3,039903(2018)]},\ \Eprint {https://arxiv.org/abs/1606.04856}
  {arXiv:1606.04856 [gr-qc]} \BibitemShut {NoStop}%
\bibitem [{\citenamefont {Abbott}\ \emph {et~al.}(2019)\citenamefont {Abbott}
  \emph {et~al.}}]{LIGOScientific:2018mvr}%
  \BibitemOpen
  \bibfield  {author} {\bibinfo {author} {\bibfnamefont {B.~P.}\ \bibnamefont
  {Abbott}} \emph {et~al.} (\bibinfo {collaboration} {{LIGO Scientific
  Collaboration and Virgo Collaboration}}),\ }\bibfield  {title} {\bibinfo
  {title} {{GWTC-1: A Gravitational-Wave Transient Catalog of Compact Binary
  Mergers Observed by LIGO and Virgo during the First and Second Observing
  Runs}},\ }\href {https://doi.org/10.1103/PhysRevX.9.031040} {\bibfield
  {journal} {\bibinfo  {journal} {Phys. Rev.}\ }\textbf {\bibinfo {volume}
  {X9}},\ \bibinfo {pages} {031040} (\bibinfo {year} {2019})},\ \Eprint
  {https://arxiv.org/abs/1811.12907} {arXiv:1811.12907 [astro-ph.HE]}
  \BibitemShut {NoStop}%
\bibitem [{\citenamefont {{Venumadhav}}\ \emph {et~al.}(2020)\citenamefont
  {{Venumadhav}}, \citenamefont {{Zackay}}, \citenamefont {{Roulet}},
  \citenamefont {{Dai}},\ and\ \citenamefont
  {{Zaldarriaga}}}]{Venumadhav:2019lyq}%
  \BibitemOpen
  \bibfield  {author} {\bibinfo {author} {\bibfnamefont {T.}~\bibnamefont
  {{Venumadhav}}}, \bibinfo {author} {\bibfnamefont {B.}~\bibnamefont
  {{Zackay}}}, \bibinfo {author} {\bibfnamefont {J.}~\bibnamefont {{Roulet}}},
  \bibinfo {author} {\bibfnamefont {L.}~\bibnamefont {{Dai}}},\ and\ \bibinfo
  {author} {\bibfnamefont {M.}~\bibnamefont {{Zaldarriaga}}},\ }\bibfield
  {title} {\bibinfo {title} {{New binary black hole mergers in the second
  observing run of Advanced LIGO and Advanced Virgo}},\ }\href
  {https://doi.org/10.1103/PhysRevD.101.083030} {\bibfield  {journal} {\bibinfo
   {journal} {\prd}\ }\textbf {\bibinfo {volume} {101}},\ \bibinfo {eid}
  {083030} (\bibinfo {year} {2020})},\ \Eprint
  {https://arxiv.org/abs/1904.07214} {arXiv:1904.07214 [astro-ph.HE]}
  \BibitemShut {NoStop}%
\bibitem [{\citenamefont {{Nitz}}\ \emph
  {et~al.}(2020{\natexlab{a}})\citenamefont {{Nitz}}, \citenamefont {{Dent}},
  \citenamefont {{Davies}}, \citenamefont {{Kumar}}, \citenamefont {{Capano}},
  \citenamefont {{Harry}}, \citenamefont {{Mozzon}}, \citenamefont {{Nuttall}},
  \citenamefont {{Lundgren}},\ and\ \citenamefont
  {{T{\'a}pai}}}]{Nitz:2019hdf}%
  \BibitemOpen
  \bibfield  {author} {\bibinfo {author} {\bibfnamefont {A.~H.}\ \bibnamefont
  {{Nitz}}}, \bibinfo {author} {\bibfnamefont {T.}~\bibnamefont {{Dent}}},
  \bibinfo {author} {\bibfnamefont {G.~S.}\ \bibnamefont {{Davies}}}, \bibinfo
  {author} {\bibfnamefont {S.}~\bibnamefont {{Kumar}}}, \bibinfo {author}
  {\bibfnamefont {C.~D.}\ \bibnamefont {{Capano}}}, \bibinfo {author}
  {\bibfnamefont {I.}~\bibnamefont {{Harry}}}, \bibinfo {author} {\bibfnamefont
  {S.}~\bibnamefont {{Mozzon}}}, \bibinfo {author} {\bibfnamefont
  {L.}~\bibnamefont {{Nuttall}}}, \bibinfo {author} {\bibfnamefont
  {A.}~\bibnamefont {{Lundgren}}},\ and\ \bibinfo {author} {\bibfnamefont
  {M.}~\bibnamefont {{T{\'a}pai}}},\ }\bibfield  {title} {\bibinfo {title}
  {{2-OGC: Open Gravitational-wave Catalog of Binary Mergers from Analysis of
  Public Advanced LIGO and Virgo Data}},\ }\href
  {https://doi.org/10.3847/1538-4357/ab733f} {\bibfield  {journal} {\bibinfo
  {journal} {\apj}\ }\textbf {\bibinfo {volume} {891}},\ \bibinfo {eid} {123}
  (\bibinfo {year} {2020}{\natexlab{a}})},\ \Eprint
  {https://arxiv.org/abs/1910.05331} {arXiv:1910.05331 [astro-ph.HE]}
  \BibitemShut {NoStop}%
\bibitem [{\citenamefont {Abbott}\ \emph
  {et~al.}(2017{\natexlab{a}})\citenamefont {Abbott} \emph
  {et~al.}}]{TheLIGOScientific:2017qsa}%
  \BibitemOpen
  \bibfield  {author} {\bibinfo {author} {\bibfnamefont {B.~P.}\ \bibnamefont
  {Abbott}} \emph {et~al.} (\bibinfo {collaboration} {{LIGO Scientific
  Collaboration and Virgo Collaboration}}),\ }\bibfield  {title} {\bibinfo
  {title} {{GW170817: Observation of Gravitational Waves from a Binary Neutron
  Star Inspiral}},\ }\href {https://doi.org/10.1103/PhysRevLett.119.161101}
  {\bibfield  {journal} {\bibinfo  {journal} {Phys. Rev. Lett.}\ }\textbf
  {\bibinfo {volume} {119}},\ \bibinfo {pages} {161101} (\bibinfo {year}
  {2017}{\natexlab{a}})},\ \Eprint {https://arxiv.org/abs/1710.05832}
  {arXiv:1710.05832 [gr-qc]} \BibitemShut {NoStop}%
\bibitem [{\citenamefont {Abbott}\ \emph
  {et~al.}(2017{\natexlab{b}})\citenamefont {Abbott} \emph
  {et~al.}}]{GBM:2017lvd}%
  \BibitemOpen
  \bibfield  {author} {\bibinfo {author} {\bibfnamefont {B.~P.}\ \bibnamefont
  {Abbott}} \emph {et~al.} (\bibinfo {collaboration} {LIGO Scientific
  Collaboration, Virgo Collaboration, Fermi GBM, INTEGRAL, IceCube, AstroSat
  Cadmium Zinc Telluride Imager Team, IPN, Insight-Hxmt, ANTARES, Swift, AGILE
  Team, 1M2H Team, Dark Energy Camera GW-EM, DES, DLT40, GRAWITA, Fermi-LAT,
  ATCA, ASKAP, Las Cumbres Observatory Group, OzGrav, DWF (Deeper Wider Faster
  Program), AST3, CAASTRO, VINROUGE, MASTER, J-GEM, GROWTH, JAGWAR,
  CaltechNRAO, TTU-NRAO, NuSTAR, Pan-STARRS, MAXI Team, TZAC Consortium, KU,
  Nordic Optical Telescope, ePESSTO, GROND, Texas Tech University, SALT Group,
  TOROS, BOOTES, MWA, CALET, IKI-GW Follow-up, H.E.S.S., LOFAR, LWA, HAWC,
  Pierre Auger, ALMA, Euro VLBI Team, Pi of Sky, Chandra Team at McGill
  University, DFN, ATLAS Telescopes, High Time Resolution Universe Survey,
  RIMAS, RATIR, SKA South Africa/MeerKAT}),\ }\bibfield  {title} {\bibinfo
  {title} {{Multi-messenger Observations of a Binary Neutron Star Merger}},\
  }\href {https://doi.org/10.3847/2041-8213/aa91c9} {\bibfield  {journal}
  {\bibinfo  {journal} {Astrophys. J.}\ }\textbf {\bibinfo {volume} {848}},\
  \bibinfo {pages} {L12} (\bibinfo {year} {2017}{\natexlab{b}})},\ \Eprint
  {https://arxiv.org/abs/1710.05833} {arXiv:1710.05833 [astro-ph.HE]}
  \BibitemShut {NoStop}%
\bibitem [{gra()}]{gracedb}%
  \BibitemOpen
  \href@noop {} {}\bibinfo {howpublished}
  {\url{https://gracedb.ligo.org/}}\BibitemShut {NoStop}%
\bibitem [{\citenamefont {Abbott}\ \emph {et~al.}(2020)\citenamefont {Abbott}
  \emph {et~al.}}]{Abbott:2020uma}%
  \BibitemOpen
  \bibfield  {author} {\bibinfo {author} {\bibfnamefont {B.~P.}\ \bibnamefont
  {Abbott}} \emph {et~al.} (\bibinfo {collaboration} {{LIGO Scientific
  Collaboration and Virgo Collaboration}}),\ }\bibfield  {title} {\bibinfo
  {title} {{GW190425: Observation of a Compact Binary Coalescence with Total
  Mass $\sim 3.4 M_{\odot}$}},\ }\href@noop {} {\  (\bibinfo {year} {2020})},\
  \Eprint {https://arxiv.org/abs/2001.01761} {arXiv:2001.01761 [astro-ph.HE]}
  \BibitemShut {NoStop}%
\bibitem [{KAG()}]{KAGRA}%
  \BibitemOpen
  \href@noop {} {}\bibinfo {howpublished}
  {\url{https://gwcenter.icrr.u-tokyo.ac.jp/en/archives/1381}}\BibitemShut
  {NoStop}%
\bibitem [{\citenamefont {{Amaro-Seoane}}\ \emph {et~al.}(2017)\citenamefont
  {{Amaro-Seoane}}, \citenamefont {{Audley}}, \citenamefont {{Babak}},\ and\
  \citenamefont {{et al}}}]{2017arXiv170200786A}%
  \BibitemOpen
  \bibfield  {author} {\bibinfo {author} {\bibfnamefont {P.}~\bibnamefont
  {{Amaro-Seoane}}}, \bibinfo {author} {\bibfnamefont {H.}~\bibnamefont
  {{Audley}}}, \bibinfo {author} {\bibfnamefont {S.}~\bibnamefont {{Babak}}},\
  and\ \bibinfo {author} {\bibnamefont {{et al}}} (\bibinfo {collaboration}
  {{LISA Team}}),\ }\bibfield  {title} {\bibinfo {title} {{Laser Interferometer
  Space Antenna}},\ }\href@noop {} {\bibfield  {journal} {\bibinfo  {journal}
  {arXiv e-prints}\ ,\ \bibinfo {eid} {arXiv:1702.00786}} (\bibinfo {year}
  {2017})}\BibitemShut {NoStop}%
\bibitem [{\citenamefont {Armano}\ \emph {et~al.}(2016)\citenamefont {Armano},
  \citenamefont {Audley}, \citenamefont {Auger} \emph
  {et~al.}}]{Armano:2016bkm}%
  \BibitemOpen
  \bibfield  {author} {\bibinfo {author} {\bibfnamefont {M.}~\bibnamefont
  {Armano}}, \bibinfo {author} {\bibfnamefont {H.}~\bibnamefont {Audley}},
  \bibinfo {author} {\bibfnamefont {G.}~\bibnamefont {Auger}}, \emph {et~al.},\
  }\bibfield  {title} {\bibinfo {title} {{Sub-Femto- g Free Fall for
  Space-Based Gravitational Wave Observatories: LISA Pathfinder Results}},\
  }\href {https://doi.org/10.1103/PhysRevLett.116.231101} {\bibfield  {journal}
  {\bibinfo  {journal} {Phys. Rev. Lett.}\ }\textbf {\bibinfo {volume} {116}},\
  \bibinfo {pages} {231101} (\bibinfo {year} {2016})},\ \bibinfo {note} {and
  references therein}\BibitemShut {NoStop}%
\bibitem [{\citenamefont {Armano}\ \emph {et~al.}(2018)\citenamefont {Armano},
  \citenamefont {Audley}, \citenamefont {Baird} \emph
  {et~al.}}]{Armano:2018kix}%
  \BibitemOpen
  \bibfield  {author} {\bibinfo {author} {\bibfnamefont {M.}~\bibnamefont
  {Armano}}, \bibinfo {author} {\bibfnamefont {H.}~\bibnamefont {Audley}},
  \bibinfo {author} {\bibfnamefont {J.}~\bibnamefont {Baird}}, \emph {et~al.},\
  }\bibfield  {title} {\bibinfo {title} {{Beyond the Required LISA Free-Fall
  Performance: New LISA Pathfinder Results down to 20 $\mu$Hz}},\ }\href
  {https://doi.org/10.1103/PhysRevLett.120.061101} {\bibfield  {journal}
  {\bibinfo  {journal} {Phys. Rev. Lett.}\ }\textbf {\bibinfo {volume} {120}},\
  \bibinfo {pages} {061101} (\bibinfo {year} {2018})},\ \bibinfo {note} {and
  references therein}\BibitemShut {NoStop}%
\bibitem [{\citenamefont {Abich}\ \emph {et~al.}(2019)\citenamefont {Abich},
  \citenamefont {Braxmaier}, \citenamefont {Gohlke} \emph
  {et~al.}}]{Abich:2019cci}%
  \BibitemOpen
  \bibfield  {author} {\bibinfo {author} {\bibfnamefont {K.}~\bibnamefont
  {Abich}}, \bibinfo {author} {\bibfnamefont {C.}~\bibnamefont {Braxmaier}},
  \bibinfo {author} {\bibfnamefont {M.}~\bibnamefont {Gohlke}}, \emph
  {et~al.},\ }\bibfield  {title} {\bibinfo {title} {{In-Orbit Performance of
  the GRACE Follow-on Laser Ranging Interferometer}},\ }\href
  {https://doi.org/10.1103/PhysRevLett.123.031101} {\bibfield  {journal}
  {\bibinfo  {journal} {Phys. Rev. Lett.}\ }\textbf {\bibinfo {volume} {123}},\
  \bibinfo {pages} {031101} (\bibinfo {year} {2019})},\ \Eprint
  {https://arxiv.org/abs/1907.00104} {arXiv:1907.00104 [astro-ph.IM]}
  \BibitemShut {NoStop}%
\bibitem [{\citenamefont {Hu}\ and\ \citenamefont {Wu}(2017)}]{Hu:2017mde}%
  \BibitemOpen
  \bibfield  {author} {\bibinfo {author} {\bibfnamefont {W.-R.}\ \bibnamefont
  {Hu}}\ and\ \bibinfo {author} {\bibfnamefont {Y.-L.}\ \bibnamefont {Wu}},\
  }\bibfield  {title} {\bibinfo {title} {{The Taiji Program in Space for
  gravitational wave physics and the nature of gravity}},\ }\href
  {https://doi.org/10.1093/nsr/nwx116} {\bibfield  {journal} {\bibinfo
  {journal} {Natl. Sci. Rev.}\ }\textbf {\bibinfo {volume} {4}},\ \bibinfo
  {pages} {685} (\bibinfo {year} {2017})}\BibitemShut {NoStop}%
\bibitem [{\citenamefont {Luo}\ \emph {et~al.}(2016)\citenamefont {Luo} \emph
  {et~al.}}]{Luo:2015ght}%
  \BibitemOpen
  \bibfield  {author} {\bibinfo {author} {\bibfnamefont {J.}~\bibnamefont
  {Luo}} \emph {et~al.} (\bibinfo {collaboration} {TianQin}),\ }\bibfield
  {title} {\bibinfo {title} {{TianQin: a space-borne gravitational wave
  detector}},\ }\href {https://doi.org/10.1088/0264-9381/33/3/035010}
  {\bibfield  {journal} {\bibinfo  {journal} {Class. Quant. Grav.}\ }\textbf
  {\bibinfo {volume} {33}},\ \bibinfo {pages} {035010} (\bibinfo {year}
  {2016})},\ \Eprint {https://arxiv.org/abs/1512.02076} {arXiv:1512.02076
  [astro-ph.IM]} \BibitemShut {NoStop}%
\bibitem [{\citenamefont {Luo}\ \emph {et~al.}(2020)\citenamefont {Luo},
  \citenamefont {Guo}, \citenamefont {Jin}, \citenamefont {Wu},\ and\
  \citenamefont {Hu}}]{Luo:2020}%
  \BibitemOpen
  \bibfield  {author} {\bibinfo {author} {\bibfnamefont {Z.}~\bibnamefont
  {Luo}}, \bibinfo {author} {\bibfnamefont {Z.}~\bibnamefont {Guo}}, \bibinfo
  {author} {\bibfnamefont {G.}~\bibnamefont {Jin}}, \bibinfo {author}
  {\bibfnamefont {Y.}~\bibnamefont {Wu}},\ and\ \bibinfo {author}
  {\bibfnamefont {W.}~\bibnamefont {Hu}},\ }\bibfield  {title} {\bibinfo
  {title} {{A brief analysis to Taiji: Science and technology}},\ }\href
  {https://doi.org/doi.org/10.1016/j.rinp.2019.102918} {\bibfield  {journal}
  {\bibinfo  {journal} {Results in Physics}\ }\textbf {\bibinfo {volume}
  {16}},\ \bibinfo {pages} {102918} (\bibinfo {year} {2020})}\BibitemShut
  {NoStop}%
\bibitem [{Tia()}]{Tianqin-1}%
  \BibitemOpen
  \href@noop {} {}\bibinfo {howpublished}
  {\url{https://en.wikipedia.org/wiki/TianQin}}\BibitemShut {NoStop}%
\bibitem [{ESA()}]{ESAnews}%
  \BibitemOpen
  \href@noop {} {}\bibinfo {howpublished}
  {\url{https://www.esa.int/Newsroom/Press_Releases/ESA_ministers_commit_to_biggest_ever_budget}
  and
  \url{https://www.lisamission.org/news/top-news/esa-ministers-commit-biggest-ever-budget}}\BibitemShut
  {NoStop}%
\bibitem [{\citenamefont {{Nandra}}\ \emph {et~al.}(2013)\citenamefont
  {{Nandra}}, \citenamefont {{Barret}}, \citenamefont {{Barcons}},\ and\
  \citenamefont {{et al}}}]{2013arXiv1306.2307N}%
  \BibitemOpen
  \bibfield  {author} {\bibinfo {author} {\bibfnamefont {K.}~\bibnamefont
  {{Nandra}}}, \bibinfo {author} {\bibfnamefont {D.}~\bibnamefont {{Barret}}},
  \bibinfo {author} {\bibfnamefont {X.}~\bibnamefont {{Barcons}}},\ and\
  \bibinfo {author} {\bibnamefont {{et al}}} (\bibinfo {collaboration} {{Athena
  Team}}),\ }\bibfield  {title} {\bibinfo {title} {{The Hot and Energetic
  Universe: A White Paper presenting the science theme motivating the Athena+
  mission}},\ }\href@noop {} {\bibfield  {journal} {\bibinfo  {journal} {arXiv
  e-prints}\ ,\ \bibinfo {eid} {arXiv:1306.2307}} (\bibinfo {year} {2013})},\
  \Eprint {https://arxiv.org/abs/1306.2307} {arXiv:1306.2307 [astro-ph.HE]}
  \BibitemShut {NoStop}%
\bibitem [{\citenamefont {Colpi}\ \emph {et~al.}()\citenamefont {Colpi},
  \citenamefont {Fabian}, \citenamefont {Guainazzi}, \citenamefont {McNamara},
  \citenamefont {Piro},\ and\ \citenamefont {Tanvir}}]{Colpi:2019}%
  \BibitemOpen
  \bibfield  {author} {\bibinfo {author} {\bibfnamefont {M.}~\bibnamefont
  {Colpi}}, \bibinfo {author} {\bibfnamefont {A.~C.}\ \bibnamefont {Fabian}},
  \bibinfo {author} {\bibfnamefont {M.}~\bibnamefont {Guainazzi}}, \bibinfo
  {author} {\bibfnamefont {P.}~\bibnamefont {McNamara}}, \bibinfo {author}
  {\bibfnamefont {L.}~\bibnamefont {Piro}},\ and\ \bibinfo {author}
  {\bibfnamefont {N.~a.}\ \bibnamefont {Tanvir}},\ }\href@noop {} {\bibinfo
  {title} {{Athena-LISA Synergies}}},\ \bibinfo {howpublished}
  {\url{https://www.cosmos.esa.int/documents/678316/1700384/Athena_LISA_Whitepaper_Iss1.0.pdf}}\BibitemShut
  {NoStop}%
\bibitem [{\citenamefont {{McGee}}\ \emph {et~al.}(2020)\citenamefont
  {{McGee}}, \citenamefont {{Sesana}},\ and\ \citenamefont
  {{Vecchio}}}]{2020NatAs.tmp....4M}%
  \BibitemOpen
  \bibfield  {author} {\bibinfo {author} {\bibfnamefont {S.}~\bibnamefont
  {{McGee}}}, \bibinfo {author} {\bibfnamefont {A.}~\bibnamefont {{Sesana}}},\
  and\ \bibinfo {author} {\bibfnamefont {A.}~\bibnamefont {{Vecchio}}},\
  }\bibfield  {title} {\bibinfo {title} {{Linking gravitational waves and X-ray
  phenomena with joint LISA and Athena observations}},\ }\href
  {https://doi.org/10.1038/s41550-019-0969-7} {\bibfield  {journal} {\bibinfo
  {journal} {Nature Astronomy}\ }\textbf {\bibinfo {volume} {4}},\ \bibinfo
  {pages} {26} (\bibinfo {year} {2020})},\ \Eprint
  {https://arxiv.org/abs/1811.00050} {arXiv:1811.00050 [astro-ph.HE]}
  \BibitemShut {NoStop}%
\bibitem [{\citenamefont {Cutler}(1998)}]{Cutler:1997ta}%
  \BibitemOpen
  \bibfield  {author} {\bibinfo {author} {\bibfnamefont {C.}~\bibnamefont
  {Cutler}},\ }\bibfield  {title} {\bibinfo {title} {{Angular resolution of the
  LISA gravitational wave detector}},\ }\href
  {https://doi.org/10.1103/PhysRevD.57.7089} {\bibfield  {journal} {\bibinfo
  {journal} {Phys. Rev.}\ }\textbf {\bibinfo {volume} {D57}},\ \bibinfo {pages}
  {7089} (\bibinfo {year} {1998})},\ \Eprint
  {https://arxiv.org/abs/gr-qc/9703068} {arXiv:gr-qc/9703068 [gr-qc]}
  \BibitemShut {NoStop}%
\bibitem [{\citenamefont {{Peterseim}}\ \emph {et~al.}(1997)\citenamefont
  {{Peterseim}}, \citenamefont {{Jennrich}}, \citenamefont {{Danzmann}},\ and\
  \citenamefont {{Schutz}}}]{1997CQGra..14.1507P}%
  \BibitemOpen
  \bibfield  {author} {\bibinfo {author} {\bibfnamefont {M.}~\bibnamefont
  {{Peterseim}}}, \bibinfo {author} {\bibfnamefont {O.}~\bibnamefont
  {{Jennrich}}}, \bibinfo {author} {\bibfnamefont {K.}~\bibnamefont
  {{Danzmann}}},\ and\ \bibinfo {author} {\bibfnamefont {B.~F.}\ \bibnamefont
  {{Schutz}}},\ }\bibfield  {title} {\bibinfo {title} {{Angular resolution of
  LISA}},\ }\href {https://doi.org/10.1088/0264-9381/14/6/019} {\bibfield
  {journal} {\bibinfo  {journal} {Classical and Quantum Gravity}\ }\textbf
  {\bibinfo {volume} {14}},\ \bibinfo {pages} {1507} (\bibinfo {year}
  {1997})}\BibitemShut {NoStop}%
\bibitem [{\citenamefont {{Cutler}}\ and\ \citenamefont
  {{Vecchio}}(1998)}]{1998AIPC..456...95C}%
  \BibitemOpen
  \bibfield  {author} {\bibinfo {author} {\bibfnamefont {C.}~\bibnamefont
  {{Cutler}}}\ and\ \bibinfo {author} {\bibfnamefont {A.}~\bibnamefont
  {{Vecchio}}},\ }\bibfield  {title} {\bibinfo {title} {{LISA's angular
  resolution for monochromatic sources}},\ }in\ \href
  {https://doi.org/10.1063/1.57427} {\emph {\bibinfo {booktitle} {Laser
  Interferometer Space Antenna, Second International LISA Symposium on the
  Detection and Observation of Gravitational Waves in Space}}},\ \bibinfo
  {series} {American Institute of Physics Conference Series}, Vol.\ \bibinfo
  {volume} {456},\ \bibinfo {editor} {edited by\ \bibinfo {editor}
  {\bibfnamefont {W.~M.}\ \bibnamefont {{Folkner}}}}\ (\bibinfo {year} {1998})\
  pp.\ \bibinfo {pages} {95--100}\BibitemShut {NoStop}%
\bibitem [{\citenamefont {Vecchio}\ and\ \citenamefont
  {Wickham}(2004)}]{Vecchio:2004ec}%
  \BibitemOpen
  \bibfield  {author} {\bibinfo {author} {\bibfnamefont {A.}~\bibnamefont
  {Vecchio}}\ and\ \bibinfo {author} {\bibfnamefont {E.~D.~L.}\ \bibnamefont
  {Wickham}},\ }\bibfield  {title} {\bibinfo {title} {{The Effect of the LISA
  response function on observations of monochromatic sources}},\ }\href
  {https://doi.org/10.1103/PhysRevD.70.082002} {\bibfield  {journal} {\bibinfo
  {journal} {Phys. Rev.}\ }\textbf {\bibinfo {volume} {D70}},\ \bibinfo {pages}
  {082002} (\bibinfo {year} {2004})},\ \Eprint
  {https://arxiv.org/abs/gr-qc/0406039} {arXiv:gr-qc/0406039 [gr-qc]}
  \BibitemShut {NoStop}%
\bibitem [{\citenamefont {{Arun}}\ \emph {et~al.}(2007)\citenamefont {{Arun}},
  \citenamefont {{Iyer}}, \citenamefont {{Sathyaprakash}}, \citenamefont
  {{Sinha}},\ and\ \citenamefont {{van den Broeck}}}]{2007PhRvD..76j4016A}%
  \BibitemOpen
  \bibfield  {author} {\bibinfo {author} {\bibfnamefont {K.~G.}\ \bibnamefont
  {{Arun}}}, \bibinfo {author} {\bibfnamefont {B.~R.}\ \bibnamefont {{Iyer}}},
  \bibinfo {author} {\bibfnamefont {B.~S.}\ \bibnamefont {{Sathyaprakash}}},
  \bibinfo {author} {\bibfnamefont {S.}~\bibnamefont {{Sinha}}},\ and\ \bibinfo
  {author} {\bibfnamefont {C.}~\bibnamefont {{van den Broeck}}},\ }\bibfield
  {title} {\bibinfo {title} {{Higher signal harmonics, LISA's angular
  resolution, and dark energy}},\ }\href
  {https://doi.org/10.1103/PhysRevD.76.104016} {\bibfield  {journal} {\bibinfo
  {journal} {\prd}\ }\textbf {\bibinfo {volume} {76}},\ \bibinfo {eid} {104016}
  (\bibinfo {year} {2007})},\ \Eprint {https://arxiv.org/abs/0707.3920}
  {arXiv:0707.3920 [astro-ph]} \BibitemShut {NoStop}%
\bibitem [{\citenamefont {{Babak}}\ \emph {et~al.}(2008)\citenamefont
  {{Babak}}, \citenamefont {{Hannam}}, \citenamefont {{Husa}},\ and\
  \citenamefont {{Schutz}}}]{2008arXiv0806.1591B}%
  \BibitemOpen
  \bibfield  {author} {\bibinfo {author} {\bibfnamefont {S.}~\bibnamefont
  {{Babak}}}, \bibinfo {author} {\bibfnamefont {M.}~\bibnamefont {{Hannam}}},
  \bibinfo {author} {\bibfnamefont {S.}~\bibnamefont {{Husa}}},\ and\ \bibinfo
  {author} {\bibfnamefont {B.}~\bibnamefont {{Schutz}}},\ }\bibfield  {title}
  {\bibinfo {title} {{Resolving Super Massive Black Holes with LISA}},\
  }\href@noop {} {\bibfield  {journal} {\bibinfo  {journal} {arXiv e-prints}\ }
  (\bibinfo {year} {2008})},\ \Eprint {https://arxiv.org/abs/0806.1591}
  {arXiv:0806.1591 [gr-qc]} \BibitemShut {NoStop}%
\bibitem [{\citenamefont {{Sesana}}(2016)}]{2016PhRvL.116w1102S}%
  \BibitemOpen
  \bibfield  {author} {\bibinfo {author} {\bibfnamefont {A.}~\bibnamefont
  {{Sesana}}},\ }\bibfield  {title} {\bibinfo {title} {{Prospects for Multiband
  Gravitational-Wave Astronomy after GW150914}},\ }\href
  {https://doi.org/10.1103/PhysRevLett.116.231102} {\bibfield  {journal}
  {\bibinfo  {journal} {\prl}\ }\textbf {\bibinfo {volume} {116}},\ \bibinfo
  {eid} {231102} (\bibinfo {year} {2016})},\ \Eprint
  {https://arxiv.org/abs/1602.06951} {arXiv:1602.06951 [gr-qc]} \BibitemShut
  {NoStop}%
\bibitem [{\citenamefont {Vallisneri}\ and\ \citenamefont
  {Galley}(2012)}]{Vallisneri:2012np}%
  \BibitemOpen
  \bibfield  {author} {\bibinfo {author} {\bibfnamefont {M.}~\bibnamefont
  {Vallisneri}}\ and\ \bibinfo {author} {\bibfnamefont {C.~R.}\ \bibnamefont
  {Galley}},\ }\bibfield  {title} {\bibinfo {title} {{Non-sky-averaged
  sensitivity curves for space-based gravitational-wave observatories}},\
  }\href {https://doi.org/10.1088/0264-9381/29/12/124015} {\bibfield  {journal}
  {\bibinfo  {journal} {Class. Quant. Grav.}\ }\textbf {\bibinfo {volume}
  {29}},\ \bibinfo {pages} {124015} (\bibinfo {year} {2012})},\ \Eprint
  {https://arxiv.org/abs/1201.3684} {arXiv:1201.3684 [gr-qc]} \BibitemShut
  {NoStop}%
\bibitem [{\citenamefont {{McWilliams}}\ \emph {et~al.}(2010)\citenamefont
  {{McWilliams}}, \citenamefont {{Thorpe}}, \citenamefont {{Baker}},\ and\
  \citenamefont {{Kelly}}}]{2010PhRvD..81f4014M}%
  \BibitemOpen
  \bibfield  {author} {\bibinfo {author} {\bibfnamefont {S.~T.}\ \bibnamefont
  {{McWilliams}}}, \bibinfo {author} {\bibfnamefont {J.~I.}\ \bibnamefont
  {{Thorpe}}}, \bibinfo {author} {\bibfnamefont {J.~G.}\ \bibnamefont
  {{Baker}}},\ and\ \bibinfo {author} {\bibfnamefont {B.~J.}\ \bibnamefont
  {{Kelly}}},\ }\bibfield  {title} {\bibinfo {title} {{Impact of mergers on
  LISA parameter estimation for nonspinning black hole binaries}},\ }\href
  {https://doi.org/10.1103/PhysRevD.81.064014} {\bibfield  {journal} {\bibinfo
  {journal} {\prd}\ }\textbf {\bibinfo {volume} {81}},\ \bibinfo {eid} {064014}
  (\bibinfo {year} {2010})},\ \Eprint {https://arxiv.org/abs/0911.1078}
  {arXiv:0911.1078 [gr-qc]} \BibitemShut {NoStop}%
\bibitem [{\citenamefont {{McWilliams}}\ \emph {et~al.}(2011)\citenamefont
  {{McWilliams}}, \citenamefont {{Lang}}, \citenamefont {{Baker}},\ and\
  \citenamefont {{Thorpe}}}]{2011PhRvD..84f4003M}%
  \BibitemOpen
  \bibfield  {author} {\bibinfo {author} {\bibfnamefont {S.~T.}\ \bibnamefont
  {{McWilliams}}}, \bibinfo {author} {\bibfnamefont {R.~N.}\ \bibnamefont
  {{Lang}}}, \bibinfo {author} {\bibfnamefont {J.~G.}\ \bibnamefont
  {{Baker}}},\ and\ \bibinfo {author} {\bibfnamefont {J.~I.}\ \bibnamefont
  {{Thorpe}}},\ }\bibfield  {title} {\bibinfo {title} {{Sky localization of
  complete inspiral-merger-ringdown signals for nonspinning massive black hole
  binaries}},\ }\href {https://doi.org/10.1103/PhysRevD.84.064003} {\bibfield
  {journal} {\bibinfo  {journal} {\prd}\ }\textbf {\bibinfo {volume} {84}},\
  \bibinfo {eid} {064003} (\bibinfo {year} {2011})},\ \Eprint
  {https://arxiv.org/abs/1104.5650} {arXiv:1104.5650 [gr-qc]} \BibitemShut
  {NoStop}%
\bibitem [{\citenamefont {Vallisneri}(2005{\natexlab{a}})}]{Vallisneri:2004bn}%
  \BibitemOpen
  \bibfield  {author} {\bibinfo {author} {\bibfnamefont {M.}~\bibnamefont
  {Vallisneri}},\ }\bibfield  {title} {\bibinfo {title} {{Synthetic LISA:
  Simulating time delay interferometry in a model LISA}},\ }\href
  {https://doi.org/10.1103/PhysRevD.71.022001} {\bibfield  {journal} {\bibinfo
  {journal} {Phys. Rev.}\ }\textbf {\bibinfo {volume} {D71}},\ \bibinfo {pages}
  {022001} (\bibinfo {year} {2005}{\natexlab{a}})},\ \Eprint
  {https://arxiv.org/abs/gr-qc/0407102} {arXiv:gr-qc/0407102 [gr-qc]}
  \BibitemShut {NoStop}%
\bibitem [{\citenamefont {{Petiteau}}\ \emph {et~al.}(2008)\citenamefont
  {{Petiteau}}, \citenamefont {{Auger}}, \citenamefont {{Halloin}},
  \citenamefont {{Jeannin}}, \citenamefont {{Plagnol}}, \citenamefont
  {{Pireaux}}, \citenamefont {{Regimbau}},\ and\ \citenamefont
  {{Vinet}}}]{2008PhRvD..77b3002P}%
  \BibitemOpen
  \bibfield  {author} {\bibinfo {author} {\bibfnamefont {A.}~\bibnamefont
  {{Petiteau}}}, \bibinfo {author} {\bibfnamefont {G.}~\bibnamefont {{Auger}}},
  \bibinfo {author} {\bibfnamefont {H.}~\bibnamefont {{Halloin}}}, \bibinfo
  {author} {\bibfnamefont {O.}~\bibnamefont {{Jeannin}}}, \bibinfo {author}
  {\bibfnamefont {E.}~\bibnamefont {{Plagnol}}}, \bibinfo {author}
  {\bibfnamefont {S.}~\bibnamefont {{Pireaux}}}, \bibinfo {author}
  {\bibfnamefont {T.}~\bibnamefont {{Regimbau}}},\ and\ \bibinfo {author}
  {\bibfnamefont {J.-Y.}\ \bibnamefont {{Vinet}}},\ }\bibfield  {title}
  {\bibinfo {title} {{LISACode: A scientific simulator of LISA}},\ }\href
  {https://doi.org/10.1103/PhysRevD.77.023002} {\bibfield  {journal} {\bibinfo
  {journal} {\prd}\ }\textbf {\bibinfo {volume} {77}},\ \bibinfo {eid} {023002}
  (\bibinfo {year} {2008})},\ \Eprint {https://arxiv.org/abs/0802.2023}
  {arXiv:0802.2023 [gr-qc]} \BibitemShut {NoStop}%
\bibitem [{\citenamefont {{Bayle}}\ \emph {et~al.}(2019)\citenamefont
  {{Bayle}}, \citenamefont {{Lilley}}, \citenamefont {{Petiteau}},\ and\
  \citenamefont {{Halloin}}}]{2019PhRvD..99h4023B}%
  \BibitemOpen
  \bibfield  {author} {\bibinfo {author} {\bibfnamefont {J.-B.}\ \bibnamefont
  {{Bayle}}}, \bibinfo {author} {\bibfnamefont {M.}~\bibnamefont {{Lilley}}},
  \bibinfo {author} {\bibfnamefont {A.}~\bibnamefont {{Petiteau}}},\ and\
  \bibinfo {author} {\bibfnamefont {H.}~\bibnamefont {{Halloin}}},\ }\bibfield
  {title} {\bibinfo {title} {{Effect of filters on the time-delay
  interferometry residual laser noise for LISA}},\ }\href
  {https://doi.org/10.1103/PhysRevD.99.084023} {\bibfield  {journal} {\bibinfo
  {journal} {\prd}\ }\textbf {\bibinfo {volume} {99}},\ \bibinfo {eid} {084023}
  (\bibinfo {year} {2019})},\ \Eprint {https://arxiv.org/abs/1811.01575}
  {arXiv:1811.01575 [astro-ph.IM]} \BibitemShut {NoStop}%
\bibitem [{\citenamefont {Crowder}\ and\ \citenamefont
  {Cornish}(2005)}]{Crowder:2005nr}%
  \BibitemOpen
  \bibfield  {author} {\bibinfo {author} {\bibfnamefont {J.}~\bibnamefont
  {Crowder}}\ and\ \bibinfo {author} {\bibfnamefont {N.~J.}\ \bibnamefont
  {Cornish}},\ }\bibfield  {title} {\bibinfo {title} {{Beyond LISA: Exploring
  future gravitational wave missions}},\ }\href
  {https://doi.org/10.1103/PhysRevD.72.083005} {\bibfield  {journal} {\bibinfo
  {journal} {Phys. Rev.}\ }\textbf {\bibinfo {volume} {D72}},\ \bibinfo {pages}
  {083005} (\bibinfo {year} {2005})},\ \Eprint
  {https://arxiv.org/abs/gr-qc/0506015} {arXiv:gr-qc/0506015 [gr-qc]}
  \BibitemShut {NoStop}%
\bibitem [{\citenamefont {{Tinto}}\ and\ \citenamefont {{de
  Araujo}}(2016)}]{2016PhRvD..94h1101T}%
  \BibitemOpen
  \bibfield  {author} {\bibinfo {author} {\bibfnamefont {M.}~\bibnamefont
  {{Tinto}}}\ and\ \bibinfo {author} {\bibfnamefont {J.~C.~N.}\ \bibnamefont
  {{de Araujo}}},\ }\bibfield  {title} {\bibinfo {title} {{Coherent
  observations of gravitational radiation with LISA and gLISA}},\ }\href
  {https://doi.org/10.1103/PhysRevD.94.081101} {\bibfield  {journal} {\bibinfo
  {journal} {\prd}\ }\textbf {\bibinfo {volume} {94}},\ \bibinfo {eid}
  {081101(R)} (\bibinfo {year} {2016})},\ \Eprint
  {https://arxiv.org/abs/1608.04790} {arXiv:1608.04790 [astro-ph.IM]}
  \BibitemShut {NoStop}%
\bibitem [{\citenamefont {Tinto}(2017)}]{Tinto:2017kre}%
  \BibitemOpen
  \bibfield  {author} {\bibinfo {author} {\bibfnamefont {M.}~\bibnamefont
  {Tinto}},\ }\bibfield  {title} {\bibinfo {title} {{Enhanced Gravitational
  Wave Science with LISA and gLISA}},\ }\bibfield  {booktitle} {\emph {\bibinfo
  {booktitle} {{Proceedings, 11th International LISA Symposium: Zurich,
  Switzerland, September 5-9, 2016}}},\ }\href
  {https://doi.org/10.1088/1742-6596/840/1/012017} {\bibfield  {journal}
  {\bibinfo  {journal} {J. Phys. Conf. Ser.}\ }\textbf {\bibinfo {volume}
  {840}},\ \bibinfo {pages} {012017} (\bibinfo {year} {2017})}\BibitemShut
  {NoStop}%
\bibitem [{\citenamefont {Ruan}\ \emph {et~al.}(2019)\citenamefont {Ruan},
  \citenamefont {Liu}, \citenamefont {Guo}, \citenamefont {Wu},\ and\
  \citenamefont {Cai}}]{Ruan:2019tje}%
  \BibitemOpen
  \bibfield  {author} {\bibinfo {author} {\bibfnamefont {W.-H.}\ \bibnamefont
  {Ruan}}, \bibinfo {author} {\bibfnamefont {C.}~\bibnamefont {Liu}}, \bibinfo
  {author} {\bibfnamefont {Z.-K.}\ \bibnamefont {Guo}}, \bibinfo {author}
  {\bibfnamefont {Y.-L.}\ \bibnamefont {Wu}},\ and\ \bibinfo {author}
  {\bibfnamefont {R.-G.}\ \bibnamefont {Cai}},\ }\bibfield  {title} {\bibinfo
  {title} {{The LISA-Taiji network: precision localization of massive black
  hole binaries}},\ }\href@noop {} {\  (\bibinfo {year} {2019})},\ \Eprint
  {https://arxiv.org/abs/1909.07104} {arXiv:1909.07104 [gr-qc]} \BibitemShut
  {NoStop}%
\bibitem [{\citenamefont {{LISA~Study~Team}}(2000)}]{LISA2000}%
  \BibitemOpen
  \bibfield  {author} {\bibinfo {author} {\bibnamefont {{LISA~Study~Team}}},\
  }\href@noop {} {\emph {\bibinfo {title} {LISA (Laser Interferometer Space
  Antenna): A Cornerstone Mission for the Observation of Gravitational
  Waves}}},\ \bibinfo {type} {Tech. Rep.}\ \bibinfo {number} {11}\ (\bibinfo
  {institution} {ESA-SCI},\ \bibinfo {year} {2000})\ \bibinfo {note} {system
  and Technology Study Report}\BibitemShut {NoStop}%
\bibitem [{\citenamefont {Folkner}\ \emph {et~al.}(1997)\citenamefont
  {Folkner}, \citenamefont {Sweetser}, \citenamefont {Vincent}, \citenamefont
  {Hechler},\ and\ \citenamefont {Bender}}]{Folkner:1997hn}%
  \BibitemOpen
  \bibfield  {author} {\bibinfo {author} {\bibfnamefont {W.~M.}\ \bibnamefont
  {Folkner}}, \bibinfo {author} {\bibfnamefont {T.~H.}\ \bibnamefont
  {Sweetser}}, \bibinfo {author} {\bibfnamefont {M.~A.}\ \bibnamefont
  {Vincent}}, \bibinfo {author} {\bibfnamefont {F.}~\bibnamefont {Hechler}},\
  and\ \bibinfo {author} {\bibfnamefont {P.~L.}\ \bibnamefont {Bender}},\
  }\bibfield  {title} {\bibinfo {title} {{LISA orbit selection and
  stability}},\ }\bibfield  {booktitle} {\emph {\bibinfo {booktitle} {{1st
  International LISA Symposium on Gravitational Waves Oxfordshire, England,
  July 9-12, 1996}}},\ }\href {https://doi.org/10.1088/0264-9381/14/6/003}
  {\bibfield  {journal} {\bibinfo  {journal} {Class. Quant. Grav.}\ }\textbf
  {\bibinfo {volume} {14}},\ \bibinfo {pages} {1405} (\bibinfo {year}
  {1997})}\BibitemShut {NoStop}%
\bibitem [{\citenamefont {Dhurandhar}\ \emph {et~al.}(2005)\citenamefont
  {Dhurandhar}, \citenamefont {Nayak}, \citenamefont {Koshti},\ and\
  \citenamefont {Vinet}}]{Dhurandhar:2004rv}%
  \BibitemOpen
  \bibfield  {author} {\bibinfo {author} {\bibfnamefont {S.~V.}\ \bibnamefont
  {Dhurandhar}}, \bibinfo {author} {\bibfnamefont {K.~R.}\ \bibnamefont
  {Nayak}}, \bibinfo {author} {\bibfnamefont {S.}~\bibnamefont {Koshti}},\ and\
  \bibinfo {author} {\bibfnamefont {J.~Y.}\ \bibnamefont {Vinet}},\ }\bibfield
  {title} {\bibinfo {title} {{Fundamentals of the LISA stable flight
  formation}},\ }\href {https://doi.org/10.1088/0264-9381/22/3/002} {\bibfield
  {journal} {\bibinfo  {journal} {Class. Quant. Grav.}\ }\textbf {\bibinfo
  {volume} {22}},\ \bibinfo {pages} {481} (\bibinfo {year} {2005})},\ \Eprint
  {https://arxiv.org/abs/gr-qc/0410093} {arXiv:gr-qc/0410093 [gr-qc]}
  \BibitemShut {NoStop}%
\bibitem [{\citenamefont {Nayak}\ \emph {et~al.}(2006)\citenamefont {Nayak},
  \citenamefont {Koshti}, \citenamefont {Dhurandhar},\ and\ \citenamefont
  {Vinet}}]{Nayak:2006zm}%
  \BibitemOpen
  \bibfield  {author} {\bibinfo {author} {\bibfnamefont {K.~R.}\ \bibnamefont
  {Nayak}}, \bibinfo {author} {\bibfnamefont {S.}~\bibnamefont {Koshti}},
  \bibinfo {author} {\bibfnamefont {S.~V.}\ \bibnamefont {Dhurandhar}},\ and\
  \bibinfo {author} {\bibfnamefont {J.~Y.}\ \bibnamefont {Vinet}},\ }\bibfield
  {title} {\bibinfo {title} {{On the minimum flexing of LISA's arms}},\ }\href
  {https://doi.org/10.1088/0264-9381/23/5/017} {\bibfield  {journal} {\bibinfo
  {journal} {Class. Quant. Grav.}\ }\textbf {\bibinfo {volume} {23}},\ \bibinfo
  {pages} {1763} (\bibinfo {year} {2006})}\BibitemShut {NoStop}%
\bibitem [{\citenamefont {Wu}\ \emph {et~al.}(2019)\citenamefont {Wu},
  \citenamefont {Huang},\ and\ \citenamefont {Qiao}}]{Wu:2019thj}%
  \BibitemOpen
  \bibfield  {author} {\bibinfo {author} {\bibfnamefont {B.}~\bibnamefont
  {Wu}}, \bibinfo {author} {\bibfnamefont {C.-G.}\ \bibnamefont {Huang}},\ and\
  \bibinfo {author} {\bibfnamefont {C.-F.}\ \bibnamefont {Qiao}},\ }\bibfield
  {title} {\bibinfo {title} {{Analytical analysis on the orbits of Taiji
  spacecrafts}},\ }\href {https://doi.org/10.1103/PhysRevD.100.122001}
  {\bibfield  {journal} {\bibinfo  {journal} {Phys. Rev.}\ }\textbf {\bibinfo
  {volume} {D100}},\ \bibinfo {pages} {122001} (\bibinfo {year} {2019})},\
  \Eprint {https://arxiv.org/abs/1907.06178} {arXiv:1907.06178 [gr-qc]}
  \BibitemShut {NoStop}%
\bibitem [{\citenamefont {Yi}\ \emph {et~al.}(2008)\citenamefont {Yi},
  \citenamefont {Li}, \citenamefont {Heinzel}, \citenamefont {Rudiger},
  \citenamefont {Jennrich}, \citenamefont {Wang}, \citenamefont {Xia},
  \citenamefont {Zeng},\ and\ \citenamefont {Zhao}}]{Yi:2008zza}%
  \BibitemOpen
  \bibfield  {author} {\bibinfo {author} {\bibfnamefont {Z.~H.}\ \bibnamefont
  {Yi}}, \bibinfo {author} {\bibfnamefont {G.}~\bibnamefont {Li}}, \bibinfo
  {author} {\bibfnamefont {G.}~\bibnamefont {Heinzel}}, \bibinfo {author}
  {\bibfnamefont {A.}~\bibnamefont {Rudiger}}, \bibinfo {author} {\bibfnamefont
  {O.}~\bibnamefont {Jennrich}}, \bibinfo {author} {\bibfnamefont
  {L.}~\bibnamefont {Wang}}, \bibinfo {author} {\bibfnamefont {Y.}~\bibnamefont
  {Xia}}, \bibinfo {author} {\bibfnamefont {F.}~\bibnamefont {Zeng}},\ and\
  \bibinfo {author} {\bibfnamefont {H.}~\bibnamefont {Zhao}},\ }\bibfield
  {title} {\bibinfo {title} {{Coorbital restricted problem and its application
  in the design of the orbits of the LISA spacecraft}},\ }\bibfield
  {booktitle} {\emph {\bibinfo {booktitle} {{Laser astrodynamics, space test of
  relativity and gravitational-wave astronomy. Proceedings, 3rd International
  ASTROD Symposium, Beijing, P.R. China, July 13-16, 2006}}},\ }\href
  {https://doi.org/10.1142/S0218271808012668} {\bibfield  {journal} {\bibinfo
  {journal} {Int. J. Mod. Phys.}\ }\textbf {\bibinfo {volume} {D17}},\ \bibinfo
  {pages} {1005} (\bibinfo {year} {2008})}\BibitemShut {NoStop}%
\bibitem [{\citenamefont {Li}\ \emph {et~al.}(2008)\citenamefont {Li},
  \citenamefont {Yi}, \citenamefont {Heinzel}, \citenamefont {Rudiger},
  \citenamefont {Jennrich}, \citenamefont {Wang}, \citenamefont {Xia},
  \citenamefont {Zeng},\ and\ \citenamefont {Zhao}}]{Li:2008al}%
  \BibitemOpen
  \bibfield  {author} {\bibinfo {author} {\bibfnamefont {G.}~\bibnamefont
  {Li}}, \bibinfo {author} {\bibfnamefont {Z.}~\bibnamefont {Yi}}, \bibinfo
  {author} {\bibfnamefont {G.}~\bibnamefont {Heinzel}}, \bibinfo {author}
  {\bibfnamefont {A.}~\bibnamefont {Rudiger}}, \bibinfo {author} {\bibfnamefont
  {O.}~\bibnamefont {Jennrich}}, \bibinfo {author} {\bibfnamefont
  {L.}~\bibnamefont {Wang}}, \bibinfo {author} {\bibfnamefont {Y.}~\bibnamefont
  {Xia}}, \bibinfo {author} {\bibfnamefont {F.}~\bibnamefont {Zeng}},\ and\
  \bibinfo {author} {\bibfnamefont {H.}~\bibnamefont {Zhao}},\ }\bibfield
  {title} {\bibinfo {title} {{Methods for orbit optimization for the LISA
  gravitational wave observatory}},\ }\bibfield  {booktitle} {\emph {\bibinfo
  {booktitle} {{Laser astrodynamics, space test of relativity and
  gravitational-wave astronomy. Proceedings, 3rd International ASTROD
  Symposium, Beijing, P.R. China, July 13-16, 2006}}},\ }\href
  {https://doi.org/10.1142/S021827180801267X} {\bibfield  {journal} {\bibinfo
  {journal} {Int. J. Mod. Phys.}\ }\textbf {\bibinfo {volume} {D17}},\ \bibinfo
  {pages} {1021} (\bibinfo {year} {2008})}\BibitemShut {NoStop}%
\bibitem [{\citenamefont {Wang}(2011)}]{Wang:2011}%
  \BibitemOpen
  \bibfield  {author} {\bibinfo {author} {\bibfnamefont {G.}~\bibnamefont
  {Wang}},\ }\href@noop {} {\bibinfo {title} {{Time-delay Interferometry for
  ASTROD-GW}}} (\bibinfo {year} {2011})\BibitemShut {NoStop}%
\bibitem [{\citenamefont {Wang}\ and\ \citenamefont {Ni}(2012)}]{Wang:2014aea}%
  \BibitemOpen
  \bibfield  {author} {\bibinfo {author} {\bibfnamefont {G.}~\bibnamefont
  {Wang}}\ and\ \bibinfo {author} {\bibfnamefont {W.-T.}\ \bibnamefont {Ni}},\
  }\bibfield  {title} {\bibinfo {title} {{Time-delay Interferometry for
  ASTROD-GW}},\ }\href {https://doi.org/10.1016/j.chinastron.2012.04.009}
  {\bibfield  {journal} {\bibinfo  {journal} {Chin. Astron. Astrophys.}\
  }\textbf {\bibinfo {volume} {36}},\ \bibinfo {pages} {211} (\bibinfo {year}
  {2012})},\ \bibinfo {note} {and references therein}\BibitemShut {NoStop}%
\bibitem [{\citenamefont {Wang}\ and\ \citenamefont
  {Ni}(2013{\natexlab{a}})}]{Wang:2012ce}%
  \BibitemOpen
  \bibfield  {author} {\bibinfo {author} {\bibfnamefont {G.}~\bibnamefont
  {Wang}}\ and\ \bibinfo {author} {\bibfnamefont {W.-T.}\ \bibnamefont {Ni}},\
  }\bibfield  {title} {\bibinfo {title} {{Numermcal simulation of time delay
  interferometry for NGO/eLISA}},\ }\href
  {https://doi.org/10.1088/0264-9381/30/6/065011} {\bibfield  {journal}
  {\bibinfo  {journal} {Class. Quant. Grav.}\ }\textbf {\bibinfo {volume}
  {30}},\ \bibinfo {pages} {065011} (\bibinfo {year} {2013}{\natexlab{a}})},\
  \Eprint {https://arxiv.org/abs/1204.2125} {arXiv:1204.2125 [gr-qc]}
  \BibitemShut {NoStop}%
\bibitem [{\citenamefont {Wang}\ and\ \citenamefont
  {Ni}(2013{\natexlab{b}})}]{Wang:2012te}%
  \BibitemOpen
  \bibfield  {author} {\bibinfo {author} {\bibfnamefont {G.}~\bibnamefont
  {Wang}}\ and\ \bibinfo {author} {\bibfnamefont {W.-T.}\ \bibnamefont {Ni}},\
  }\bibfield  {title} {\bibinfo {title} {{Orbit optimization for ASTROD-GW and
  its time delay interferometry with two arms using CGC ephemeris}},\ }\href
  {https://doi.org/10.1088/1674-1056/22/4/049501} {\bibfield  {journal}
  {\bibinfo  {journal} {Chin. Phys.}\ }\textbf {\bibinfo {volume} {B22}},\
  \bibinfo {pages} {049501} (\bibinfo {year} {2013}{\natexlab{b}})},\ \Eprint
  {https://arxiv.org/abs/1205.5175} {arXiv:1205.5175 [gr-qc]} \BibitemShut
  {NoStop}%
\bibitem [{\citenamefont {Dhurandhar}\ \emph {et~al.}(2013)\citenamefont
  {Dhurandhar}, \citenamefont {Ni},\ and\ \citenamefont
  {Wang}}]{Dhurandhar:2011ik}%
  \BibitemOpen
  \bibfield  {author} {\bibinfo {author} {\bibfnamefont {S.~V.}\ \bibnamefont
  {Dhurandhar}}, \bibinfo {author} {\bibfnamefont {W.~T.}\ \bibnamefont {Ni}},\
  and\ \bibinfo {author} {\bibfnamefont {G.}~\bibnamefont {Wang}},\ }\bibfield
  {title} {\bibinfo {title} {{Numerical simulation of time delay interferometry
  for a LISA-like mission with the simplification of having only one
  interferometer}},\ }\href {https://doi.org/10.1016/j.asr.2012.09.017}
  {\bibfield  {journal} {\bibinfo  {journal} {Adv. Space Res.}\ }\textbf
  {\bibinfo {volume} {51}},\ \bibinfo {pages} {198} (\bibinfo {year} {2013})},\
  \Eprint {https://arxiv.org/abs/1102.4965} {arXiv:1102.4965 [gr-qc]}
  \BibitemShut {NoStop}%
\bibitem [{\citenamefont {Wang}\ and\ \citenamefont {Ni}(2015)}]{Wang:2014cla}%
  \BibitemOpen
  \bibfield  {author} {\bibinfo {author} {\bibfnamefont {G.}~\bibnamefont
  {Wang}}\ and\ \bibinfo {author} {\bibfnamefont {W.-T.}\ \bibnamefont {Ni}},\
  }\bibfield  {title} {\bibinfo {title} {{Orbit optimization and time delay
  interferometry for inclined ASTROD-GW formation with half-year
  precession-period}},\ }\href {https://doi.org/10.1088/1674-1056/24/5/059501}
  {\bibfield  {journal} {\bibinfo  {journal} {Chin. Phys.}\ }\textbf {\bibinfo
  {volume} {B24}},\ \bibinfo {pages} {059501} (\bibinfo {year} {2015})},\
  \Eprint {https://arxiv.org/abs/1409.4162} {arXiv:1409.4162 [gr-qc]}
  \BibitemShut {NoStop}%
\bibitem [{\citenamefont {Wang}\ and\ \citenamefont {Ni}(2019)}]{Wang:2017aqq}%
  \BibitemOpen
  \bibfield  {author} {\bibinfo {author} {\bibfnamefont {G.}~\bibnamefont
  {Wang}}\ and\ \bibinfo {author} {\bibfnamefont {W.-T.}\ \bibnamefont {Ni}},\
  }\bibfield  {title} {\bibinfo {title} {{Numerical simulation of time delay
  interferometry for TAIJI and new LISA}},\ }\href
  {https://doi.org/10.1088/1674-4527/19/4/58} {\bibfield  {journal} {\bibinfo
  {journal} {Res. Astron. Astrophys.}\ }\textbf {\bibinfo {volume} {19}},\
  \bibinfo {pages} {058} (\bibinfo {year} {2019})},\ \Eprint
  {https://arxiv.org/abs/1707.09127} {arXiv:1707.09127 [astro-ph.IM]}
  \BibitemShut {NoStop}%
\bibitem [{\citenamefont {Wang}\ \emph {et~al.}(2020)\citenamefont {Wang},
  \citenamefont {Ni},\ and\ \citenamefont {Wu}}]{Wang:2019ipi}%
  \BibitemOpen
  \bibfield  {author} {\bibinfo {author} {\bibfnamefont {G.}~\bibnamefont
  {Wang}}, \bibinfo {author} {\bibfnamefont {W.-T.}\ \bibnamefont {Ni}},\ and\
  \bibinfo {author} {\bibfnamefont {A.-M.}\ \bibnamefont {Wu}},\ }\bibfield
  {title} {\bibinfo {title} {{Orbit design and thruster requirement for various
  constant-arm space mission concepts for gravitational-wave observation}},\
  }\bibfield  {journal} {\bibinfo  {journal} {Int. J. Mod. Phys.}\ }\href
  {https://doi.org/10.1142/S0218271819400066} {10.1142/S0218271819400066}
  (\bibinfo {year} {2020}),\ \Eprint {https://arxiv.org/abs/1908.05444}
  {arXiv:1908.05444 [gr-qc]} \BibitemShut {NoStop}%
\bibitem [{\citenamefont {Folkner}\ \emph {et~al.}(2014)\citenamefont
  {Folkner}, \citenamefont {Williams}, \citenamefont {Boggs}, \citenamefont
  {Park},\ and\ \citenamefont {Kuchynka}}]{DE430}%
  \BibitemOpen
  \bibfield  {author} {\bibinfo {author} {\bibfnamefont {W.~M.}\ \bibnamefont
  {Folkner}}, \bibinfo {author} {\bibfnamefont {J.~G.}\ \bibnamefont
  {Williams}}, \bibinfo {author} {\bibfnamefont {D.~H.}\ \bibnamefont {Boggs}},
  \bibinfo {author} {\bibfnamefont {R.~S.}\ \bibnamefont {Park}},\ and\
  \bibinfo {author} {\bibfnamefont {P.}~\bibnamefont {Kuchynka}},\ }\bibfield
  {title} {\bibinfo {title} {{The Planetary and Lunar Ephemerides DE430 and
  DE431}}} (\bibinfo {year} {2014}),\ \bibinfo {note} {iPN Progress Report
  {42-196}}\BibitemShut {NoStop}%
\bibitem [{\citenamefont {Bender}\ and\ \citenamefont
  {Welter}(2013)}]{Bender:2013mma}%
  \BibitemOpen
  \bibfield  {author} {\bibinfo {author} {\bibfnamefont {P.~L.}\ \bibnamefont
  {Bender}}\ and\ \bibinfo {author} {\bibfnamefont {G.~L.}\ \bibnamefont
  {Welter}},\ }\bibfield  {title} {\bibinfo {title} {{Possible Periodic Orbit
  Control Maneuvers for an eLISA Mission}},\ }\bibfield  {booktitle} {\emph
  {\bibinfo {booktitle} {{Proceedings, 9th International LISA Symposium (LISA
  2012): Paris, France, May 21-25, 2012}}},\ }\href@noop {} {\bibfield
  {journal} {\bibinfo  {journal} {ASP Conf. Ser.}\ }\textbf {\bibinfo {volume}
  {467}},\ \bibinfo {pages} {203} (\bibinfo {year} {2013})}\BibitemShut
  {NoStop}%
\bibitem [{\citenamefont {Halloin}(2017)}]{Halloin:2017tzi}%
  \BibitemOpen
  \bibfield  {author} {\bibinfo {author} {\bibfnamefont {H.}~\bibnamefont
  {Halloin}},\ }\bibfield  {title} {\bibinfo {title} {{Optimizing orbits for
  (e)LISA}},\ }\bibfield  {booktitle} {\emph {\bibinfo {booktitle}
  {{Proceedings, 11th International LISA Symposium: Zurich, Switzerland,
  September 5-9, 2016}}},\ }\href
  {https://doi.org/10.1088/1742-6596/840/1/012048} {\bibfield  {journal}
  {\bibinfo  {journal} {J. Phys. Conf. Ser.}\ }\textbf {\bibinfo {volume}
  {840}},\ \bibinfo {pages} {012048} (\bibinfo {year} {2017})}\BibitemShut
  {NoStop}%
\bibitem [{\citenamefont {{Armstrong}}\ \emph {et~al.}(1999)\citenamefont
  {{Armstrong}}, \citenamefont {{Estabrook}},\ and\ \citenamefont
  {{Tinto}}}]{1999ApJ...527..814A}%
  \BibitemOpen
  \bibfield  {author} {\bibinfo {author} {\bibfnamefont {J.~W.}\ \bibnamefont
  {{Armstrong}}}, \bibinfo {author} {\bibfnamefont {F.~B.}\ \bibnamefont
  {{Estabrook}}},\ and\ \bibinfo {author} {\bibfnamefont {M.}~\bibnamefont
  {{Tinto}}},\ }\bibfield  {title} {\bibinfo {title} {{Time-Delay
  Interferometry for Space-based Gravitational Wave Searches}},\ }\href
  {https://doi.org/10.1086/308110} {\bibfield  {journal} {\bibinfo  {journal}
  {\apj}\ }\textbf {\bibinfo {volume} {527}},\ \bibinfo {pages} {814} (\bibinfo
  {year} {1999})}\BibitemShut {NoStop}%
\bibitem [{\citenamefont {{Estabrook}}\ \emph {et~al.}(2000)\citenamefont
  {{Estabrook}}, \citenamefont {{Tinto}},\ and\ \citenamefont
  {{Armstrong}}}]{2000PhRvD..62d2002E}%
  \BibitemOpen
  \bibfield  {author} {\bibinfo {author} {\bibfnamefont {F.~B.}\ \bibnamefont
  {{Estabrook}}}, \bibinfo {author} {\bibfnamefont {M.}~\bibnamefont
  {{Tinto}}},\ and\ \bibinfo {author} {\bibfnamefont {J.~W.}\ \bibnamefont
  {{Armstrong}}},\ }\bibfield  {title} {\bibinfo {title} {{Time-delay analysis
  of LISA gravitational wave data: Elimination of spacecraft motion effects}},\
  }\href {https://doi.org/10.1103/PhysRevD.62.042002} {\bibfield  {journal}
  {\bibinfo  {journal} {\prd}\ }\textbf {\bibinfo {volume} {62}},\ \bibinfo
  {eid} {042002} (\bibinfo {year} {2000})}\BibitemShut {NoStop}%
\bibitem [{\citenamefont {{Armstrong}}\ \emph {et~al.}(2001)\citenamefont
  {{Armstrong}}, \citenamefont {{Estabrook}},\ and\ \citenamefont
  {{Tinto}}}]{2001CQGra..18.4059A}%
  \BibitemOpen
  \bibfield  {author} {\bibinfo {author} {\bibfnamefont {J.~W.}\ \bibnamefont
  {{Armstrong}}}, \bibinfo {author} {\bibfnamefont {F.~B.}\ \bibnamefont
  {{Estabrook}}},\ and\ \bibinfo {author} {\bibfnamefont {M.}~\bibnamefont
  {{Tinto}}},\ }\bibfield  {title} {\bibinfo {title} {{Sensitivities of
  alternate LISA configurations}},\ }\href
  {https://doi.org/10.1088/0264-9381/18/19/313} {\bibfield  {journal} {\bibinfo
   {journal} {Classical and Quantum Gravity}\ }\textbf {\bibinfo {volume}
  {18}},\ \bibinfo {pages} {4059} (\bibinfo {year} {2001})}\BibitemShut
  {NoStop}%
\bibitem [{\citenamefont {Prince}\ \emph {et~al.}(2002)\citenamefont {Prince},
  \citenamefont {Tinto}, \citenamefont {Larson},\ and\ \citenamefont
  {Armstrong}}]{Prince:2002hp}%
  \BibitemOpen
  \bibfield  {author} {\bibinfo {author} {\bibfnamefont {T.~A.}\ \bibnamefont
  {Prince}}, \bibinfo {author} {\bibfnamefont {M.}~\bibnamefont {Tinto}},
  \bibinfo {author} {\bibfnamefont {S.~L.}\ \bibnamefont {Larson}},\ and\
  \bibinfo {author} {\bibfnamefont {J.~W.}\ \bibnamefont {Armstrong}},\
  }\bibfield  {title} {\bibinfo {title} {{The LISA optimal sensitivity}},\
  }\href {https://doi.org/10.1103/PhysRevD.66.122002} {\bibfield  {journal}
  {\bibinfo  {journal} {Phys. Rev.}\ }\textbf {\bibinfo {volume} {D66}},\
  \bibinfo {pages} {122002} (\bibinfo {year} {2002})},\ \Eprint
  {https://arxiv.org/abs/gr-qc/0209039} {arXiv:gr-qc/0209039 [gr-qc]}
  \BibitemShut {NoStop}%
\bibitem [{\citenamefont {Dhurandhar}\ \emph {et~al.}(2002)\citenamefont
  {Dhurandhar}, \citenamefont {Nayak},\ and\ \citenamefont
  {Vinet}}]{Dhurandhar:2002zcl}%
  \BibitemOpen
  \bibfield  {author} {\bibinfo {author} {\bibfnamefont {S.~V.}\ \bibnamefont
  {Dhurandhar}}, \bibinfo {author} {\bibfnamefont {K.~R.}\ \bibnamefont
  {Nayak}},\ and\ \bibinfo {author} {\bibfnamefont {J.~Y.}\ \bibnamefont
  {Vinet}},\ }\bibfield  {title} {\bibinfo {title} {{Algebraic approach to
  time-delay data analysis for LISA}},\ }\href
  {https://doi.org/10.1103/PhysRevD.65.102002} {\bibfield  {journal} {\bibinfo
  {journal} {\prd}\ }\textbf {\bibinfo {volume} {65}},\ \bibinfo {pages}
  {102002} (\bibinfo {year} {2002})},\ \Eprint
  {https://arxiv.org/abs/gr-qc/0112059} {arXiv:gr-qc/0112059 [gr-qc]}
  \BibitemShut {NoStop}%
\bibitem [{\citenamefont {Cornish}\ and\ \citenamefont
  {Hellings}(2003)}]{Cornish:2003tz}%
  \BibitemOpen
  \bibfield  {author} {\bibinfo {author} {\bibfnamefont {N.~J.}\ \bibnamefont
  {Cornish}}\ and\ \bibinfo {author} {\bibfnamefont {R.~W.}\ \bibnamefont
  {Hellings}},\ }\bibfield  {title} {\bibinfo {title} {{The Effects of orbital
  motion on LISA time delay interferometry}},\ }\href
  {https://doi.org/10.1088/0264-9381/20/22/009} {\bibfield  {journal} {\bibinfo
   {journal} {Class. Quant. Grav.}\ }\textbf {\bibinfo {volume} {20}},\
  \bibinfo {pages} {4851} (\bibinfo {year} {2003})},\ \Eprint
  {https://arxiv.org/abs/gr-qc/0306096} {arXiv:gr-qc/0306096 [gr-qc]}
  \BibitemShut {NoStop}%
\bibitem [{\citenamefont {{Tinto}}\ \emph {et~al.}(2003)\citenamefont
  {{Tinto}}, \citenamefont {{Shaddock}}, \citenamefont {{Sylvestre}},\ and\
  \citenamefont {{Armstrong}}}]{2003PhRvD..67l2003T}%
  \BibitemOpen
  \bibfield  {author} {\bibinfo {author} {\bibfnamefont {M.}~\bibnamefont
  {{Tinto}}}, \bibinfo {author} {\bibfnamefont {D.~A.}\ \bibnamefont
  {{Shaddock}}}, \bibinfo {author} {\bibfnamefont {J.}~\bibnamefont
  {{Sylvestre}}},\ and\ \bibinfo {author} {\bibfnamefont {J.~W.}\ \bibnamefont
  {{Armstrong}}},\ }\bibfield  {title} {\bibinfo {title} {{Implementation of
  time-delay interferometry for LISA}},\ }\href
  {https://doi.org/10.1103/PhysRevD.67.122003} {\bibfield  {journal} {\bibinfo
  {journal} {\prd}\ }\textbf {\bibinfo {volume} {67}},\ \bibinfo {eid} {122003}
  (\bibinfo {year} {2003})},\ \Eprint {https://arxiv.org/abs/gr-qc/0303013}
  {arXiv:gr-qc/0303013 [gr-qc]} \BibitemShut {NoStop}%
\bibitem [{\citenamefont {Vallisneri}(2005{\natexlab{b}})}]{Vallisneri:2005ji}%
  \BibitemOpen
  \bibfield  {author} {\bibinfo {author} {\bibfnamefont {M.}~\bibnamefont
  {Vallisneri}},\ }\bibfield  {title} {\bibinfo {title} {{Geometric time delay
  interferometry}},\ }\href {https://doi.org/10.1103/PhysRevD.76.109903,
  10.1103/PhysRevD.72.042003} {\bibfield  {journal} {\bibinfo  {journal} {Phys.
  Rev.}\ }\textbf {\bibinfo {volume} {D72}},\ \bibinfo {pages} {042003}
  (\bibinfo {year} {2005}{\natexlab{b}})},\ \bibinfo {note} {[Erratum: Phys.
  Rev.D76,109903(2007)]},\ \Eprint {https://arxiv.org/abs/gr-qc/0504145}
  {arXiv:gr-qc/0504145 [gr-qc]} \BibitemShut {NoStop}%
\bibitem [{\citenamefont {Tinto}\ and\ \citenamefont
  {Dhurandhar}(2014)}]{Tinto:2014lxa}%
  \BibitemOpen
  \bibfield  {author} {\bibinfo {author} {\bibfnamefont {M.}~\bibnamefont
  {Tinto}}\ and\ \bibinfo {author} {\bibfnamefont {S.~V.}\ \bibnamefont
  {Dhurandhar}},\ }\bibfield  {title} {\bibinfo {title} {{Time-Delay
  Interferometry}},\ }\href {https://doi.org/10.12942/lrr-2014-6} {\bibfield
  {journal} {\bibinfo  {journal} {Living Rev. Rel.}\ }\textbf {\bibinfo
  {volume} {17}},\ \bibinfo {pages} {6} (\bibinfo {year} {2014})}\BibitemShut
  {NoStop}%
\bibitem [{\citenamefont {Vallisneri}\ \emph {et~al.}(2008)\citenamefont
  {Vallisneri}, \citenamefont {Crowder},\ and\ \citenamefont
  {Tinto}}]{Vallisneri:2007xa}%
  \BibitemOpen
  \bibfield  {author} {\bibinfo {author} {\bibfnamefont {M.}~\bibnamefont
  {Vallisneri}}, \bibinfo {author} {\bibfnamefont {J.}~\bibnamefont
  {Crowder}},\ and\ \bibinfo {author} {\bibfnamefont {M.}~\bibnamefont
  {Tinto}},\ }\bibfield  {title} {\bibinfo {title} {{Sensitivity and
  parameter-estimation precision for alternate LISA configurations}},\ }\href
  {https://doi.org/10.1088/0264-9381/25/6/065005} {\bibfield  {journal}
  {\bibinfo  {journal} {Class. Quant. Grav.}\ }\textbf {\bibinfo {volume}
  {25}},\ \bibinfo {pages} {065005} (\bibinfo {year} {2008})},\ \Eprint
  {https://arxiv.org/abs/0710.4369} {arXiv:0710.4369 [gr-qc]} \BibitemShut
  {NoStop}%
\bibitem [{\citenamefont {{Estabrook}}\ and\ \citenamefont
  {{Wahlquist}}(1975)}]{1975GReGr...6..439E}%
  \BibitemOpen
  \bibfield  {author} {\bibinfo {author} {\bibfnamefont {F.~B.}\ \bibnamefont
  {{Estabrook}}}\ and\ \bibinfo {author} {\bibfnamefont {H.~D.}\ \bibnamefont
  {{Wahlquist}}},\ }\bibfield  {title} {\bibinfo {title} {{Response of Doppler
  spacecraft tracking to gravitational radiation.}},\ }\href
  {https://doi.org/10.1007/BF00762449} {\bibfield  {journal} {\bibinfo
  {journal} {General Relativity and Gravitation}\ }\textbf {\bibinfo {volume}
  {6}},\ \bibinfo {pages} {439} (\bibinfo {year} {1975})}\BibitemShut {NoStop}%
\bibitem [{\citenamefont {{Wang}}\ \emph {et~al.}(tion)\citenamefont {{Wang}},
  \citenamefont {{Ni}},\ and\ \citenamefont {{Han}}}]{Wang:2020ongoing}%
  \BibitemOpen
  \bibfield  {author} {\bibinfo {author} {\bibfnamefont {G.}~\bibnamefont
  {{Wang}}}, \bibinfo {author} {\bibfnamefont {W.-T.}\ \bibnamefont {{Ni}}},\
  and\ \bibinfo {author} {\bibfnamefont {W.-B.}\ \bibnamefont {{Han}}},\
  }\href@noop {} {\  (\bibinfo {year} {in preparation})}\BibitemShut {NoStop}%
\bibitem [{\citenamefont {Colpi}(2014)}]{Colpi:2014poa}%
  \BibitemOpen
  \bibfield  {author} {\bibinfo {author} {\bibfnamefont {M.}~\bibnamefont
  {Colpi}},\ }\bibfield  {title} {\bibinfo {title} {{Massive binary black holes
  in galactic nuclei and their path to coalescence}},\ }\href
  {https://doi.org/10.1007/s11214-014-0067-1} {\bibfield  {journal} {\bibinfo
  {journal} {Space Sci. Rev.}\ }\textbf {\bibinfo {volume} {183}},\ \bibinfo
  {pages} {189} (\bibinfo {year} {2014})},\ \Eprint
  {https://arxiv.org/abs/1407.3102} {arXiv:1407.3102 [astro-ph.GA]}
  \BibitemShut {NoStop}%
\bibitem [{\citenamefont {Ade}\ \emph {et~al.}(2016)\citenamefont {Ade},
  \citenamefont {Aghanim}, \citenamefont {Arnaud},\ and\ \citenamefont
  {et~al}}]{Ade:2015xua}%
  \BibitemOpen
  \bibfield  {author} {\bibinfo {author} {\bibfnamefont {P.~A.~R.}\
  \bibnamefont {Ade}}, \bibinfo {author} {\bibfnamefont {N.}~\bibnamefont
  {Aghanim}}, \bibinfo {author} {\bibfnamefont {M.}~\bibnamefont {Arnaud}},\
  and\ \bibinfo {author} {\bibnamefont {et~al}} (\bibinfo {collaboration}
  {{Planck Collaboration}}),\ }\bibfield  {title} {\bibinfo {title} {{Planck
  2015 results. XIII. Cosmological parameters}},\ }\href
  {https://doi.org/10.1051/0004-6361/201525830} {\bibfield  {journal} {\bibinfo
   {journal} {Astron. Astrophys.}\ }\textbf {\bibinfo {volume} {594}},\
  \bibinfo {pages} {A13} (\bibinfo {year} {2016})},\ \Eprint
  {https://arxiv.org/abs/1502.01589} {arXiv:1502.01589 [astro-ph.CO]}
  \BibitemShut {NoStop}%
\bibitem [{\citenamefont {{Khan}}\ \emph {et~al.}(2016)\citenamefont {{Khan}},
  \citenamefont {{Husa}}, \citenamefont {{Hannam}}, \citenamefont {{Ohme}},
  \citenamefont {{P{\"u}rrer}}, \citenamefont {{Forteza}},\ and\ \citenamefont
  {{Boh{\'e}}}}]{Khan:2015jqa}%
  \BibitemOpen
  \bibfield  {author} {\bibinfo {author} {\bibfnamefont {S.}~\bibnamefont
  {{Khan}}}, \bibinfo {author} {\bibfnamefont {S.}~\bibnamefont {{Husa}}},
  \bibinfo {author} {\bibfnamefont {M.}~\bibnamefont {{Hannam}}}, \bibinfo
  {author} {\bibfnamefont {F.}~\bibnamefont {{Ohme}}}, \bibinfo {author}
  {\bibfnamefont {M.}~\bibnamefont {{P{\"u}rrer}}}, \bibinfo {author}
  {\bibfnamefont {X.~J.}\ \bibnamefont {{Forteza}}},\ and\ \bibinfo {author}
  {\bibfnamefont {A.}~\bibnamefont {{Boh{\'e}}}},\ }\bibfield  {title}
  {\bibinfo {title} {{Frequency-domain gravitational waves from nonprecessing
  black-hole binaries. II. A phenomenological model for the advanced detector
  era}},\ }\href {https://doi.org/10.1103/PhysRevD.93.044007} {\bibfield
  {journal} {\bibinfo  {journal} {\prd}\ }\textbf {\bibinfo {volume} {93}},\
  \bibinfo {eid} {044007} (\bibinfo {year} {2016})},\ \Eprint
  {https://arxiv.org/abs/1508.07253} {arXiv:1508.07253 [gr-qc]} \BibitemShut
  {NoStop}%
\bibitem [{\citenamefont {{LIGO Scientific Collaboration}}(2018)}]{lalsuite}%
  \BibitemOpen
  \bibfield  {author} {\bibinfo {author} {\bibnamefont {{LIGO Scientific
  Collaboration}}},\ }\href {https://doi.org/10.7935/GT1W-FZ16} {\bibinfo
  {title} {{LIGO} {A}lgorithm {L}ibrary - {LALS}uite}},\ \bibinfo
  {howpublished} {free software (GPL)} (\bibinfo {year} {2018})\BibitemShut
  {NoStop}%
\bibitem [{\citenamefont {{Nitz}}\ \emph
  {et~al.}(2020{\natexlab{b}})\citenamefont {{Nitz}}, \citenamefont {{Harry}},
  \citenamefont {{Brown}}, \citenamefont {{Biwer}}, \citenamefont {{Willis}},
  \citenamefont {{Dal Canton}}, \citenamefont {{Capano}}, \citenamefont
  {{Pekowsky}}, \citenamefont {{Dent}}, \citenamefont {{Williamson}},
  \citenamefont {{De}}, \citenamefont {{Davies}}, \citenamefont {{Cabero}},
  \citenamefont {{Macleod}}, \citenamefont {{Machenschalk}}, \citenamefont
  {{Reyes}}, \citenamefont {{Kumar}}, \citenamefont {{Massinger}},
  \citenamefont {{Pannarale}}, \citenamefont {{dfinstad}}, \citenamefont
  {{T{\'a}pai}}, \citenamefont {{Fairhurst}}, \citenamefont {{Khan}},
  \citenamefont {{Singer}}, \citenamefont {{Nielsen}}, \citenamefont {{Kumar}},
  \citenamefont {{shasvath}}, \citenamefont {{idorrington92}}, \citenamefont
  {{Gabbard}},\ and\ \citenamefont {{Varsha Uday Gadre}}}]{pycbc}%
  \BibitemOpen
  \bibfield  {author} {\bibinfo {author} {\bibfnamefont {A.}~\bibnamefont
  {{Nitz}}}, \bibinfo {author} {\bibfnamefont {I.}~\bibnamefont {{Harry}}},
  \bibinfo {author} {\bibfnamefont {D.}~\bibnamefont {{Brown}}}, \bibinfo
  {author} {\bibfnamefont {C.~M.}\ \bibnamefont {{Biwer}}}, \bibinfo {author}
  {\bibfnamefont {J.}~\bibnamefont {{Willis}}}, \bibinfo {author}
  {\bibfnamefont {T.}~\bibnamefont {{Dal Canton}}}, \bibinfo {author}
  {\bibfnamefont {C.}~\bibnamefont {{Capano}}}, \bibinfo {author}
  {\bibfnamefont {L.}~\bibnamefont {{Pekowsky}}}, \bibinfo {author}
  {\bibfnamefont {T.}~\bibnamefont {{Dent}}}, \bibinfo {author} {\bibfnamefont
  {A.~R.}\ \bibnamefont {{Williamson}}}, \bibinfo {author} {\bibfnamefont
  {S.}~\bibnamefont {{De}}}, \bibinfo {author} {\bibfnamefont {G.}~\bibnamefont
  {{Davies}}}, \bibinfo {author} {\bibfnamefont {M.}~\bibnamefont {{Cabero}}},
  \bibinfo {author} {\bibfnamefont {D.}~\bibnamefont {{Macleod}}}, \bibinfo
  {author} {\bibfnamefont {B.}~\bibnamefont {{Machenschalk}}}, \bibinfo
  {author} {\bibfnamefont {S.}~\bibnamefont {{Reyes}}}, \bibinfo {author}
  {\bibfnamefont {P.}~\bibnamefont {{Kumar}}}, \bibinfo {author} {\bibfnamefont
  {T.}~\bibnamefont {{Massinger}}}, \bibinfo {author} {\bibfnamefont
  {F.}~\bibnamefont {{Pannarale}}}, \bibinfo {author} {\bibnamefont
  {{dfinstad}}}, \bibinfo {author} {\bibfnamefont {M.}~\bibnamefont
  {{T{\'a}pai}}}, \bibinfo {author} {\bibfnamefont {S.}~\bibnamefont
  {{Fairhurst}}}, \bibinfo {author} {\bibfnamefont {S.}~\bibnamefont {{Khan}}},
  \bibinfo {author} {\bibfnamefont {L.}~\bibnamefont {{Singer}}}, \bibinfo
  {author} {\bibfnamefont {A.}~\bibnamefont {{Nielsen}}}, \bibinfo {author}
  {\bibfnamefont {S.}~\bibnamefont {{Kumar}}}, \bibinfo {author} {\bibnamefont
  {{shasvath}}}, \bibinfo {author} {\bibnamefont {{idorrington92}}}, \bibinfo
  {author} {\bibfnamefont {H.}~\bibnamefont {{Gabbard}}},\ and\ \bibinfo
  {author} {\bibfnamefont {B.}~\bibnamefont {{Varsha Uday Gadre}}},\ }\href
  {https://doi.org/10.5281/zenodo.3596447} {\bibinfo {title} {{gwastro/pycbc
  v1.15.3}}} (\bibinfo {year} {2020}{\natexlab{b}})\BibitemShut {NoStop}%
\bibitem [{\citenamefont {{Cutler}}\ and\ \citenamefont
  {{Flanagan}}(1994)}]{1994PhRvD..49.2658C}%
  \BibitemOpen
  \bibfield  {author} {\bibinfo {author} {\bibfnamefont {C.}~\bibnamefont
  {{Cutler}}}\ and\ \bibinfo {author} {\bibfnamefont {{\'E}.~E.}\ \bibnamefont
  {{Flanagan}}},\ }\bibfield  {title} {\bibinfo {title} {{Gravitational waves
  from merging compact binaries: How accurately can one extract the binary's
  parameters from the inspiral waveform?}},\ }\href
  {https://doi.org/10.1103/PhysRevD.49.2658} {\bibfield  {journal} {\bibinfo
  {journal} {\prd}\ }\textbf {\bibinfo {volume} {49}},\ \bibinfo {pages} {2658}
  (\bibinfo {year} {1994})},\ \Eprint {https://arxiv.org/abs/gr-qc/9402014}
  {arXiv:gr-qc/9402014 [gr-qc]} \BibitemShut {NoStop}%
\bibitem [{\citenamefont {Vallisneri}(2008)}]{Vallisneri:2007ev}%
  \BibitemOpen
  \bibfield  {author} {\bibinfo {author} {\bibfnamefont {M.}~\bibnamefont
  {Vallisneri}},\ }\bibfield  {title} {\bibinfo {title} {{Use and abuse of the
  Fisher information matrix in the assessment of gravitational-wave
  parameter-estimation prospects}},\ }\href
  {https://doi.org/10.1103/PhysRevD.77.042001} {\bibfield  {journal} {\bibinfo
  {journal} {Phys. Rev.}\ }\textbf {\bibinfo {volume} {D77}},\ \bibinfo {pages}
  {042001} (\bibinfo {year} {2008})},\ \Eprint
  {https://arxiv.org/abs/gr-qc/0703086} {arXiv:gr-qc/0703086 [GR-QC]}
  \BibitemShut {NoStop}%
\bibitem [{\citenamefont {Kuns}\ \emph {et~al.}(2019)\citenamefont {Kuns},
  \citenamefont {Yu}, \citenamefont {Chen},\ and\ \citenamefont
  {Adhikari}}]{Kuns:2019upi}%
  \BibitemOpen
  \bibfield  {author} {\bibinfo {author} {\bibfnamefont {K.~A.}\ \bibnamefont
  {Kuns}}, \bibinfo {author} {\bibfnamefont {H.}~\bibnamefont {Yu}}, \bibinfo
  {author} {\bibfnamefont {Y.}~\bibnamefont {Chen}},\ and\ \bibinfo {author}
  {\bibfnamefont {R.~X.}\ \bibnamefont {Adhikari}},\ }\bibfield  {title}
  {\bibinfo {title} {{Astrophysics and cosmology with a deci-hertz
  gravitational-wave detector: TianGO}},\ }\href@noop {} {\  (\bibinfo {year}
  {2019})},\ \Eprint {https://arxiv.org/abs/1908.06004} {arXiv:1908.06004
  [gr-qc]} \BibitemShut {NoStop}%
\bibitem [{\citenamefont {Nissanke}\ \emph {et~al.}(2012)\citenamefont
  {Nissanke}, \citenamefont {Vallisneri}, \citenamefont {Nelemans},\ and\
  \citenamefont {Prince}}]{Nissanke:2012eh}%
  \BibitemOpen
  \bibfield  {author} {\bibinfo {author} {\bibfnamefont {S.}~\bibnamefont
  {Nissanke}}, \bibinfo {author} {\bibfnamefont {M.}~\bibnamefont
  {Vallisneri}}, \bibinfo {author} {\bibfnamefont {G.}~\bibnamefont
  {Nelemans}},\ and\ \bibinfo {author} {\bibfnamefont {T.~A.}\ \bibnamefont
  {Prince}},\ }\bibfield  {title} {\bibinfo {title} {{Gravitational-wave
  emission from compact Galactic binaries}},\ }\href
  {https://doi.org/10.1088/0004-637X/758/2/131} {\bibfield  {journal} {\bibinfo
   {journal} {Astrophys. J.}\ }\textbf {\bibinfo {volume} {758}},\ \bibinfo
  {pages} {131} (\bibinfo {year} {2012})},\ \Eprint
  {https://arxiv.org/abs/1201.4613} {arXiv:1201.4613 [astro-ph.GA]}
  \BibitemShut {NoStop}%
\bibitem [{\citenamefont {Maggiore}(2007)}]{Maggiore:2007}%
  \BibitemOpen
  \bibfield  {author} {\bibinfo {author} {\bibfnamefont {M.}~\bibnamefont
  {Maggiore}},\ }\href@noop {} {\emph {\bibinfo {title} {{Gravitational Waves.
  Vol. 1: Theory and Experiments}}}}\ (\bibinfo  {publisher} {Oxford University
  Press},\ \bibinfo {year} {2007})\BibitemShut {NoStop}%
\bibitem [{\citenamefont {{London}}\ \emph {et~al.}(2018)\citenamefont
  {{London}}, \citenamefont {{Khan}}, \citenamefont {{Fauchon-Jones}},
  \citenamefont {{Garc{\'\i}a}}, \citenamefont {{Hannam}}, \citenamefont
  {{Husa}}, \citenamefont {{Jim{\'e}nez-Forteza}}, \citenamefont
  {{Kalaghatgi}}, \citenamefont {{Ohme}},\ and\ \citenamefont
  {{Pannarale}}}]{London:2017bcn}%
  \BibitemOpen
  \bibfield  {author} {\bibinfo {author} {\bibfnamefont {L.}~\bibnamefont
  {{London}}}, \bibinfo {author} {\bibfnamefont {S.}~\bibnamefont {{Khan}}},
  \bibinfo {author} {\bibfnamefont {E.}~\bibnamefont {{Fauchon-Jones}}},
  \bibinfo {author} {\bibfnamefont {C.}~\bibnamefont {{Garc{\'\i}a}}}, \bibinfo
  {author} {\bibfnamefont {M.}~\bibnamefont {{Hannam}}}, \bibinfo {author}
  {\bibfnamefont {S.}~\bibnamefont {{Husa}}}, \bibinfo {author} {\bibfnamefont
  {X.}~\bibnamefont {{Jim{\'e}nez-Forteza}}}, \bibinfo {author} {\bibfnamefont
  {C.}~\bibnamefont {{Kalaghatgi}}}, \bibinfo {author} {\bibfnamefont
  {F.}~\bibnamefont {{Ohme}}},\ and\ \bibinfo {author} {\bibfnamefont
  {F.}~\bibnamefont {{Pannarale}}},\ }\bibfield  {title} {\bibinfo {title}
  {{First Higher-Multipole Model of Gravitational Waves from Spinning and
  Coalescing Black-Hole Binaries}},\ }\href
  {https://doi.org/10.1103/PhysRevLett.120.161102} {\bibfield  {journal}
  {\bibinfo  {journal} {\prl}\ }\textbf {\bibinfo {volume} {120}},\ \bibinfo
  {eid} {161102} (\bibinfo {year} {2018})},\ \Eprint
  {https://arxiv.org/abs/1708.00404} {arXiv:1708.00404 [gr-qc]} \BibitemShut
  {NoStop}%
\bibitem [{\citenamefont {{Brumberg}}(1991)}]{Brumberg1991}%
  \BibitemOpen
  \bibfield  {author} {\bibinfo {author} {\bibfnamefont {V.~A.}\ \bibnamefont
  {{Brumberg}}},\ }\href@noop {} {\emph {\bibinfo {title} {{Essential
  relativistic celestial mechanics.}}}}\ (\bibinfo  {publisher} {CRC Press},\
  \bibinfo {year} {1991})\BibitemShut {NoStop}%
\end{thebibliography}%

\end{document}